%% file: paper.tex
\documentstyle[10pt,epsf,epsfig,graphics,rotating,amssymb,dp_delphititle,my_hack]{dp_delphi}
\makeindex
\input{references.tex}    

\begin{document}
\makeatletter
\input{coll.sty}
\makeatother

\begin{titlepage}
\pagenumbering{roman}

\CERNpreprint{\DpPaperGroup}{\DpPaperRef}   
\date{{\small\DpDate}}                      
\title{\DpTitle}                            
\address{\DpAuthors}                        

\begin{shortabs}                            
\noindent
\input{abstract.tex}    
\end{shortabs}

\vfill

\begin{center}
\DpSubmit  \ \\      
\DpComment \ \\
\DpEMail   \ \\
\end{center}

\vfill
\clearpage

\headsep 10.0pt

\addtolength{\textheight}{10mm}
\addtolength{\footskip}{-5mm}
\begingroup
%
\newcommand{\DpName}[2]{\hbox{#1$^{\ref{#2}}$},\hfill}
\newcommand{\DpNameTwo}[3]{\hbox{#1$^{\ref{#2},\ref{#3}}$},\hfill}
\newcommand{\DpNameThree}[4]{\hbox{#1$^{\ref{#2},\ref{#3},\ref{#4}}$},\hfill}
\newskip\Bigfill \Bigfill = 0pt plus 1000fill
\newcommand{\DpNameLast}[2]{\hbox{#1$^{\ref{#2}}$}\hspace{\Bigfill}}
\small
\noindent
\input{authors.tex} 
\normalsize
\endgroup

\input{instituts.tex}   
\addtolength{\textheight}{-10mm}
\addtolength{\footskip}{5mm}
\clearpage

\headsep 30.0pt
\end{titlepage}

%
\pagenumbering{arabic}                  
\setcounter{footnote}{0}                %
\large
\input{document.tex}    

\end{document}

%% file: references.tex
\def\DpPaperGroup{EP} 
\def\DpPaperRef{99-133} 
\def\DpDate{26 January 2000 \\ revised} 
\def\DpAuthors{DELPHI Collaboration} 
\def\DpSubmit{(Submitted to E. Phys. J. C)} 
\def\DpTitle{{Consistent Measurements of \boldmath $\alpha_s$ 
              from Precise Oriented Event Shape Distributions}} 
\def\DpComment{} 
\def\DpEMail{} 

%% file: abstract.tex
%
\noindent

An updated analysis using about 1.5 million events recorded at $ \sqrt{s} = 
M_Z $ with the DELPHI detector in 1994 is presented. Eighteen infrared and 
collinear safe event shape observables are measured as a function of the 
polar angle of the thrust axis. The data are compared to theoretical 
calculations in $ {\cal O} ( \alpha_s^2 )$ including the event orientation. A 
combined fit of $\alpha_s$ and of the renormalization scale $x_{\mu}$ in 
$\cal O$($\alpha_s^2$) yields an excellent description of the high statistics 
data. \\ 

The weighted average from 18 observables including quark mass effects and 
correlations is $ \rm \alpha_s(M_Z^2) = 0.1174 \pm 0.0026 $. The final 
result, derived from the jet cone energy fraction, the observable with 
the smallest theoretical and experimental uncertainty, is 

\begin{center} 
 $ \rm \alpha_s(M_Z^2) = 0.1180 \pm 0.0006 (exp.) \pm 0.0013 (hadr.) 
                                \pm 0.0008 (scale) \pm 0.0007 (mass)$. 
\end{center}

Further studies include an $\alpha_s$ determination using theoretical 
predictions in the next-to-leading log approximation (NLLA), 
matched NLLA and $\cal O$($\alpha_s^2$) predictions as well as theoretically 
motivated optimized scale setting methods. The influence of higher order 
contributions was also investigated by using the method of Pad\'{e} approximants. 
Average $\alpha_s$ values derived from the different approaches are in 
good agreement.

%% file: authors.tex
\DpName{P.Abreu}{LIP}
\DpName{W.Adam}{VIENNA}
\DpName{T.Adye}{RAL}
\DpName{P.Adzic}{DEMOKRITOS}
\DpName{Z.Albrecht}{KARLSRUHE}
\DpName{T.Alderweireld}{AIM}
\DpName{G.D.Alekseev}{JINR}
\DpName{R.Alemany}{VALENCIA}
\DpName{T.Allmendinger}{KARLSRUHE}
\DpName{P.P.Allport}{LIVERPOOL}
% ...

%% file: instituts.tex
\titlefoot{Department of Physics and Astronomy, Iowa State
     University, Ames IA 50011-3160, USA
    \label{AMES}}
\titlefoot{Physics Department, Univ. Instelling Antwerpen,
     Universiteitsplein 1, BE-2610 Wilrijk, Belgium \\
     \indent~~and IIHE, ULB-VUB,
     Pleinlaan 2, BE-1050 Brussels, Belgium \\
     \indent~~and Facult\'e des Sciences,
     Univ. de l'Etat Mons, Av. Maistriau 19, BE-7000 Mons, Belgium
    \label{AIM}}
\titlefoot{Physics Laboratory, University of Athens, Solonos Str.
     104, GR-10680 Athens, Greece
    \label{ATHENS}}
\titlefoot{Department of Physics, University of Bergen,
     All\'egaten 55, NO-5007 Bergen, Norway
    \label{BERGEN}}
\titlefoot{Dipartimento di Fisica, Universit\`a di Bologna and INFN,
     Via Irnerio 46, IT-40126 Bologna, Italy
    \label{BOLOGNA}}
\titlefoot{Centro Brasileiro de Pesquisas F\'{\i}sicas, rua Xavier Sigaud 150,
     BR-22290 Rio de Janeiro, Brazil \\
     \indent~~and Depto. de F\'{\i}sica, Pont. Univ. Cat\'olica,
     C.P. 38071 BR-22453 Rio de Janeiro, Brazil \\
     \indent~~and Inst. de F\'{\i}sica, Univ. Estadual do Rio de Janeiro,
     rua S\~{a}o Francisco Xavier 524, Rio de Janeiro, Brazil
    \label{BRASIL}}
\titlefoot{Comenius University, Faculty of Mathematics and Physics,
     Mlynska Dolina, SK-84215 Bratislava, Slovakia
    \label{BRATISLAVA}}
\titlefoot{Coll\`ege de France, Lab. de Physique Corpusculaire, IN2P3-CNRS,
     FR-75231 Paris Cedex 05, France
    \label{CDF}}
\titlefoot{CERN, CH-1211 Geneva 23, Switzerland
    \label{CERN}}
\titlefoot{Institut de Recherches Subatomiques, IN2P3 - CNRS/ULP - BP20,
     FR-67037 Strasbourg Cedex, France
    \label{CRN}}
\titlefoot{Institute of Nuclear Physics, N.C.S.R. Demokritos,
     P.O. Box 60228, GR-15310 Athens, Greece
    \label{DEMOKRITOS}}
\titlefoot{INFN, Largo E. Fermi 2, IT-50125 Firenze, Italy 
    \label{FIRENZE}}
\titlefoot{FZU, Inst. of Phys. of the C.A.S. High Energy Physics Division,
     Na Slovance 2, CZ-180 40, Praha 8, Czech Republic
    \label{FZU}}
\titlefoot{Dipartimento di Fisica, Universit\`a di Genova and INFN,
     Via Dodecaneso 33, IT-16146 Genova, Italy
    \label{GENOVA}}
\titlefoot{Institut des Sciences Nucl\'eaires, IN2P3-CNRS, Universit\'e
     de Grenoble 1, FR-38026 Grenoble Cedex, France
    \label{GRENOBLE}}
\titlefoot{Helsinki Institute of Physics, HIP,
     P.O. Box 9, FI-00014 Helsinki, Finland
    \label{HELSINKI}}
\titlefoot{Joint Institute for Nuclear Research, Dubna, Head Post
     Office, P.O. Box 79, RU-101 000 Moscow, Russian Federation
    \label{JINR}}
\titlefoot{Institut f\"ur Experimentelle Kernphysik,
     Universit\"at Karlsruhe, Postfach 6980, DE-76128 Karlsruhe,
     Germany
    \label{KARLSRUHE}}
\titlefoot{Institute of Nuclear Physics and University of Mining and Metalurgy,
     Ul. Kawiory 26a, PL-30055 Krakow, Poland
    \label{KRAKOW}}
\titlefoot{Universit\'e de Paris-Sud, Lab. de l'Acc\'el\'erateur
     Lin\'eaire, IN2P3-CNRS, B\^{a}t. 200, FR-91405 Orsay Cedex, France
    \label{LAL}}
\titlefoot{School of Physics and Chemistry, University of Lancaster,
     Lancaster LA1 4YB, UK
    \label{LANCASTER}}
\titlefoot{LIP, IST, FCUL - Av. Elias Garcia, 14-$1^{o}$,
     PT-1000 Lisboa Codex, Portugal
    \label{LIP}}
\titlefoot{Department of Physics, University of Liverpool, P.O.
     Box 147, Liverpool L69 3BX, UK
    \label{LIVERPOOL}}
\titlefoot{LPNHE, IN2P3-CNRS, Univ.~Paris VI et VII, Tour 33 (RdC),
     4 place Jussieu, FR-75252 Paris Cedex 05, France
    \label{LPNHE}}
\titlefoot{Department of Physics, University of Lund,
     S\"olvegatan 14, SE-223 63 Lund, Sweden
    \label{LUND}}
\titlefoot{Universit\'e Claude Bernard de Lyon, IPNL, IN2P3-CNRS,
     FR-69622 Villeurbanne Cedex, France
    \label{LYON}}
\titlefoot{Univ. d'Aix - Marseille II - CPP, IN2P3-CNRS,
     FR-13288 Marseille Cedex 09, France
    \label{MARSEILLE}}
\titlefoot{Dipartimento di Fisica, Universit\`a di Milano and INFN,
     Via Celoria 16, IT-20133 Milan, Italy
    \label{MILANO}}
\titlefoot{Niels Bohr Institute, Blegdamsvej 17,
     DK-2100 Copenhagen {\O}, Denmark
    \label{NBI}}
\titlefoot{NC, Nuclear Centre of MFF, Charles University, Areal MFF,
     V Holesovickach 2, CZ-180 00, Praha 8, Czech Republic
    \label{NC}}
\titlefoot{NIKHEF, Postbus 41882, NL-1009 DB
     Amsterdam, The Netherlands
    \label{NIKHEF}}
\titlefoot{National Technical University, Physics Department,
     Zografou Campus, GR-15773 Athens, Greece
    \label{NTU-ATHENS}}
\titlefoot{Physics Department, University of Oslo, Blindern,
     NO-1000 Oslo 3, Norway
    \label{OSLO}}
\titlefoot{Dpto. Fisica, Univ. Oviedo, Avda. Calvo Sotelo
     s/n, ES-33007 Oviedo, Spain
    \label{OVIEDO}}
\titlefoot{Department of Physics, University of Oxford,
     Keble Road, Oxford OX1 3RH, UK
    \label{OXFORD}}
\titlefoot{Dipartimento di Fisica, Universit\`a di Padova and
     INFN, Via Marzolo 8, IT-35131 Padua, Italy
    \label{PADOVA}}
\titlefoot{Rutherford Appleton Laboratory, Chilton, Didcot
     OX11 OQX, UK
    \label{RAL}}
\titlefoot{Dipartimento di Fisica, Universit\`a di Roma II and
     INFN, Tor Vergata, IT-00173 Rome, Italy
    \label{ROMA2}}
\titlefoot{Dipartimento di Fisica, Universit\`a di Roma III and
     INFN, Via della Vasca Navale 84, IT-00146 Rome, Italy
    \label{ROMA3}}
\titlefoot{DAPNIA/Service de Physique des Particules,
     CEA-Saclay, FR-91191 Gif-sur-Yvette Cedex, France
    \label{SACLAY}}
\titlefoot{Instituto de Fisica de Cantabria (CSIC-UC), Avda.
     los Castros s/n, ES-39006 Santander, Spain
    \label{SANTANDER}}
\titlefoot{Dipartimento di Fisica, Universit\`a degli Studi di Roma
     La Sapienza, Piazzale Aldo Moro 2, IT-00185 Rome, Italy
    \label{SAPIENZA}}
\titlefoot{J. Stefan Institute, Jamova 39, SI-1000 Ljubljana, Slovenia
     and Laboratory for Astroparticle Physics,\\
     \indent~~Nova Gorica Polytechnic, Kostanjeviska 16a, SI-5000 Nova Gorica, Slovenia, \\
     \indent~~and Department of Physics, University of Ljubljana,
     SI-1000 Ljubljana, Slovenia
    \label{SLOVENIJA}}
\titlefoot{Fysikum, Stockholm University,
     Box 6730, SE-113 85 Stockholm, Sweden
    \label{STOCKHOLM}}
\titlefoot{Dipartimento di Fisica Sperimentale, Universit\`a di
     Torino and INFN, Via P. Giuria 1, IT-10125 Turin, Italy
    \label{TORINO}}
\titlefoot{Dipartimento di Fisica, Universit\`a di Trieste and
     INFN, Via A. Valerio 2, IT-34127 Trieste, Italy \\
     \indent~~and Istituto di Fisica, Universit\`a di Udine,
     IT-33100 Udine, Italy
    \label{TU}}
\titlefoot{Univ. Federal do Rio de Janeiro, C.P. 68528
     Cidade Univ., Ilha do Fund\~ao
     BR-21945-970 Rio de Janeiro, Brazil
    \label{UFRJ}}
\titlefoot{Department of Radiation Sciences, University of
     Uppsala, P.O. Box 535, SE-751 21 Uppsala, Sweden
    \label{UPPSALA}}
\titlefoot{IFIC, Valencia-CSIC, and D.F.A.M.N., U. de Valencia,
     Avda. Dr. Moliner 50, ES-46100 Burjassot (Valencia), Spain
    \label{VALENCIA}}
\titlefoot{Institut f\"ur Hochenergiephysik, \"Osterr. Akad.
     d. Wissensch., Nikolsdorfergasse 18, AT-1050 Vienna, Austria
    \label{VIENNA}}
\titlefoot{Inst. Nuclear Studies and University of Warsaw, Ul.
     Hoza 69, PL-00681 Warsaw, Poland
    \label{WARSZAWA}}
\titlefoot{Fachbereich Physik, University of Wuppertal, Postfach
     100 127, DE-42097 Wuppertal, Germany
    \label{WUPPERTAL}}
\titlefoot{On leave of absence from IHEP Serpukhov
    \label{MILAN-SERPOU}}
\titlefoot{Now at University of Florida
    \label{FLORIDA}}

%% file: document.tex
\def\nin{\noindent}

\newcommand{\ecm}{E_{cm}} 
\newcommand{\ww}{{\rm WW}} 
\newcommand{\zg}{{\rm Z}/\gamma} 
\newcommand{\eps}{\varepsilon} 
\newcommand{\fig}{Fig.~\ref} 
\newcommand{\tab}{Table~\ref} 
\newcommand{\das}{ $\Delta \alpha_s$  } 
\newcommand{\asmz} {$\alpha_s(M_{\rm Z}^2)$ } 
\newcommand{\asmzx} {$\alpha_s(M_{\rm Z}^2).$ } 
\newcommand{\chindf} { $ \chi^2 / n_{df} $ } 
\newcommand{\dasmz}{ $\Delta \alpha_s(M_Z^2)$ } 
\newcommand{\as}{$\alpha_s$\hspace{0.1cm}} 
\newcommand{\bfas}{\protect{ \boldmath $ \alpha_s $ \unboldmath } } 
\newcommand{\bfasmz}{\protect{ \boldmath $ \alpha_s(M_Z^2) $ \unboldmath } } 
\newcommand{\oas}{$\cal O$($\alpha_s$)\hspace{0.1cm}}
\newcommand{\oass}{$\cal O$($\alpha_s^2$)\hspace{0.1cm}}
\newcommand{\oassx}{$\cal O$($\alpha_s^2$)\hspace{0.05cm}.\hspace{0.1cm}}
\newcommand{\oasss}{$\cal O$($\alpha_s^3$)\hspace{0.1cm}}
\newcommand{\oasssx}{$\cal O$($\alpha_s^3$)\hspace{0.05cm}.\hspace{0.1cm}}
\newcommand{\rhoe}{\rho_{{\rm eff}}}
\newcommand{\gev}{\mbox{\,\,Ge\kern-0.2exV }}
\newcommand{\mev}{\mbox{\,\,Me\kern-0.2exV }}
\newcommand{\ee} {$\mbox{e}^+\mbox{e}^-$ } 
\newcommand{\emi} {$\mbox{e}^-$ } 
\newcommand{\ddsigma} { \frac {1} {\sigma}
                        \frac {d^2\sigma} { dY d\cos\vartheta_T} }  

\newcommand{\spa}{\hspace{-0.2 cm}}
\newcommand{\spb}{\hspace{-0.4 cm}}
\newcommand{\spc}{\hspace{-0.6 cm}}
\newcommand{\spd}{\hspace{-0.8 cm}}
\newcommand{\spm}{\hspace{-0.1 cm}}
\newcommand{\spv}{\hspace{-0.05 cm}}

\renewcommand{\textfraction}{0.02}
\renewcommand{\bottomfraction}{0.98}
\renewcommand{\topfraction}{0.98}
\renewcommand{\marginparwidth}{2.6 cm}
\renewcommand{\marginparsep}{0.2 cm}


\section{Introduction}

This paper presents a highly  improved test of second order 
perturbation theory and an improved measurement of \asmzx
It is based on progress in 
next-to-leading order QCD calculations of oriented event shape 
distributions \cite{GenAlg}. Furthermore, the DELPHI data 
used in this analysis are much improved in both their statistical 
and systematic precision compared with those of previous DELPHI publications
\cite{DelPubs1,DelPubs2}. The distributions of 18 different 
infrared and collinear safe hadronic event observables are determined 
from 1.4 million hadronic Z-decays at various values of the polar angle 
$ \vartheta_T $ of the thrust axis.
The $ \vartheta_T $ dependence of all detector properties are  
taken into account, thus achieving the best possible experimental
precision. \\

The precise experimental data are fully consistent with the expectation 
from second order QCD. A two parameter fit to each of the distributions 
measured at different polar angles $ \vartheta_T $ allows an experimental
optimization of the \oass renormalization scale giving a  
consistent set of eighteen \asmz values. For most of the distributions the 
largest uncertainty on the \asmz values is due to hadronization corrections
and not to renormalization scale errors. Any artificial increase 
\cite{scaledisk} of the uncertainty of \asmz due to a large variation 
of the renormalization scale is avoided so that the degree of
precision to which QCD can be tested remains transparent. 
An average value of \asmz is derived taking account of the correlations 
between the values obtained from the 18 distributions. \\

A number of additional studies have been performed to check the reliability 
of the \asmz results obtained from experimentally optimized scales. In one of 
these studies `optimized' renormalization scales as discussed in the 
literature are used to determine \asmz in second order pertubation theory. 
The different methods applied for choosing an optimized scale lead to 
consistent results for the average value of \asmzx However, the scatter among 
\asmz values from the individual distributions is smaller for the experimentally
optimized scales than that obtained using theoretically 
motivated scale evaluation methods.  The correlation between the renormalization 
scales obtained with the different methods is also investigated. \\ 

Further determinations of \asmz are performed by using all orders resummed
calculations in the next-to-leading logarithmic approximation (NLLA). Here
two different methods are applied. In the first case the pure NLLA
predictions are confronted with the data in a limited fit range. In
the second method \as is determined using matched NLLA and \oass 
calculations. For both methods the renormalization scale is chosen to 
be $ \mu = M_Z $. Both methods lead to average \as values consistent
with the average value obtained in \oass with experimentally optimized
renormalization scales. The agreement between the results of
the pure NLLA fits and those of the \oass is emphasized. A closer
inspection of the fits in matched NLLA and \oass to the very
precise data reveals a so far unreported problem with this method in that
the trend of the data deviates systematically from the 
expectation of the matched theory. \\

The selection of hadronic events and the correction procedures applied 
to the data are described in Section \ref{selhadrsect}. Section 
\ref{shapedefsect} introduces the investigated event shapes and compares
the expectations from various fragmentation models. 
Section \ref{exoptsect} contains the comparison with angular dependent 
second order QCD and a detailed discussion of the determination of \asmzx
Section \ref{theooptsect} summarises determinations of \asmz using
the renormalization scales discussed in the literature. Section \ref{padesect}
discusses results obtained by applying Pad\'{e} approximants for the
extrapolation of the pertubative predictions to higher orders. 
Section \ref{NLLAsect} discusses results from applying  NLLA. 
The heavy quark mass correction of the \asmz values derived 
from experimentally optimized renormalization scales is described in Section
\ref{qmassect}. The final results are summarized in the last section.
                   

\section{Detector and Data Analysis}
\label{selhadrsect}
                   

In this analysis the final data measured with the DELPHI detector
in 1994 at a centre-of-mass energy of $ \sqrt{s} = M_{\rm Z} $ are used.
The statistics of the 1994 data is fully sufficient for the
accurate QCD studies described in this paper.
DELPHI is a hermetic detector with a solenoidal magnetic field of 1.2 T.
The detector and its performance have been described in detail in \cite{Detector}.
The following components are relevant to this analysis:

\begin{itemize}

\item
the Vertex Detector, VD, measuring charged particle track coordinates in the
plane perpendicular to the beam with three layers of silicon micro-strip
detectors at radii between 6.3 and 11 cm and covering polar angles, $ \vartheta $,
with respect to the \emi beam between $ 37^{\circ} $ and $ 143^{\circ} $; 

\item
the Inner Detector, ID, a cylindrical jet chamber with a polar angle
coverage from $ 17^{\circ} $ to $ 163^{\circ} $;

\item
the Time Projection Chamber, TPC, the principal tracking detector of DELPHI,
which has 6 sector plates in both the forward and backward hemispheres, 
each with 16 pad rows and 192 sense wires, 
inner and outer radii of 30 cm and 122 cm
and covers polar angles from $ 20^{\circ} $ to $ 160^{\circ} $;

\item
the Outer Detector, OD, a five layer drift chamber at 198 to 206 cm radius 
covering polar angles between $ 43^{\circ} $ and $ 137^{\circ} $; 

\item
two sets of forward planar drift chambers, FCA and FCB, with 6 and 12 layers 
respectively and overall polar angle coverages of $ 11^{\circ} $ to 
$ 35^{\circ} $ and $ 145^{\circ} $ to $ 169^{\circ} $;

\item
the High Density Projection Chamber, HPC, a lead-glass electromagnetic
calorimeter with a very good spatial resolution located inside the DELPHI
coil between 208 cm and 260 cm radius and covering polar angles between
$ 43^{\circ} $ and $ 137^{\circ} $;

\item
the Forward Electromagnetic Calorimeter, FEMC, comprising two lead-glass
arrays, one in each endcap, each consisting of 4500 lead-glass blocks
with a projective geometry, and covering polar angles from $ 10^{\circ} $ to 
$ 36.5^{\circ} $ and from $ 143.5^{\circ} $ to $ 170^{\circ} $;    
 
\item
The hadron calorimeter, HAC, an iron-gas hadronic calorimeter outside the coil,
consisting of 19 to 20 layers of streamer tubes and 5 cm thick iron plates
also used as flux return, whose overall angular coverage is from 
$ 11.2^{\circ} $ to $ 168.8^{\circ} $.

\end{itemize}

\subsection{Event Selection}

Only charged particles in hadronic events were used.
They were required to pass the following selection criteria:

\begin{itemize}

\item
momentum, $ p $, greater than 0.4 GeV/c,

\item
$\Delta p / p $ less than 100\%,

\item
measured track length greater than 30 cm,

\item
track polar angle between $ 16^{\circ} $ and $  164^{\circ} $,

\item
impact parameter with respect to the nominal interaction point
within 4 cm perpendicular to and 10 cm along the beam.

\end{itemize}

\pagebreak

\nin Hadronic events were selected by requiring:

\begin{itemize}

\item
at least 5 charged particles,

\item
the total energy of charged particles greater than $ 12\% \sqrt{s} $
where the pion mass has been assumed for all particles,
\item
the charged energy in each hemisphere of the detector, defined by
the plane perpendicular to the beam, $ E_{hemis} $, 
greater than $ 3\% \sqrt{s} $,

\item
the polar angle of the thrust axis \footnote{Thrust and the thrust axis 
have been calculated applying the algorithm included in the JETSET package \cite{Jetset}.}, 
$ \vartheta_T $,  between $ 90.0^{\circ}$ and $ 16.3^{\circ} $. 

\end{itemize}

In total about 1.4 million events satisfy these cuts. The selection 
efficiency is 92\%. Since the thrust axis does not distinguish 
between forward and backward directions, it is chosen such that
$ \cos \vartheta_T \ge 0 $. $ \vartheta_T $ is called 
the event orientation. The data are binned 
according to the event orientation into eight equal bins
\footnote{For the comparison of event shape observables with QCD predictions 
          in all orders resummed next-to-leading-log approximation, the 
          distributions have been integrated over $ \vartheta_T $. Differing 
          from the event selection criteria listed above, the hadronic 
          events were selected if the polar angle of the thrust axis satisfied 
          $\rm 40.0^{\circ} < \vartheta_T < 90.0^{\circ} $ for these 
          angular integrated distributions.} 
of $ \cos \vartheta_T $ between 0.0 and 0.96. 
With the exception of the eighth bin, the thrust axis is well contained 
within the detector acceptance. 
 

\subsection{Correction Procedure}

The contamination of beam gas events, $\gamma \gamma $ events and leptonic
events other than $ \tau^+ \tau^- $, is expected to be less than 0.1 \%
and has been neglected. The influence of $ \tau^+ \tau^- $ events which 
have a pronounced 2-jet topology and contain high momentum particles has been
determined by the KORALZ model \cite{Koralz} treated by the full simulation
of the DELPHI detector DELSIM \cite{Dperf} and the standard data 
reconstruction chain. The $ \tau^+ \tau^- $ contributions have been
subtracted from the measured data according to the relative rate of
$\tau^+ \tau^- $ $ ( 0.46 \% \pm 0.03\%) $ and hadronic events. \\  

The observed data distributions were corrected for kinematic cuts, 
limited acceptance and resolution of the detector as well
as effects due to reinteractions of particles inside the detector material.
The simulated data were processed in the same way as the real data.
The correction for initial state photon radiation has been determined
using events generated by JETSET 7.3 PS \cite{Jetset} with and
without initial state radiation as predicted by DYMU3 \cite{Dymu}.
For any given observable Y the bin-by-bin correction factor 
$ C(Y,\cos \vartheta_T) $  is calculated as:

\begin{equation}
    C(Y,\cos \vartheta_T) =
    \frac { \left( \ddsigma \right)^{DELSIM}_{generated} } 
          { \left( \ddsigma  \right)^{DELSIM}_{reconstructed} }
    \cdot
    \frac { \left( \ddsigma \right)^{noISR} } 
          { \left( \ddsigma \right)^{ISR} } ~~~.
\end{equation}

\hspace{0.1 cm} \\

Particles with a lifetime larger than 1 ns were considered as stable in the 
generated distributions. The bin widths were chosen on the basis of the 
estimated experimental resolution so as to minimize bin-to-bin migration 
effects. \\

For the evaluation of systematic errors the cuts for the track and event
selections were varied over a wide range, including additional cuts on 
the momentum imbalance, etc. Variation of the tracking efficiency
has been considered by discarding 2\% of the accepted tracks at random. 
The influence of uncertainties in the 
momentum resolution was estimated by applying an additional 
Gaussian smearing of the inverse momenta of the simulated tracks.  
From the stability of the measured 
distributions a systematic uncertainty has been computed as the variance 
with respect to the central value. As the systematic error is expected to grow 
in proportion to the deviation of the overall correction factor from unity,
an additional relative systematic uncertainty of 10\% of this deviation 
has been added quadratically to the above value.  


\section{Measured Event Shapes and Comparison with Fragmentation Models}
\label{shapedefsect}

This analysis includes all commonly used shape observables which have a 
perturbative expansion known at least to next to leading order.
 

\subsection{Definition of the Observables}

Thrust $ \rm T$ is defined by \cite{Thrust} :
                  
\begin{equation} 
T = \max_{\vec{n}_{T}} \frac 
     {\sum_{i} \left| \vec{p}_{i} \cdot \vec{n}_{T} \right| } 
     {\sum_{i} \left| \vec{p}_{i} \right| } ~~~,
\end{equation}

\nin where $ \vec{p}_{i}$ is the momentum vector of particle $ i$, and 
$ \vec{n}_{T}$ is the thrust axis to be determined. \\

Major $ \rm M$ and Minor $\rm m$ are defined similarly, replacing
$ \vec{n}_{T}$ in the expression above by the Major axis
$ \vec{n}_{Maj}$, which maximizes the momentum sum transverse to
$ \vec{n}_{T}$ or the Minor axis
$ \vec{n}_{Min} \, = \, \vec{n}_{Maj} \times \vec{n}_{T}$
respectively. \\

The oblateness $ \rm O$ is then defined by \cite{Obl}: 

\begin{equation} 
O=M-m ~~~.
\end{equation} 

The C-parameter $ \rm C $ is derived from the eigenvalues $ \lambda$
of the infrared-safe linear momentum tensor $ \Theta^{i,j}$ \cite{CPar}:

\begin{equation} 
   \Theta^{i,j} = \frac
      {1}
      {\sum_{k} \left|\vec{p}_{k}\right|}
      \cdot \sum_{k} \frac
      {p^{i}_{k}p^{j}_{k}} 
      {\left|\vec{p}_{k}\right|}      
\end{equation} 

\begin{equation} 
   C=3\cdot(\lambda_{1}\lambda_{2}+\lambda_{2}\lambda_{3}+
   \lambda_{3}\lambda_{1}) ~~~.
\end{equation} 

\nin Here $ p^{i}_{k} $ denotes the {\it i}-component of $ \vec{p}_{k} $.\\

Events can be divided into two hemispheres, $ a$ and $ b$, 
by a plane perpendicular to the thrust axis $\vec{n}_{T}$. With 
$ M_{a}$ and $ M_{b}$ denoting the invariant masses of the two
hemispheres, the normalized heavy jet mass $\rm \rho_{H}$, light jet mass 
$\rm \rho_{L}$, the sum of the jet masses $\rm \rho_{S}$ 
and their difference $\rm \rho_{D}$ can be defined as

\begin{equation}    
   \rho_{H}= \frac 
   {\max(M_{a}^{2},M_{b}^{2})} {E_{vis}^{2}} 
\end{equation} 

\begin{equation}    
   \rho_{L}= \frac 
      {\min(M_{a}^{2},M_{b}^{2})} {E_{vis}^{2}} 
\end{equation}

\begin{equation}  
   \rho_{S}= \rho_{H} + \rho_{L} 
\end{equation} 

\begin{equation}
\label{rhod}  
   \rho_{D}= \rho_{H} - \rho_{L} 
\end{equation} 

\nin where

\begin{equation} 
\label{equ-evis}
    E_{vis} = \sum_{i} E_{i}
\end{equation} 
 
\nin and the energy of the particles $i$ has been calculated assuming
pion mass for charged and zero mass for neutral particles. \\

Jet broadening measures have been proposed in \cite{JBroad}. In both
hemispheres $ a $ and $ b $ the transverse momenta of the particles 
are summed thus: 

\begin{equation} 
    B_{a,b} = \frac
   { \sum_{i \in a,b } \left| \vec{p}_{i}\times \vec{n}_{T}\right|   }
   { 2 \sum_{i}\left| \vec{p}_{i}\right| }  ~~~.
\end{equation} 

The wide jet broadening $\rm B_{max} $, the narrow jet broadening 
$\rm B_{min} $, and the total jet broadening $\rm B_{sum} $
are then defined by

\begin{equation} 
    B_{max}=\max(B_{a},B_{b})
\end{equation} 

\begin{equation} 
    B_{min}=\min(B_{a},B_{b})
\end{equation} 

\begin{equation} 
    B_{sum}= B_{max} + B_{min} ~~~.
\end{equation} 

The first order prediction in perturbative QCD vanishes for both
$\rm \rho_{L} $ and $\rm B_{min} $. Therefore these observables
cannot be used for the determination of $\rm \alpha_{s} $. \\

Jet rates are commonly obtained using iterative clustering algorithms
\cite{clusalg} in which a distance criterion or a metric $ y_{ij}$, 
such as the scaled invariant mass, is computed for all pairs of particles 
$ i $ and $ j $. The pair with the smallest $ y_{ij} $ is 
combined into a pseudoparticle (cluster) according to one of several
recombination schemes. The clustering procedure is repeated until all of
the $ y_{ij} $ are greater than a given threshold, the jet resolution
parameter $ y_{cut} $. The jet multiplicity of the event is defined 
as the number of clusters remaining; the n-jet rate 
$ R_{n}(y_{cut}) $ is the fraction of events classified as 
n-jet, and the differential two-jet rate is defined as 

\begin{equation} 
   D_{2}(y_{cut})=\frac
   {R_{2}(y_{cut})-R_{2}(y_{cut}-\Delta y_{cut}) } {\Delta y_{cut} } ~~~.
\end{equation} 

Several algorithms have been proposed differing from each other in 
their definition of $ y_{ij} $ and their recombination procedure.
We apply the $\rm E0 $, $\rm P$, $\rm P0 $, JADE \cite{Jade}, 
Durham \cite{Durham}, Geneva \cite{clusalg} and the Cambridge 
algorithms \cite{Camjet}. 
The definitions of the metrics $ y_{ij} $ and the recombination 
schemes for the different algorithms are given below. \\

In the $\rm E0 $ algorithm $ y_{ij} $ is defined as the square of   
the scaled invariant mass of the pair of particles 
$ i $ and $ j $:

\begin{equation} 
\label{invmass}
   y_{ij} = \frac
   {(p_{i} + p_{j})^2 } {E_{vis}^2} ~~~.
\end{equation} 

\nin The recombination is defined by:

\begin{equation} 
    E_{k} = E_{i} + E_{j} 
   \hspace{.2 cm} , \hspace{.2 cm}
    \vec{p}_{k} = \frac
       {E_{k}}{\left| \vec{p}_{i} + \vec{p}_{j} \right|} 
       (\vec{p}_{i} + \vec{p}_{j})  ~~~,
\end{equation}

\nin where $ E_{i} $ and $ E_{j} $ are the energies and $ \vec{p}_{i} $
and $ \vec{p}_{j} $ are the momenta of the particles.  \\

In the $\rm P $ algorithm $ y_{ij} $ is defined by Eq. (\ref{invmass}), 
and the recombination is defined by
 
\begin{equation} 
\label{mrecomb}
    \vec{p}_{k} = \vec{p}_{i} + \vec{p}_{j} 
    \hspace{.2 cm} , \hspace{.2 cm}
    E_{k} = \left| \vec{p}_{k} \right|  ~~~.
\end{equation} 

The $\rm P0 $ algorithm is defined similarly to the $\rm P $ algorithm,
however the total energy $ E_{vis} $ (Eq. \ref{equ-evis}) is recalculated 
at each iteration for the remaining pseudoparticles.  \\

In the JADE algorithm, the definition of $ y_{ij} $ is

\begin{equation}
\label{jadeyij} 
   y_{ij}= \frac
   {2 E_{i}  E_{j}(1 - \cos \theta_{ij} ) } 
   {E_{vis}^2} ~~~,
\end{equation} 

\nin where $ \theta_{ij} $ is the angle between the pair of particles
$ i$ and $ j $. \\

For the  Durham algorithm $ y_{ij} $ is given by

\begin{equation} 
\label{durhamyij}
   y_{ij}= \frac
   {2 \min (E_{i}^2 ,E_{j}^2)(1 - \cos \theta_{ij} ) } 
   {E_{vis}^2}
\end{equation} 

\nin and for the Geneva algorithm by 

\begin{equation}
\label{genevayij} 
   y_{ij}= \frac
   {8 E_{i}  E_{j}(1 - \cos \theta_{ij} ) }   
   {9(E_{i}+E_{j})^2} ~~~.
\end{equation}

For the algorithms given by Equations (\ref{jadeyij}), (\ref{durhamyij}) and
(\ref{genevayij}) the recombination is done by adding the particles
four-momenta.  \\

The recently proposed Cambridge algorithm \cite{Camjet} introduces an ordering 
of the particles $ i $ and $ j $ according to their opening angle,
using the ordering variable 

\begin{equation} 
   \nu_{ij}= 2 ( 1 - \cos \theta_{ij} ) 
\end{equation} 

\nin and $ y_{ij} $ is defined by Eq. (\ref{durhamyij}). The algorithm starts 
clustering from a table of $ N_{obj} $ primary objects, which are 
the particles' four-momenta, and proceeds as follows:

\begin{enumerate}

\item
If only one object remains, store this as a jet and stop.

\item
Select the pair of objects $ i $ and $ j $ that have the minimal
value of the ordering variable $ \nu_{ij} $ and calculate $ y_{ij} $
for that pair.

\item
If $ y_{ij} < y_{cut} $ then remove the objects $ i $ and $ j $
from the table and add the combined object with four-momentum 
$ p_i + p_j $. If $ y_{ij} \ge y_{cut} $ then store the object 
$ i $ or $ j $ with the smaller energy as a separated jet and remove
it from the table. The higher energy object remains in the table.

\item
go to 1.
 
\end{enumerate}

The energy-energy correlation EEC \cite{EEC} is defined in terms of the 
angle $ \chi_{ij} $ between two particles $ i $ and $ j $ in an 
hadronic event:

\begin{equation} 
   EEC(\chi)= \frac {1}{N} 
      \frac {1}{\Delta \chi}
      \sum\limits_{N} \sum\limits_{i,j} 
      \frac {E_{i} E_{j}} {E_{vis}^2}
      \int\limits_{\chi-\frac{\Delta\chi}{2}}^{\chi+\frac{\Delta\chi}{2}}
      \delta(\chi' - \chi_{ij}) d \chi' ~~~,
\end{equation} 

\nin where $ N $ is the total number of events, $ \Delta \chi $ is the
angular bin width and the angle $ \chi $ is taken from 
$ \chi = 0^{\circ} $ to $ \chi = 180^{\circ} $. \\

The asymmetry of the energy-energy correlation AEEC is defined as

\begin{equation}
   AEEC(\chi) = EEC(180^{\circ}-\chi) - EEC(\chi) ~~~.      
\end{equation} 

The jet cone energy fraction JCEF \cite{JCEF} integrates the energy within
a conical shell of an opening angle $ \chi $ about the thrust axis. It
is defined as 

\begin{equation} 
   JCEF(\chi)= \frac {1}{N} 
      \frac {1}{\Delta \chi}
      \sum\limits_{N} \sum\limits_{i} 
      \frac {E_{i}} {E_{vis}}
      \int\limits_{\chi-\frac{\Delta\chi}{2}}^{\chi+\frac{\Delta\chi}{2}}
      \delta(\chi' - \chi_{i}) d \chi' ~~~,
\end{equation} 

\nin where $\chi_{i}$ is the opening angle between a particle and the thrust axis vector 
$ \vec{n}_{T} $, whose direction is defined here to point from
the heavy jet mass hemisphere to the light jet mass hemisphere. 
Although the JCEF is a particularly simple and excellent observable for the determination of
\as, it has been rarely used until now in experimental measurements. 
Within an \oass analysis, the region $ 90^{\circ} < \chi \le 180^{\circ} $, 
corresponding to the heavy jet mass hemisphere, can be used for the measurement of \as.
The distribution of the JCEF is shown in Figure \ref{MCData}. Hadronization corrections 
and detector corrections as well as the next-to-leading order perturbative
corrections are small. This allows a specially wide fit range to be used.


\subsection{Fragmentation Models}
\label{fragmod}

\begin{sidewaysfigure} 
\begin{center}
\mbox{\epsfig{file=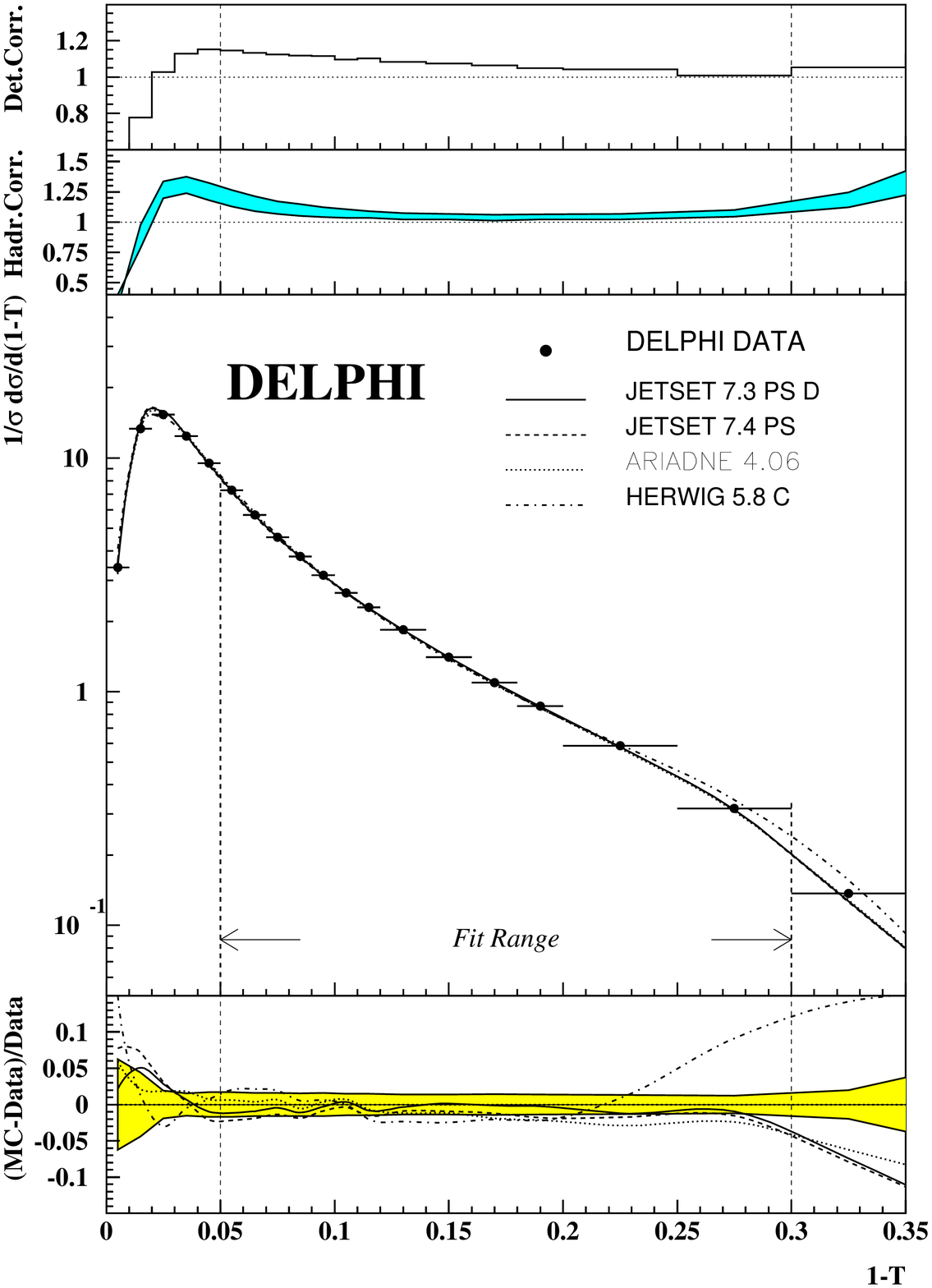, width=8.8cm}}
\hspace{1. cm}
\mbox{\epsfig{file=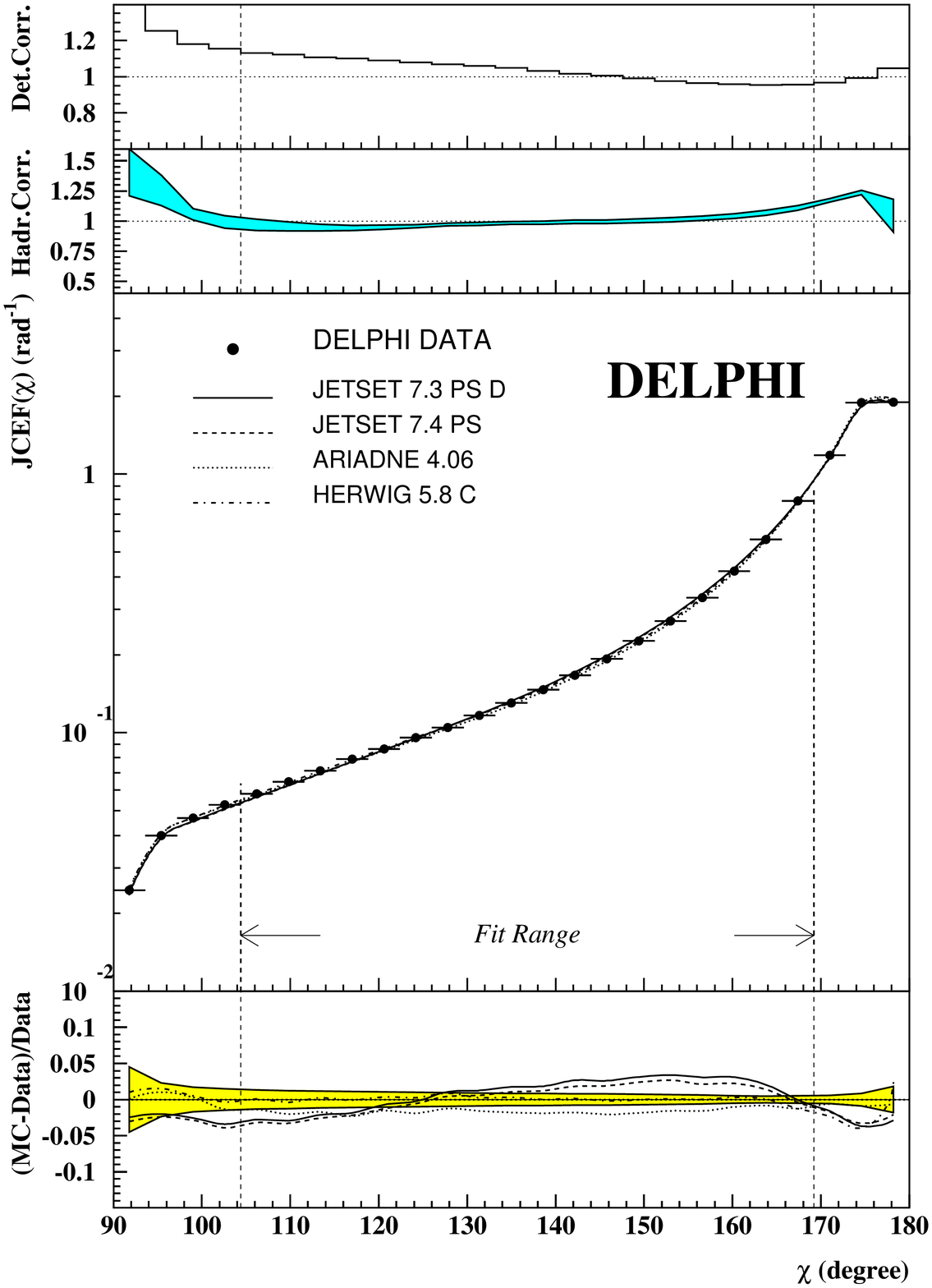, width=8.8cm}}
\caption[]{ {\it left part:}
          Measured 1-T distribution integrated over $ \rm \cos \vartheta_T $.
          The upper part shows the detector correction including effects due 
          to initial state radiation.
          The part below shows the size of the hadronization correction.
          The width of the band indicates the uncertainty of the 
          correction. In the central part the measured 1-T distribution
          is compared to the expectation from four hadronization
          generators, JETSET 7.3 PS D with DELPHI modification of heavy
          particle decays, JETSET 7.4 PS, ARIADNE 4.06 and HERWIG 5.8c.
          Also shown is the 1-T range used in the QCD fit. The lower part
          shows the ratio (Monte Carlo simulation-data)/data for the
          four hadronization generators. The width of the band indicates
          the size of the experimental errors. 
          {\it right part:} 
          Same curves as shown in the left part but for JCEF 
          integrated over $ \rm \cos \vartheta_T $.               }   
\label{MCData}
\end{center}
\end{sidewaysfigure}

\begin{sidewaysfigure} 
\begin{center}
\mbox{\epsfig{file=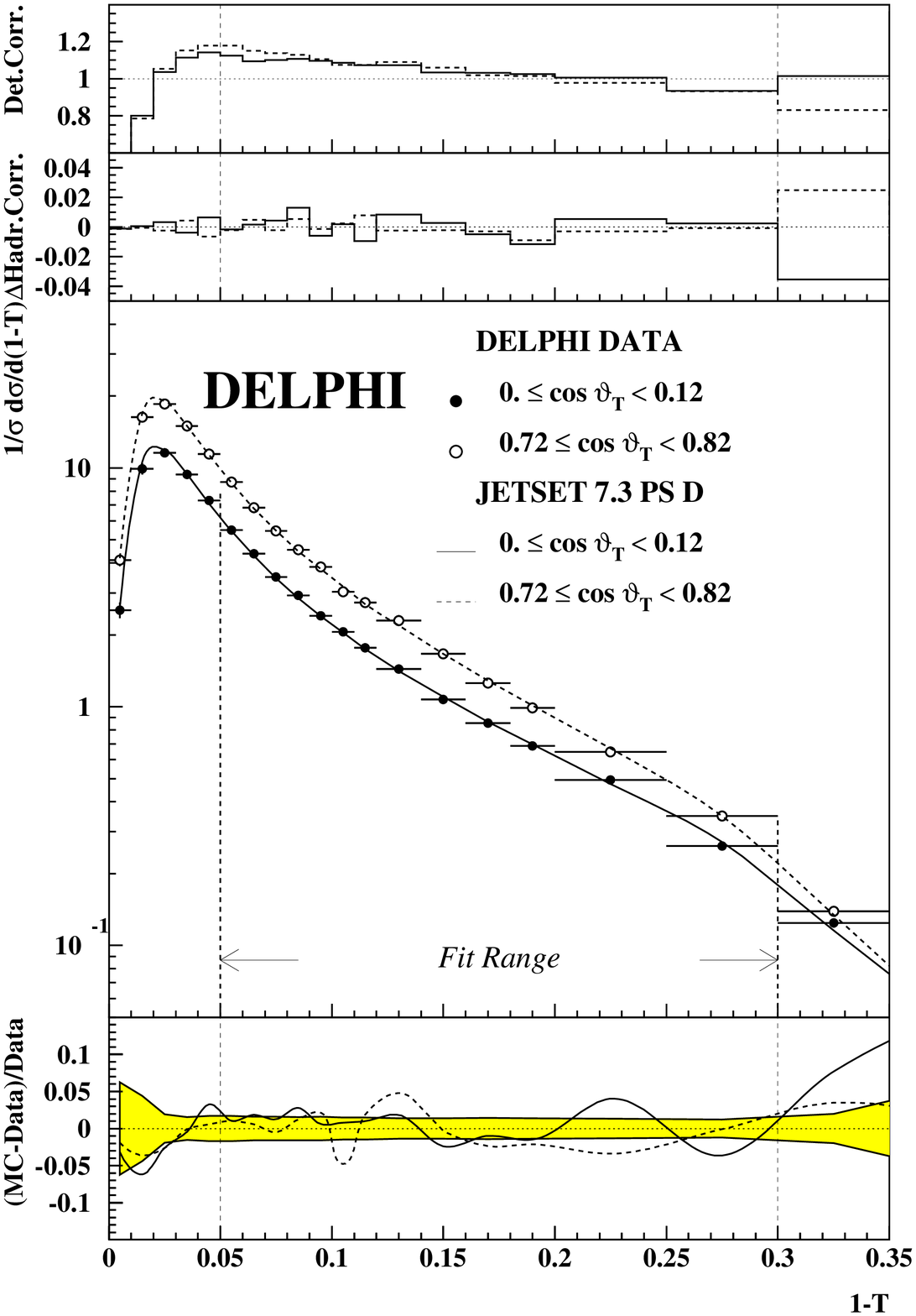, width=8.8cm}}
\hspace{1. cm}
\mbox{\epsfig{file=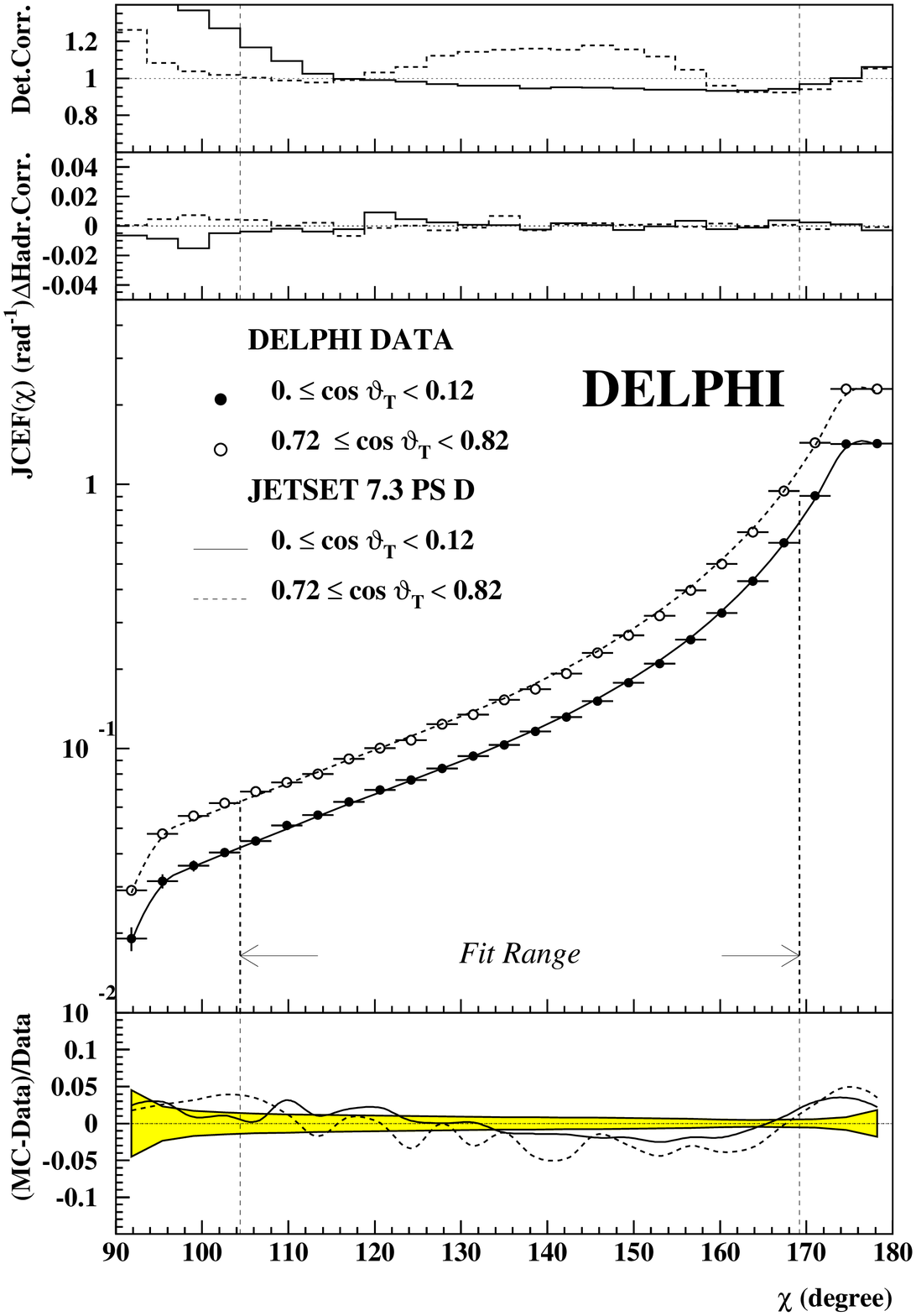, width=8.8cm}}
\caption[]{ {\it left part:} 
          Measured 1-T distribution in two bins of
          $ \rm \cos \vartheta_T $. The upper part shows the detector
          corrections in the two $ \rm \cos \vartheta_T $ bins. 
          The part below shows the size of the relative hadronization 
          correction in the two $ \rm \cos \vartheta_T $ bins with 
          respect to the average correction. In the central part 
          the measured 1-T distributions are compared to JETSET 7.3 PS D.
          The lower part shows the ratio 
          (Monte Carlo simulation-data)/data for the two 
          $ \rm \cos \vartheta_T $ bins. 
          {\it right part:} 
          Same curves as shown in the left part but for
          JCEF in two bins of $ \rm \cos \vartheta_T $.              }
\label{MCDaThe}
\end{center}
\end{sidewaysfigure}

QCD based hadronization models, which describe well the distributions 
of the event shape observables in the hadronic final state of \ee 
annihilation, are commonly used for modelling the transition from the 
primary quarks to the hadronic final state. Perturbative QCD can describe 
only a part of this transition, the radiation of hard gluons and the 
evolution of a parton shower. 
For a determination of the strong coupling constant $\rm \alpha_s $ 
one has to take account of the so-called fragmentation or
hadronization process, which is characterized by a small momentum transfer
and hence a breakdown of perturbation theory. 
Several Monte Carlo models are in use to estimate the size of the 
hadronization effects and the corresponding uncertainty. 
The most frequently used fragmentation models, namely JETSET 7.3 PS
\cite{Jetset}, ARIADNE 4.06 \cite{Ariadne} and HERWIG 5.8c \cite{Herwig}  
have been extensively studied and tuned to DELPHI data and to identified 
particle spectra from all LEP experiments in \cite{Tuning}. As discussed
in detail in \cite{Tuning} all models describe the data well.
Examples of the measured hadron distributions are presented in Figures 
\ref{MCData} and \ref{MCDaThe}. The increased systematic accuracy of 
the data is partially due to the fact that the $ \rm \vartheta_T $ 
dependence of the detector corrections is explicitly taken into account.
The $ \vartheta_T $ dependence of the detector corrections is shown, 
for instance, in Figure \ref{MCDaThe} for two of the observables studied.
Detailed tables of the individual event shape distributions including
their statistical and systematic errors will be made available in the HEPDATA
database \cite{HEPDB}. For figures of the distributions see 
also \cite{DelNoteData}. \\


Before theoretical expressions describing parton distributions can be
compared with experimental data, corrections have to be made for
hadronization effects, i.e. effects resulting from the transition of the
parton state into the observed hadronic state. 
For the global event shape observables this transition is performed 
by a matrix P, where $\rm P_{ij} $ is the probability that an event 
contributing to the bin $\rm j $ of the partonic distribution will 
contribute to the bin $\rm i $ in the hadronic distribution and is computed 
from a Monte Carlo model. This probability matrix has been applied 
to the distributions from \oass perturbative theory 
$\rm D_{pert.} (Y,\cos\vartheta_T) $ to obtain the distributions for 
the predictions of the observed final state
$\rm D_{hadr.} (Y,\cos\vartheta_T) $:

\begin{equation}
    D_{hadr.} (Y,\cos\vartheta_T)_i = 
    \sum_{j} P_{ij}(Y,\cos\vartheta_T) D_{pert.} (Y,\cos\vartheta_T)_j  ~~~.
\end{equation}

In the case of the $\rm JCEF $, $\rm EEC $ and $\rm AEEC $, which are
defined in terms of single particles and pairs of particles, respectively,
bin-by-bin correction factors $\rm C_{Hadr.} $ similar to that described
above for the detector effects have been computed such as:

\begin{equation}
    D_{hadr.} (Y,\cos\vartheta_T)_i = 
    C_{Hadr.}(Y,\cos\vartheta_T)_i D_{pert.} (Y,\cos\vartheta_T)_i ~~~.
\end{equation}

Our reference model for evaluating hadronization effects is the JETSET 7.3
Parton Shower (PS) Generator, which has been modified with respect to the heavy
particle decays to obtain a better description of the heavy particle 
branching fraction. This modified version is denoted by JETSET 7.3 PS D
in the following. 
The tuned parameters have been taken from \cite{Tuning},
where the updated tuning procedure is described in detail. \\

In order to estimate the systematic error of the hadronization correction, 
the analysis was repeated using alternative  Monte Carlo generators with 
different hadronization models. In addition, the parameters for the 
JETSET PS were varied. A description of the models and 
their differences can be found for example in \cite{Tuning}. 
The alternative models used are ARIADNE 4.06, HERWIG 5.8c as well as 
version 7.4 JETSET PS \cite{Jetset}. 
All these models have been tuned to DELPHI data \cite{Tuning}. 
For our standard Monte Carlo program we applied also an alternative tuning
to the DELPHI data which includes Bose-Einstein correlations, 
not included in the reference tuning. Whereas the number of hard gluons
predicted by second order QCD matrix elements is simulated by the hadronization
models \cite{Fuerst}, additional soft gluons are produced within
the parton shower cascade, controlled by the JETSET PS parameter $\rm Q_{0} $,   
which describes the parton virtuality at which the parton shower is stopped.
To account for the sensitivity of the shape observables with respect to the
additional soft gluons,
$\rm Q_{0} $ has been varied from 0.5 $\rm GeV $ to 4.0 $\rm GeV $. \\

The systematic error of $ \alpha_s $ originating from hadronization 
corrections is then estimated as the variance of the fitted 
$\rm \alpha_{s} $ values obtained by using all the hadronization corrections 
mentioned above. Further studies have been made to investigate the influence 
of the main fragmentation parameters of the JETSET PS model by varying them
within their experimental uncertainty. It has been found that this
contribution to the uncertainty of \as in general is less than one per mille, 
and has been neglected. \\

\pagebreak

\section{Comparison with Angular Dependent Second Order QCD using
         Experimentally Optimized Scales}
\label{exoptsect}

The evaluation of the $\rm {\cal O} ( \alpha_s^2 )$ coefficients is
performed by using EVENT2 \cite{EVENT2}, a program for the integration of 
the \oass matrix elements. The algorithm is described in \cite{GenAlg}. \\

Using this program, one can calculate the double differential 
cross-section for any infrared and collinear safe observable $ Y $ in 
\ee annihilation as a function of the event orientation:

\begin{eqnarray} 
     \frac { 1 } { \sigma_{tot} } 
     \frac { d^{2} \sigma (Y, \cos \vartheta_T ) } 
           { dY d \cos \vartheta_T }                       & 
                                                     =     &
     \bar{\alpha}_s ( \mu^2 ) \cdot  
     A (Y, \cos \vartheta_T )                        \nonumber \\
                                                            &
                                                        +   &
     \bar{\alpha}_s^2 ( \mu^2 ) \cdot 
     \Bigg[ B (Y, \cos \vartheta_{T} ) + 
            \Big( 2 \pi \beta_{0} \ln ( x_{\mu} ) -2 \Big) 
            A(Y,\cos\vartheta_{T}) \Bigg] , 
\label{thseco}
\end{eqnarray} 

\nin where $\rm \bar{\alpha}_s  = \alpha_s / 2\pi $ and 
$\beta_0 = (33 - 2n_f) / 12 \pi $, $ n_f $ is the number of
active quark flavours and
$ \sigma_{tot} $ is the one loop corrected cross-section for the process 
\ee $\rm \to $ hadrons. The event orientation enters via $\rm \vartheta_T $,
which denotes the polar angle of the thrust axis with respect to the 
$\mbox{e}^+\mbox{e}^-$ beam direction. 
The renormalization scale factor $x_{\mu} $ is defined by 
$\mu^2 = x_{\mu} Q^2 $ where $\rm Q = M_Z $ is the 
centre-of-mass energy. $\rm A $ and $\rm B $ denote the \oas and \oass 
QCD coefficients, respectively.      
Alternatively, the double differential cross-section can be normalized
to the partial cross-section in each $\rm \cos \vartheta_T $ interval:

\begin{equation}
     R(Y,\cos \vartheta_T) = 
     \left( \frac  {d\sigma} { d \cos \vartheta_T }  \right)^{-1}
     \frac { d^2 \sigma (Y,\cos \vartheta_T) } {dY d \cos \vartheta_T }
\end{equation}

\nin which is more appropriate for the study of residual QCD effects. \\

The strong coupling $\rm \alpha_s $ at the renormalization
scale $\rm \mu $ is in second order perturbative QCD expressed as

\begin{equation} 
    \alpha_s( \mu ) = \frac
    {1}   { \beta_0 \ln \frac { \mu^2 } { \Lambda^2 } }
    \left(  1 - \frac { \beta_1 } { \beta_0^2 }
                \frac { \ln \ln \frac { \mu^2 } { \Lambda^2 } }
                      { \ln \frac { \mu^2 } { \Lambda^2 } }           
    \right)  ~~~,
\end{equation} 

\nin where $\rm \Lambda \equiv \Lambda_{ \overline{MS}}^{(5)} $ is the QCD scale 
parameter computed in the Modified Minimal Subtraction 
$\rm (\overline{MS}) $ scheme for $ n_f = 5 $ flavours and 
$\beta_1 = (153-19 n_f) / 24 \pi^2 $. \\

The renormalization scale $\rm \mu $ is a formally unphysical parameter
and should not enter at all into an exact infinite order calculation 
\cite{ApplicPQCD}. Within the context of a truncated finite order 
perturbative expansion for any particular process under consideration,
the definition of $\rm \mu $ depends on the renormalization scheme
employed, and its value is in principle completely arbitrary. This
renormalization scale problem has been discussed extensively in the 
literature \cite{scaledisk,ApplicPQCD,theoambig}. \\

The traditional experimental approach to account for this problem has been
to measure all observables at the same fixed scale value, the so-called
physical scale $ x_{\mu} = 1 $ or equivalently $\rm \mu^2 = Q^2 $. The 
scale dependence has been taken into account by varying $\rm \mu $
over some wide ad hoc range, quoting the resulting change in the QCD 
predictions as theoretical uncertainty. \\

However, the approach of choosing $\ x_{\mu} =1 $ has a severe disadvantage.
If we consider the ratio of the  \oas  and the  \oass  contributions to the
cross-section, defined as

\begin{equation}
\label{asratio}
     r_{NLO}(Y)  = \frac 
                   { \alpha_s ( x_{\mu} ) 
                     \bigg[ B(Y) + A(Y) 
                     \big( 2 \pi \beta_0 \ln ( x_{\mu} ) - 2 
                     \big) 
                     \bigg] 
                   }
                   { A(Y) } ~~~,
\end{equation}

\nin we find for many observables quite large values for the 
second order contributions. In some cases this ratio can have 
a magnitude approaching unity, indicating a poor convergence
behavior of the \oass  predictions in the $\rm \overline{MS} $ scheme 
which would quite naturally result in a wide spread of the measured 
$\rm \alpha_s $ values. This has indeed been observed in previous analyses
using \oass QCD \cite{DelPubs1,OpalPap,NLLAPubs}.  \\

Several proposals have been made which resolve the problem by choosing
optimized scales according to different theoretical prescriptions
\cite{PMSScale,ECHScale,BLMScale}. These methods have been discussed 
with some controversy \cite{ApplicPQCD} and until now no consensus 
has been achieved. The only approach  for determining 
an optimized scale value, which does not rely on specific theoretical 
assumptions, is the experimental evaluation of an optimized \oass scale value 
$ x_{\mu} $ for each measured observable separately. This strategy has
therefore been chosen to be the primary method. For previous analyses 
including experimentally optimized renormalization scales 
see for example \cite{OpalPap,Expscales}. The theoretical
approaches for choosing an optimized renormalization scale value are
studied in detail in Section \ref{theooptsect}. As will be shown later,
the approach of applying experimentally optimized scales yields an impressive 
consistency of the \asmz measurements from different observables. \\

The procedure applied here, is a combined fit of $\rm \alpha_s $
and the scale parameter $ x_{\mu} $. In the past this strategy suffered
from a poor sensitivity of the fit with respect to $ x_{\mu} $ for most
of the observables. Due to the high statistics and high precision data 
now available, one may expect a better sensitivity at least for some of
the observables under consideration.  \\

We determined $\alpha_s(M_Z^2) $ and the renormalization scale factor 
$x_{\mu}$ simultaneously by comparing the corrected distributions 
for each observable $ Y $ with the perturbative QCD calculations corrected
for hadronization effects as described in the previous section. The 
theoretical predictions have been fitted to the measured distributions
$ R(Y,\cos\vartheta_T) $ by minimizing $ \chi^2 $, defined by using the 
sum of the squares of the statistical and systematic experimental errors, 
with respect to the variation of 
$\Lambda_{\overline{MS}} $ and $ x_{\mu} $. \\ 

The fit range for the central analysis, i.e. including the experimental
optimization of $ x_{\mu} $,  was chosen according to the following 
considerations:

\begin{itemize}

\item
Requiring a detector acceptance larger than 80\%, the last bin in 
$\rm \cos \vartheta_T $ was excluded in general, i.e. the fit range 
was restricted to the interval $\rm 0 \le  \cos \vartheta_T < 0.84 $ 
which corresponds to the polar angle interval 
$\rm 32.9^{\circ} < \vartheta_T \le 90.0^{\circ}. $ 

\item 
Acceptance corrections were required to be below about 25\% 
and the hadronization corrections to be below $ \sim $ 40\%.

\item
The contribution of the absolute value of the second order term 
$ r_{NLO}(Y) $ as defined in Eq. (\ref{asratio}) was required  
to be less than one\footnote{This requirement restricts the 
fit interval only for the total jet broadening observable 
$ B_{sum} $, which yields rather large \oass contributions 
for any choice of the renormalization scale value.} 
over the whole fit range. 

\item
The requirement that the data can be well described by the theoretical
prediction, i.e.  $ \chi^2 / n_{df} $ is approximately 1 and stable
over  the fit range.

\item
Stability of the $\rm \alpha_s $ - measurement with respect to the variation
of the fit range.
 
\end{itemize}

For the analysis with a fixed renormalization scale value $ x_{\mu} = 1 $,
the requirement that $ \chi^2 / n_{df} $ is approximately 1 can in general
not be applied, since it would cause an unreasonably large reduction
of the fit range for many observables.
The thrust distribution for example  
could be fitted only over a range of at most three bins. 
For further details see also Section \ref{exoptressubsect}. 
The fit ranges for this analysis as well as for the analyses
applying theoretically motivated scale setting methods have therefore 
been chosen identical to the analysis with experimentally optimized 
scale values, regardless of the $\chi^{2}$ values of the fits.
  

\subsection{Systematic and Statistical Uncertainties}

For each observable the uncertainties from the fit of 
$ \alpha_s(M_Z^2) $ and of $ x_{\mu} $ have been determined by
changing the parameters corresponding to a unit increase of $ \chi^2 $. 
In the case of asymmetric errors the higher value was taken.  \\

The systematic experimental uncertainty was estimated by repeating the 
analysis with different selections to calculate the acceptance corrections
as described in Section \ref{selhadrsect}.  
Additionally, an analysis was 
performed including neutral clusters measured with the hadronic and/or 
electromagnetic calorimeters. The overall uncertainty was taken as the 
variance of the individual $\alpha_s(M_Z^2) $ measurements. \\

An additional source of experimental uncertainty arises from the determination
of the fit range, which has been estimated by varying the lower and
the upper edge of the fit range by $ \pm 1 $ bin, respectively, while the
other edge is kept fixed. Half of the maximum deviation in $ \alpha_s(M_Z^2) $ 
has been taken as the error due to the variation of the fit range and has
been added in quadrature.  \\

The hadronization uncertainty was determined as described in 
Section \ref{shapedefsect}. \\
 
The total uncertainty on $ \alpha_s(M_Z^2) $ is determined from the sum
of the squares of the errors listed above. 

\subsubsection*{Uncertainties due to Missing Higher Order Calculations} 

An additional source of theoretical uncertainty arises due to the
missing higher order calculations of perturbative QCD.
It is commonly assumed, that the size of these uncertainties 
can be estimated by varying the renormalization
scale value applied for the determination of \asmz within some 
`reasonable' range \cite{asstatus}. The choice of a
`reasonable' range involves subjective jugdement and so far
no common agreement about the size of this range has been achieved. 
Furthermore, this commonly used approach has been criticized in the 
literature \cite{scaledisk}. According to \cite{scaledisk}
any artificial increase of the uncertainty of \asmz due to a 
large variation of the renormalization scale should be avoided 
so that the degree of precision to which QCD can be tested 
remains transparent. It should be pointed 
out that no such additional uncertainty is required to understand 
the scatter of the measurements from a large number of observables 
if experimentally optimized renormalization scale values are applied. 
This will be demonstrated in the following section. 
Other procedures for estimating uncertainties due to  missing
higher order corrections have been suggested, in particular the comparison
of \asmz values obtained by applying different reasonable renormalization
schemes or by replacing the missing higher order terms by their 
Pad\'{e} Approximants \cite{asstatus}. Both strategies have been studied.
By comparing the size of the uncertainties derived applying these
methods (see e.g. Table \ref{jcefsummar}), a variation of 
$ x_{\mu} $ between $ 0.5 \cdot x_{\mu}^{exp} $ and $ 2 \cdot x_{\mu}^{exp} $
seems justified to obtain an estimate of these uncertainties.
Similar or identical ranges have for example also been chosen 
in \cite{DelPubs2,taures}. 


\subsection{Results}    
\label{exoptressubsect}

\begin{figure}
\vspace{-1.5 cm}
\hspace{-1.5 cm}
\begin{center}
\mbox{\epsfig{file=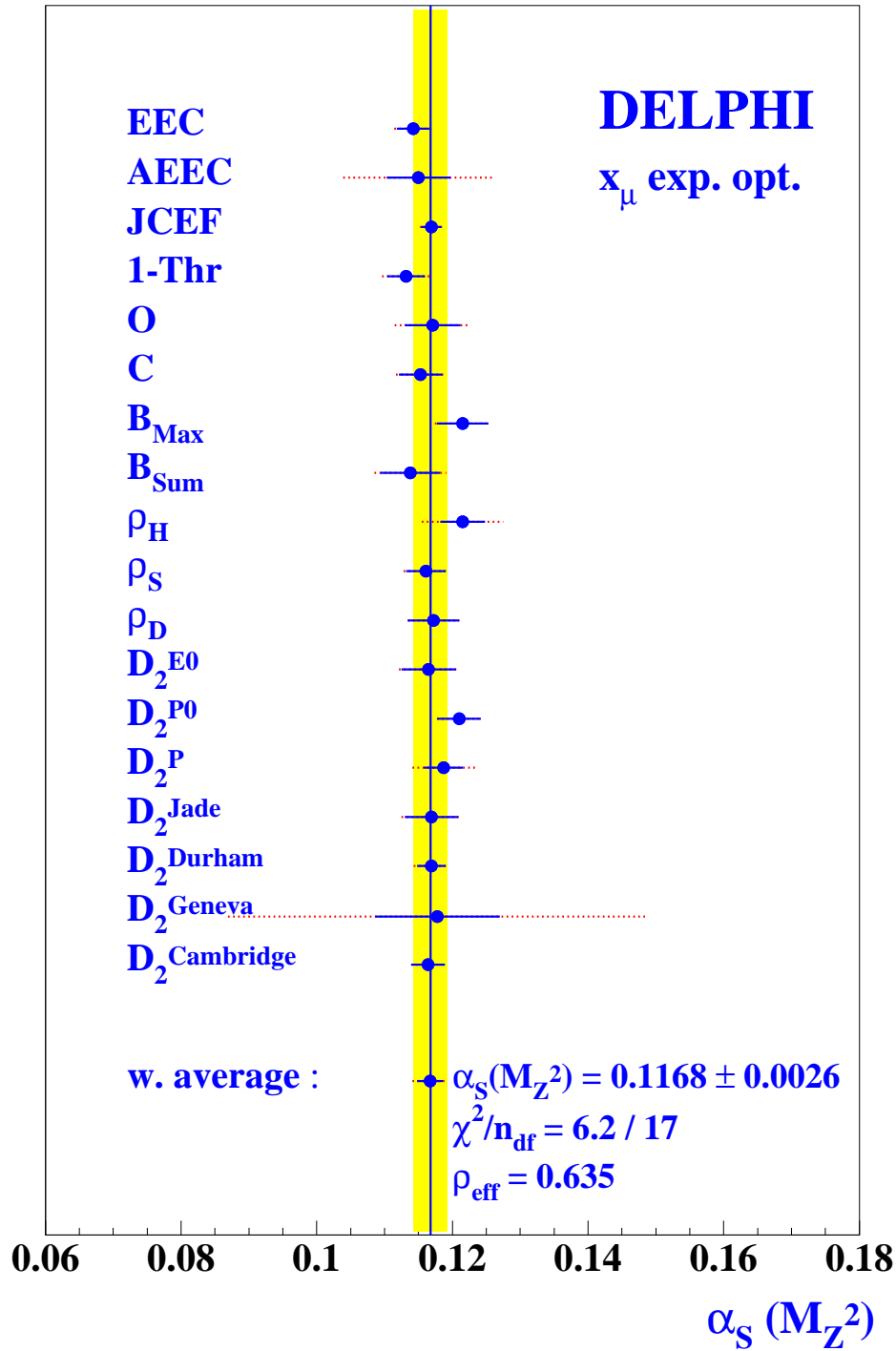,width=14.0cm}}

\caption[]{Results of the QCD fits applying experimentally optimized scales
           for  18 event shape distributions. 
           The error bars indicated by the solid lines
           are the quadratic sum of the experimental and the hadronization 
           uncertainty. The error bars indicated by the dotted lines include
           also the additional uncertainty due to the variation of the 
           renormalization scale due to scale variation around the central 
           value $ x_{\mu}^{exp} $ in the range 
           $ 0.5 \cdot x_{\mu}^{exp} \le x_{\mu} \le 2 \cdot x_{\mu}^{exp} $. 
           Also shown is the correlated weighted average (see text). The 
           $ \chi^{2}$-value is given before readjusting according to 
           Eq. \ref{chisquare}. 
           }
\label{uaverexp}
\end{center}
\end{figure}

The results of the fits to the 18 event shape 
distributions are summarized in tables \ref{results} and \ref{expres} 
and shown in Figure \ref{uaverexp}. It should be noted that for all
observables the normalized $ \chi^{2} $ is about one for a typically 
large number of degrees of freedom $ (n_{df}=16 - 236 $, see Table \ref{results}). 
The individual errors contributing 
to the total error on the value of $\alpha_s $ are listed in Table \ref{expres}. 
Among the observables considered the $ \rm JCEF $ yields the most precise result. \\


\begin{table} [b]
\begin{center}
\begin{tabular} { l c c c c c}
\hline
\hline

Observable        & Fit Range  & $ \cos\vartheta_T $ Range  &
                    $ x_{\mu} $         & $\chi^2 / n_{df} $ & $ n_{df} $    \\
\hline

$\rm EEC             $   &  $28.8^{\circ} - 151.2^{\circ}$ & 0.0 - 0.84  &
                         $ 0.0112 \pm 0.0006 $    & 1.02   & 236        \\ 

$\rm AEEC            $   &  $25.2^{\circ} -  64.8^{\circ}$ & 0.0 - 0.84  &
                         $ 0.0066 \pm 0.0018   $    & 0.98 & 75         \\ 

$\rm JCEF            $   & $104.4^{\circ} - 169.2^{\circ}$ & 0.0 - 0.84  &
                         $ 0.0820 \pm 0.0046 $    & 1.05   & 124        \\ 

$\rm 1-T             $   &  0.05 - 0.30              & 0.0 - 0.84     & 
                         $ 0.0033 \pm 0.0002 $    & 1.24   & 89      \\ 

$\rm O               $   &  0.24 - 0.44              & 0.0 - 0.84     &
                         $ 2.30 \pm 0.40     $    & 0.90   & 33      \\ 

$\rm C               $   &  0.24 - 0.72              & 0.0 - 0.84     &
                         $ 0.0068 \pm 0.0006 $    & 1.02   & 82      \\ 

$\rm B_{Max}         $   &  0.10 - 0.24              & 0.0 - 0.84     &  
                         $ 0.0204 \pm 0.0090  $   & 0.89   & 47      \\ 

$\rm B_{Sum}         $   &  0.12 - 0.24              & 0.0 - 0.84     &  
                         $ 0.0092 \pm 0.0022 $    & 1.19   & 40      \\ 

$\rm \rho_H          $   &  0.03 - 0.14              & 0.0 - 0.84     &
                         $ 0.0036 \pm 0.0004 $    & 0.63   & 54      \\ 

$\rm \rho_S          $   &  0.10 - 0.30              & 0.0 - 0.36     & 
                         $ 0.0027 \pm 0.0019 $    & 0.82   & 16      \\ 

$\rm \rho_D          $   &  0.05 - 0.30              & 0.0 - 0.84     &
                         $ 2.21  \pm 0.38  $      & 1.02   & 68   \\ 

$\rm D_2^{E0}        $   &  0.07 - 0.25              & 0.0 - 0.84     &  
                         $ 0.048  \pm 0.020   $   & 0.85   & 68     \\ 

$\rm D_2^{P0}        $   &  0.05 - 0.18              & 0.0 - 0.84     &
                         $ 0.112  \pm 0.048   $   & 1.02   & 68     \\ 

$\rm D_2^{P}         $   &  0.10 - 0.25              & 0.0 - 0.84     &
                         $ 0.0044 \pm 0.0004  $   & 1.00   & 47     \\ 

$\rm D_2^{Jade}      $   &  0.06 - 0.25              & 0.0 - 0.84     & 
                         $ 0.126 \pm 0.049   $    & 1.05   & 75      \\ 

$\rm D_2^{Durham}    $   &  0.015 - 0.16             & 0.0 - 0.84     & 
                         $ 0.0126 \pm 0.0015 $    & 0.92   & 96      \\ 

$\rm D_2^{Geneva}    $   &  0.015 - 0.03             & 0.0 - 0.84     &
                         $ 7.10 \pm 0.28     $    & 0.84   & 19      \\ 

$\rm D_2^{Cambridge} $   &  0.011 - 0.18             & 0.0 - 0.84     &
                         $ 0.066 \pm 0.019   $    & 0.98   & 145     \\ 

\hline
\hline
\end{tabular}
\end{center}
\caption[]{ Observables used in the \oass QCD fits. For each of the observables 
            the fit range, the range in $ \cos \vartheta_T $, 
            the measured renormalization scale factor $  x_{\mu} $ together 
            with the uncertainty as determined from the 
            fit, the $\chi^2 / n_{df} $ and the
            number of degrees of freedom $  n_{df} $ are shown.    
            In the case of asymmetric errors the higher value is given.
           }
\label{results}
\end{table}

\begin{figure}
\vspace{-1.5 cm}
\hspace{-1.5 cm}
\begin{center}
\mbox{\epsfig{file=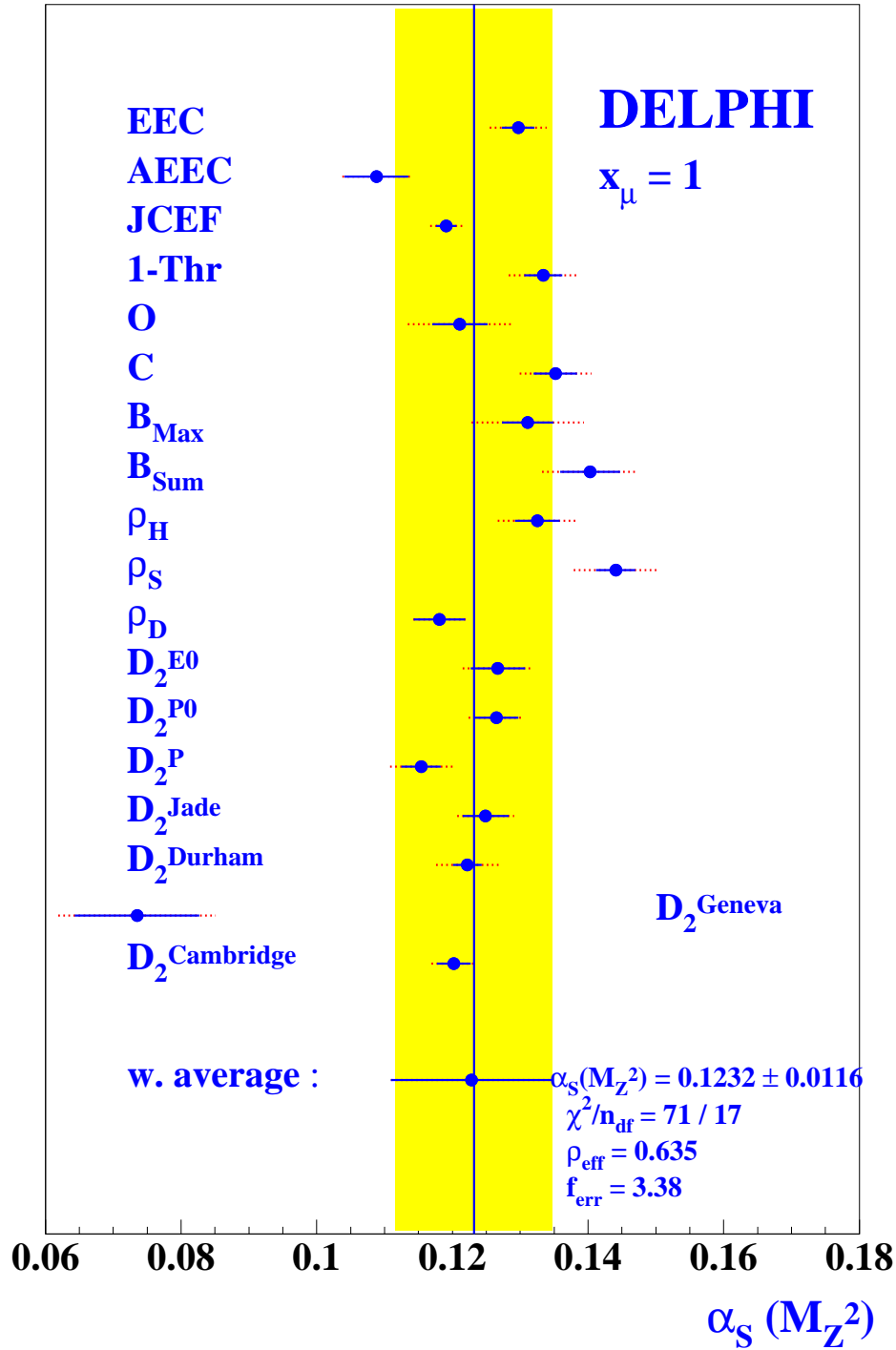,width=14.0cm}}
\caption[]{Results of the QCD fits applying a fixed renormalization scale
           $ x_\mu = 1 $ .           
           The error bars indicated by the solid lines
           are the quadratic sum of the experimental and the hadronization 
           uncertainty. The error bars indicated by the dotted lines include
           also the additional uncertainty due to the variation of the 
           renormalization scale around the central value $ x_{\mu}^{exp} $ from 
           $ 0.5 \cdot x_{\mu}^{exp} \le x_{\mu} \le 2 \cdot x_{\mu}^{exp} $. 
           Also shown is the correlated weighted average. It has been
           calculated assuming the same effective correlation  
           $ \rhoe = 0.635 $ as for the fit results applying experimentally
           optimized scales. The \chindf for the weighted average is 71/17, where
           the $ \chi^{2} $ given corresponds to the value before adjusting $ \rhoe $.
           In order to yield $ \chi^2 / n_{df} = 1 $, the errors have to be scaled 
           by a factor $\rm f_{err} =3.38 $. }
\label{uaverfix}
\end{center}
\end{figure}

For comparison, the data have also been fitted in \oass applying 
a fixed renormalization scale value $ x_{\mu} = 1 $. The results of these fits 
are summarized in table \ref{fixres} and shown in Figure \ref{uaverfix}. 
As can be seen from the $ \chi^2 / n_{df} $ values of the fits, 
the choice of $ x_{\mu} = 1 $ yields only a poor description of the data 
for most of the observables, for many observables the description is even 
unacceptable. \\ 


\begin{table} [b]
\begin{center}
\begin{tabular} { l c c c c c }
\hline
\hline

Observable        &   \asmz  &   \das (Exp.) &  \das (Hadr.)       
                  &   \das (Scale.) & \das (Tot.)                    \\
\hline

$\rm EEC          $
                      &  0.1142       &  $\pm$ 0.0007 &  $\pm$ 0.0023  
                      &  $\pm$ 0.0014 &  $\pm$ 0.0028                    \\  

$\rm AEEC         $
                      &  0.1150       &  $\pm$ 0.0037 &  $\pm$ 0.0029  
                      &  $\pm$ 0.0100 &  $\pm$ 0.0111                    \\  

$\rm JCEF         $
                      &  0.1169       &  $\pm$ 0.0006 &  $\pm$ 0.0013  
                      &  $\pm$ 0.0008 &  $\pm$ 0.0017                    \\  

$\rm 1-T          $  
                      &  0.1132       &  $\pm$ 0.0009 &  $\pm$ 0.0026  
                      &  $\pm$ 0.0023 &  $\pm$ 0.0036                    \\  

$\rm O            $
                      &  0.1171       &  $\pm$ 0.0028 &  $\pm$ 0.0030  
                      &  $\pm$ 0.0038 &  $\pm$ 0.0056                    \\  

$\rm C            $
                      &  0.1153       &  $\pm$ 0.0021 &  $\pm$ 0.0023  
                      &  $\pm$ 0.0017 &  $\pm$ 0.0036                    \\  

$\rm B_{Max}      $
                      &  0.1215       &  $\pm$ 0.0022 &  $\pm$ 0.0031  
                      &  $\pm$ 0.0013 &  $\pm$ 0.0041                    \\  

$\rm B_{Sum}      $
                      &  0.1138       &  $\pm$ 0.0030 &  $\pm$ 0.0032  
                      &  $\pm$ 0.0030 &  $\pm$ 0.0053                    \\  

$\rm \rho_H       $
                      &  0.1215       &  $\pm$ 0.0014 &  $\pm$ 0.0029  
                      &  $\pm$ 0.0050 &  $\pm$ 0.0060                    \\  

$\rm \rho_S       $
                      &  0.1161       &  $\pm$ 0.0014 &  $\pm$ 0.0018  
                      &  $\pm$ 0.0016 &  $\pm$ 0.0033                    \\  

$\rm \rho_D       $ 
                      &  0.1172       &  $\pm$ 0.0013 &  $\pm$ 0.0034  
                      &  $\pm$ 0.0007 &  $\pm$ 0.0038                    \\  

$\rm D_2^{E0}     $
                      &  0.1165       &  $\pm$ 0.0027 &  $\pm$ 0.0029  
                      &  $\pm$ 0.0017 &  $\pm$ 0.0044                    \\  

$\rm D_2^{P0}     $
                      &  0.1210       &  $\pm$ 0.0018 &  $\pm$ 0.0026  
                      &  $\pm$ 0.0009 &  $\pm$ 0.0033                    \\  

$\rm D_2^{P}      $
                      &  0.1187       &  $\pm$ 0.0019 &  $\pm$ 0.0021  
                      &  $\pm$ 0.0036 &  $\pm$ 0.0046                    \\  

$\rm D_2^{Jade}   $
                      &  0.1169       &  $\pm$ 0.0011 &  $\pm$ 0.0020  
                      &  $\pm$ 0.0028 &  $\pm$ 0.0040                    \\  

$\rm D_2^{Durham} $
                      &  0.1169       &  $\pm$ 0.0013 &  $\pm$ 0.0016  
                      &  $\pm$ 0.0015 &  $\pm$ 0.0026                    \\  

$\rm D_2^{Geneva} $
                      &  0.1178       &  $\pm$ 0.0052 &  $\pm$ 0.0075  
                      &  $\pm$ 0.0295 &  $\pm$ 0.0309                    \\  

$\rm D_2^{Cambridge} $
                      &  0.1164       &  $\pm$ 0.0008 &  $\pm$ 0.0023  
                      &  $\pm$ 0.0004 &  $\pm$ 0.0025                    \\  

\hline
\hline
\end{tabular}
\end{center}
\caption[]{ Individual sources of errors of the \asmz measurement. For each 
            observable, the value of \asmz, the experimental uncertainty 
            (statistical and systematic), the uncertainty resulting
            from hadronization corrections, the theoretical uncertainty due to
            scale variation around the central value $ x_{\mu}^{exp} $
            in the range 
            $ 0.5 \cdot x_{\mu}^{exp} \le x_{\mu} \le 2 \cdot x_{\mu}^{exp} $ 
            and the total uncertainty are shown. }
\label{expres}
\end{table}
\hspace{-1. cm}


\begin{table} [b]
\begin{center}
\begin{tabular} { l c c c c }
\hline
\hline

Observable  & \asmz & \das(Scale) & \das (Tot.) & $ \chi^2 / n_{df} $ \\

\hline

$\rm EEC             $  &  0.1297 &  $\pm$ 0.0037 &  $\pm$ 0.0042 &  10.7  \\ 
$\rm AEEC            $  &  0.1088 &  $\pm$ 0.0015 &  $\pm$ 0.0050 &  2.04  \\
$\rm JCEF            $  &  0.1191 &  $\pm$ 0.0012 &  $\pm$ 0.0024 &  7.7   \\  
$\rm 1-T             $  &  0.1334 &  $\pm$ 0.0042 &  $\pm$ 0.0051 &  25.9  \\  
$\rm O               $  &  0.1211 &  $\pm$ 0.0065 &  $\pm$ 0.0077 &  2.38  \\  
$\rm C               $  &  0.1352 &  $\pm$ 0.0043 &  $\pm$ 0.0053 &  12.0  \\  
$\rm B_{Max}         $  &  0.1311 &  $\pm$ 0.0073 &  $\pm$ 0.0083 &  1.67  \\  
$\rm B_{Sum}         $  &  0.1403 &  $\pm$ 0.0056 &  $\pm$ 0.0071 &  8.1   \\  
$\rm \rho_H          $  &  0.1325 &  $\pm$ 0.0036 &  $\pm$ 0.0049 &  5.1   \\  
$\rm \rho_S          $  &  0.1441 &  $\pm$ 0.0055 &  $\pm$ 0.0062 &  2.16  \\  
$\rm \rho_D          $  &  0.1181 &  $\pm$ 0.0012 &  $\pm$ 0.0039 &  1.54  \\  
$\rm D_2^{E0}        $  &  0.1267 &  $\pm$ 0.0033 &  $\pm$ 0.0052 &  1.35  \\  
$\rm D_2^{P0}        $  &  0.1265 &  $\pm$ 0.0026 &  $\pm$ 0.0041 &  1.31  \\  
$\rm D_2^{P}         $  &  0.1154 &  $\pm$ 0.0019 &  $\pm$ 0.0036 &  5.35  \\  
$\rm D_2^{Jade}      $  &  0.1249 &  $\pm$ 0.0030 &  $\pm$ 0.0042 &  1.53  \\  
$\rm D_2^{Durham}    $  &  0.1222 &  $\pm$ 0.0034 &  $\pm$ 0.0046 &  3.47  \\  
$\rm D_2^{Geneva}    $  &  0.0735 &  $\pm$ 0.0071 &  $\pm$ 0.0116 &  120.  \\  
$\rm D_2^{Cambridge} $  &  0.1202 &  $\pm$ 0.0021 &  $\pm$ 0.0033 &  1.32  \\  

\hline
\hline
\end{tabular}
\end{center}
\caption[]{ Results of the \asmz measurements using a fixed 
            renormalization scale $ x_\mu = 1 $. For each observable, 
            the value of \asmz,  the uncertainty from the variation 
            of the scale between $ 0.5 \le x_\mu \le 2 $, the total
            uncertainty and the $\chi^2 / n_{df} $ of the fit are shown. }
\label{fixres}
\end{table}

\begin{sidewaysfigure}      
\begin{center}
\hspace{-1. cm}
\mbox{\epsfig{file=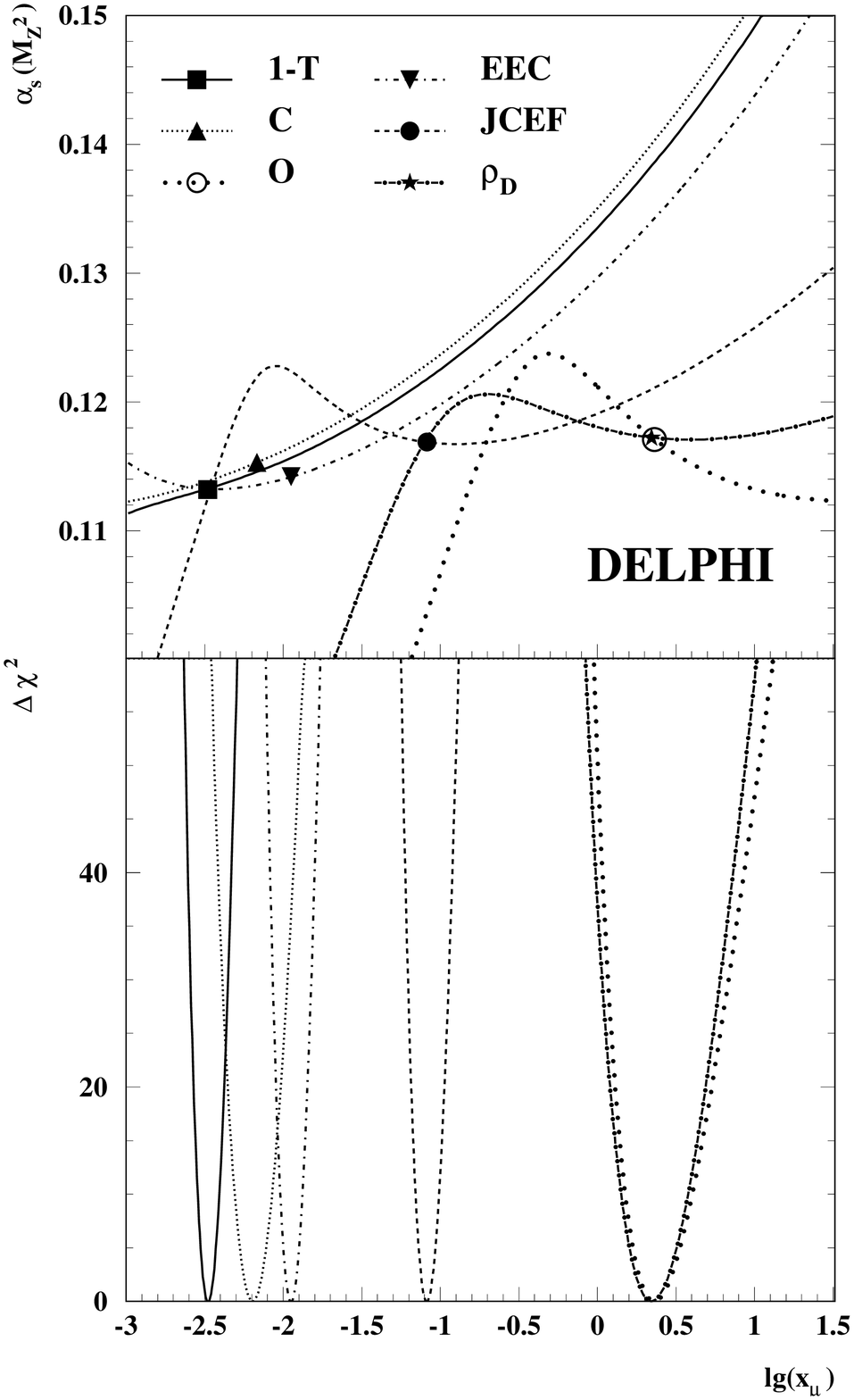, width=12.0cm, height=16.0cm }}
\hspace{-1. cm}
\mbox{\epsfig{file=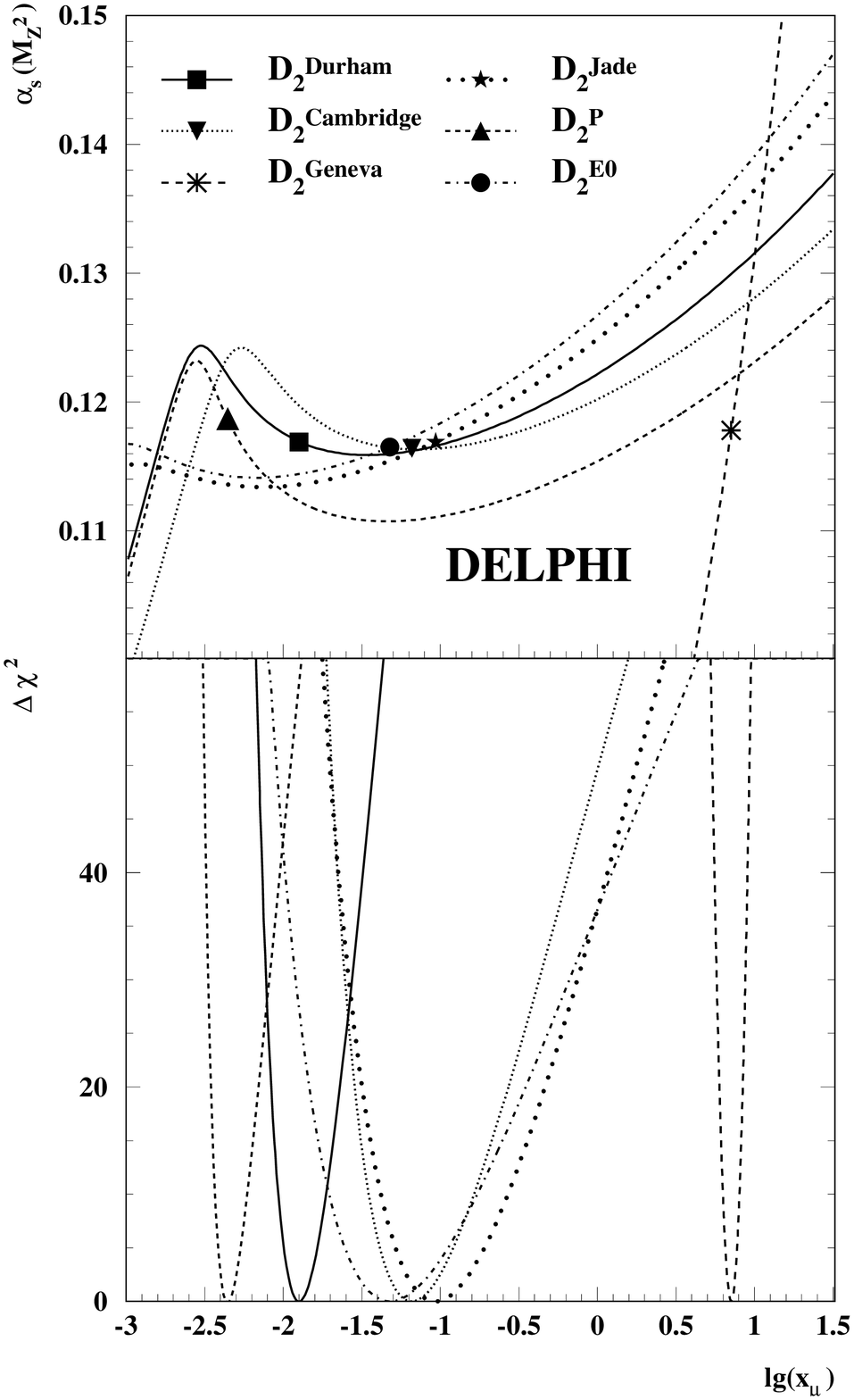, width=12.0cm, height=16.0cm }}
\hspace{-1. cm}
\caption[]{ ({\it left side})  \asmz and $\rm \Delta \chi^2 = \chi^2 - \chi^2_{min} $ 
                               from \oass fits to the double differential distributions
                               in $\rm \cos \vartheta_T $ and $\rm 1-T $, $\rm C $, 
                               $\rm O $, $ \rm EEC $, $ \rm JCEF $, $\rho_D $ . 
                               Additionally, the $ \chi^2 $ minima 
                               are indicated in the \asmz curves.  
            ({\it right side}) The same for the double differential
                               distributions in $\rm \cos \vartheta_T $ and the 
                               differential  2-jet rate applying the Durham, Cambridge,
                               Geneva, Jade, P and E0 jet algorithm. }
\label{Scale}
\end{center}
\end{sidewaysfigure}
\begin{sidewaysfigure}
\begin{center}
\hspace{1. cm}  
\vspace{1. cm}
\mbox{\epsfig{file=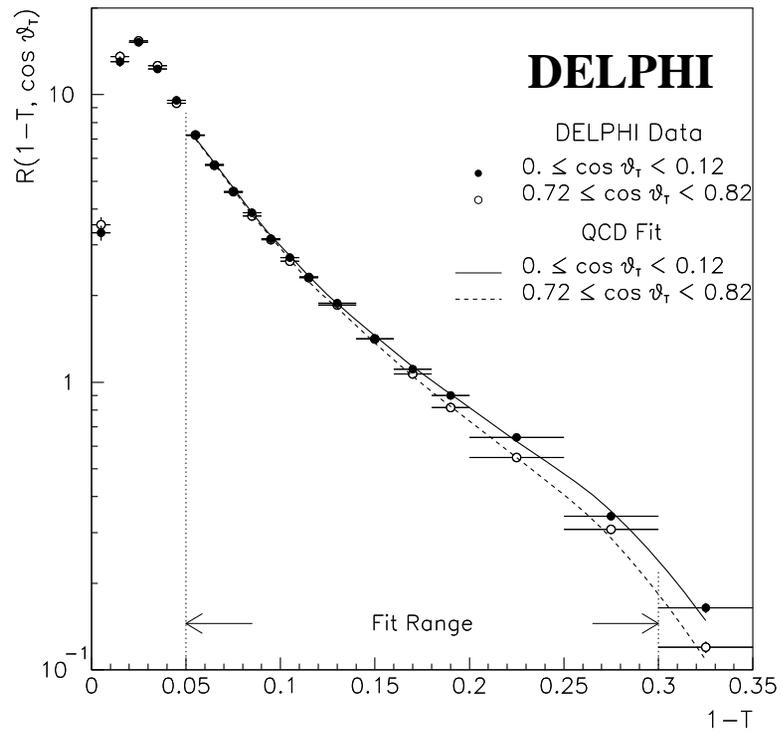, width=11.0cm}}
\mbox{\epsfig{file=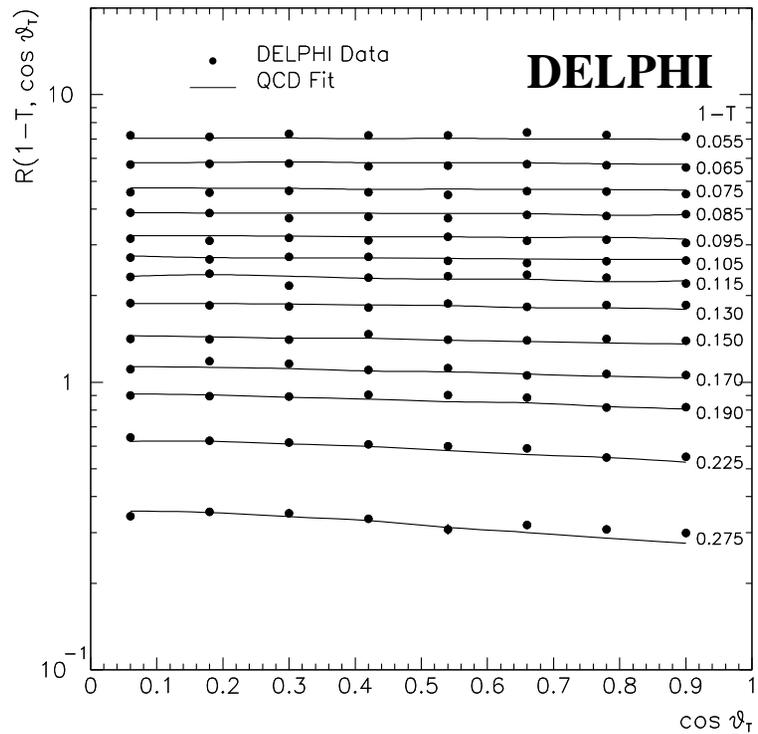, width=11.0cm}}  
\caption[]{({\it left side})  QCD fits to the measured thrust distribution 
                              for two bins in $ \rm \cos \vartheta_T $. 
           ({\it right side}) Measured thrust distribution at various fixed 
                              values of $\rm 1-T $ as a function of 
                              $ \rm \cos \vartheta_T $. The solid lines represent 
                              the QCD fit. } 
\label{DataThr}
\end{center}
\end{sidewaysfigure}
\begin{sidewaysfigure} 
\begin{center}
\hspace{1. cm}
\vspace{1. cm}
\mbox{\epsfig{file=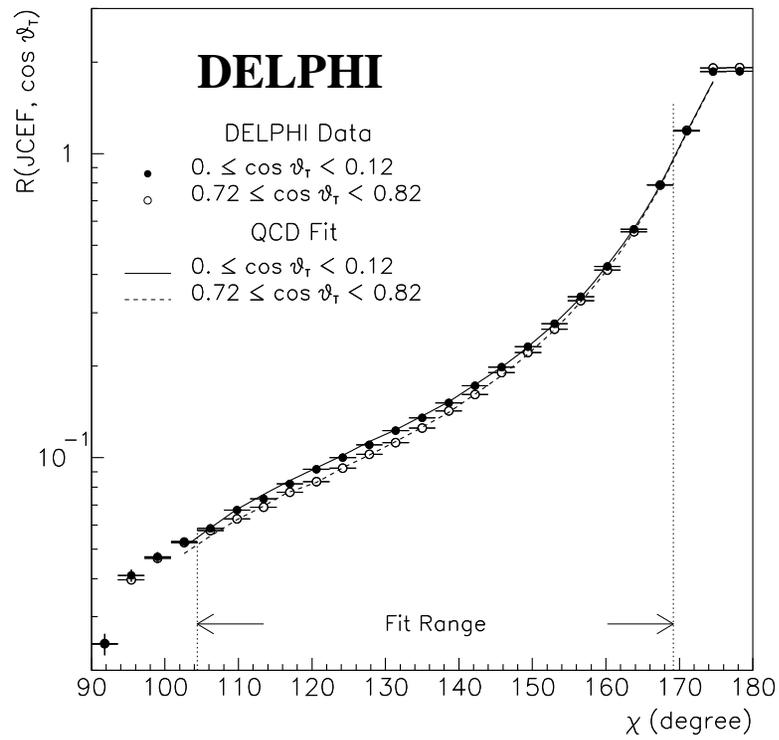, width=11.0cm}}
\mbox{\epsfig{file=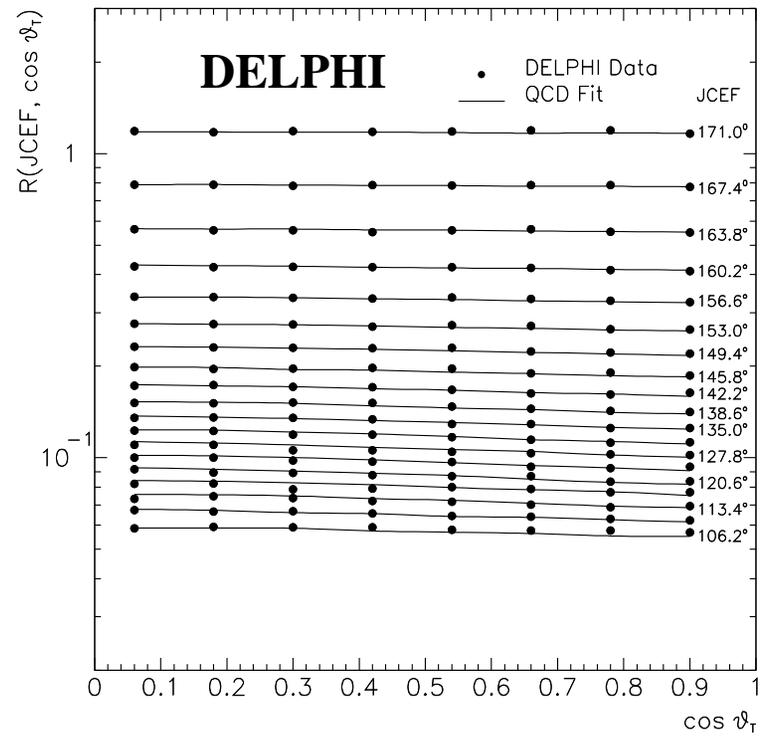, width=11.0cm}}
\caption[]{Same as Figure 6 but for the jet cone energy fraction $\rm JCEF $ }
\label{DataJCEF}
\end{center}
\end{sidewaysfigure}
\begin{sidewaysfigure} 
\begin{center}
\hspace{1. cm}
\vspace{1. cm}  
\mbox{\epsfig{file=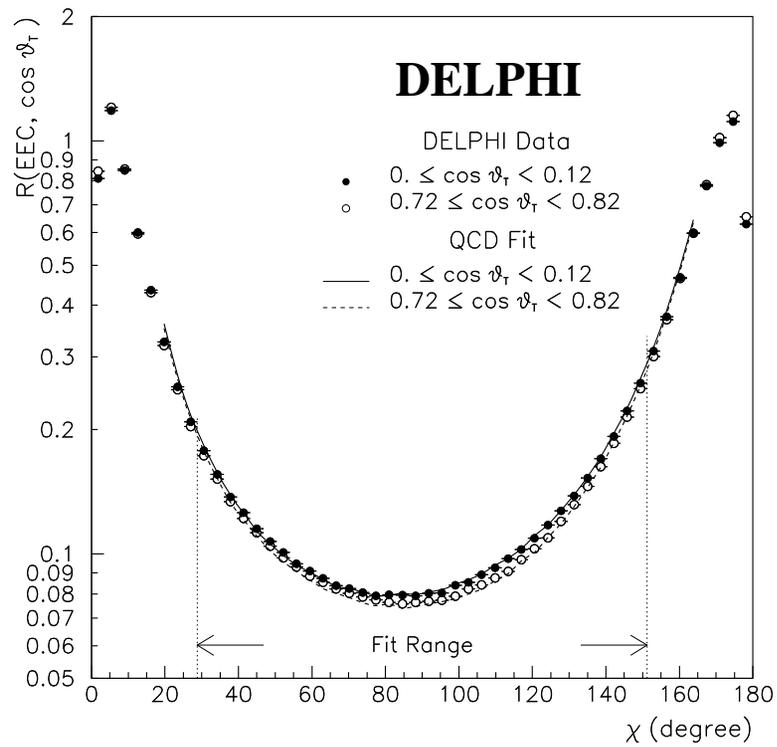, width=11.0cm}}
\mbox{\epsfig{file=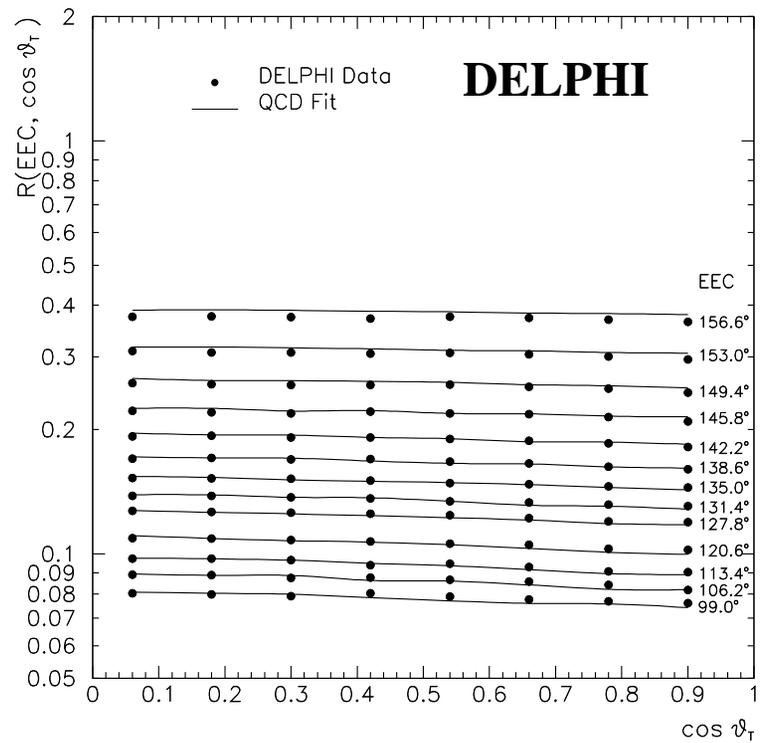, width=11.0cm}}  
\caption[]{Same as Figure 6 but for the energy energy correlation $\rm EEC $.}
\label{DataEEC}
\end{center}
\end{sidewaysfigure}
\begin{sidewaysfigure}
\begin{center}
\hspace{1. cm}
\vspace{1. cm}  
\mbox{\epsfig{file=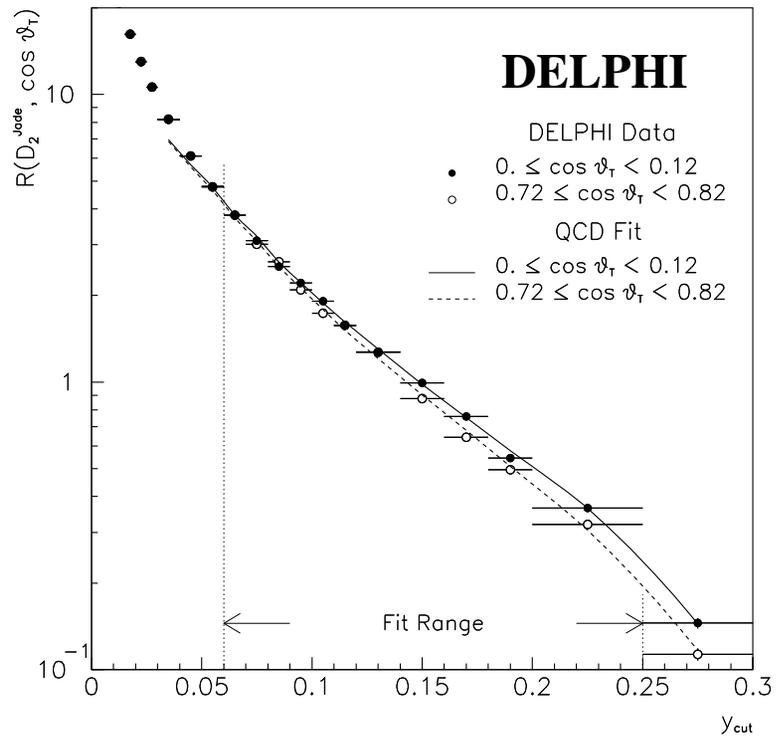, width=11.0cm}}
\mbox{\epsfig{file=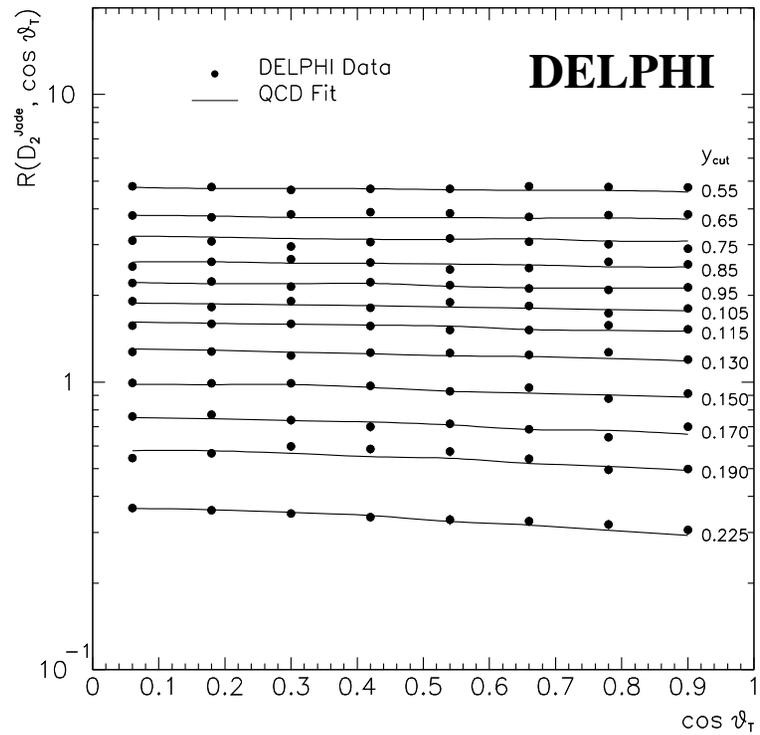, width=11.0cm}}  
\caption[]{Same as Figure 6 but for the differential two jet rate 
         with the Jade Algorithm $\rm D_2^{Jade} $.}
\label{DataD2Jad}
\end{center}
\end{sidewaysfigure}

More details concerning the QCD fits are presented in Figures \ref{Scale}
to \ref{DataD2Jad}. Figure \ref{Scale} shows the values 
of \asmz and the corresponding values of $ \rm \Delta \chi^2 $, i.e. the
change of $ \chi^{2} $ with respect to the optimal value, for the fits 
as a function of the scale $ \rm \lg (x_{\mu}) $ for some of the investigated 
observables. The shape of the $ \rm \Delta \chi^2 $ curves indicates 
that for most distributions the renormalization scale has to be fixed to a 
rather narrow range of values in order to be consistent with the data. 
For most of the observables the renormalization scale dependence of \asmz 
is significantly smaller in the region of the scale value for the minimum 
in $ \chi^2 / n_{df} $ than for the region around $ x_{\mu} = 1 $. 
It should however be noted, that even for observables
exhibiting a strong scale dependence of \asmz, e.g. $ D_2^{Geneva} $,
the \asmz value for the experimentally optimized scale value is perfectly
consistent with the average value.  \\
       
Figures \ref{DataThr} to \ref{DataD2Jad} contain a direct comparison of the 
data measured at various bins in $ \rm \cos \vartheta_T $ with the 
results of the QCD fits. The measured dependence on both 
$ \rm \cos \vartheta_T $ and the studied observable are precisely 
reproduced by the fits. \\

\begin{sidewaysfigure} 
\begin{center}
\hspace{-1. cm}
\mbox{\epsfig{file=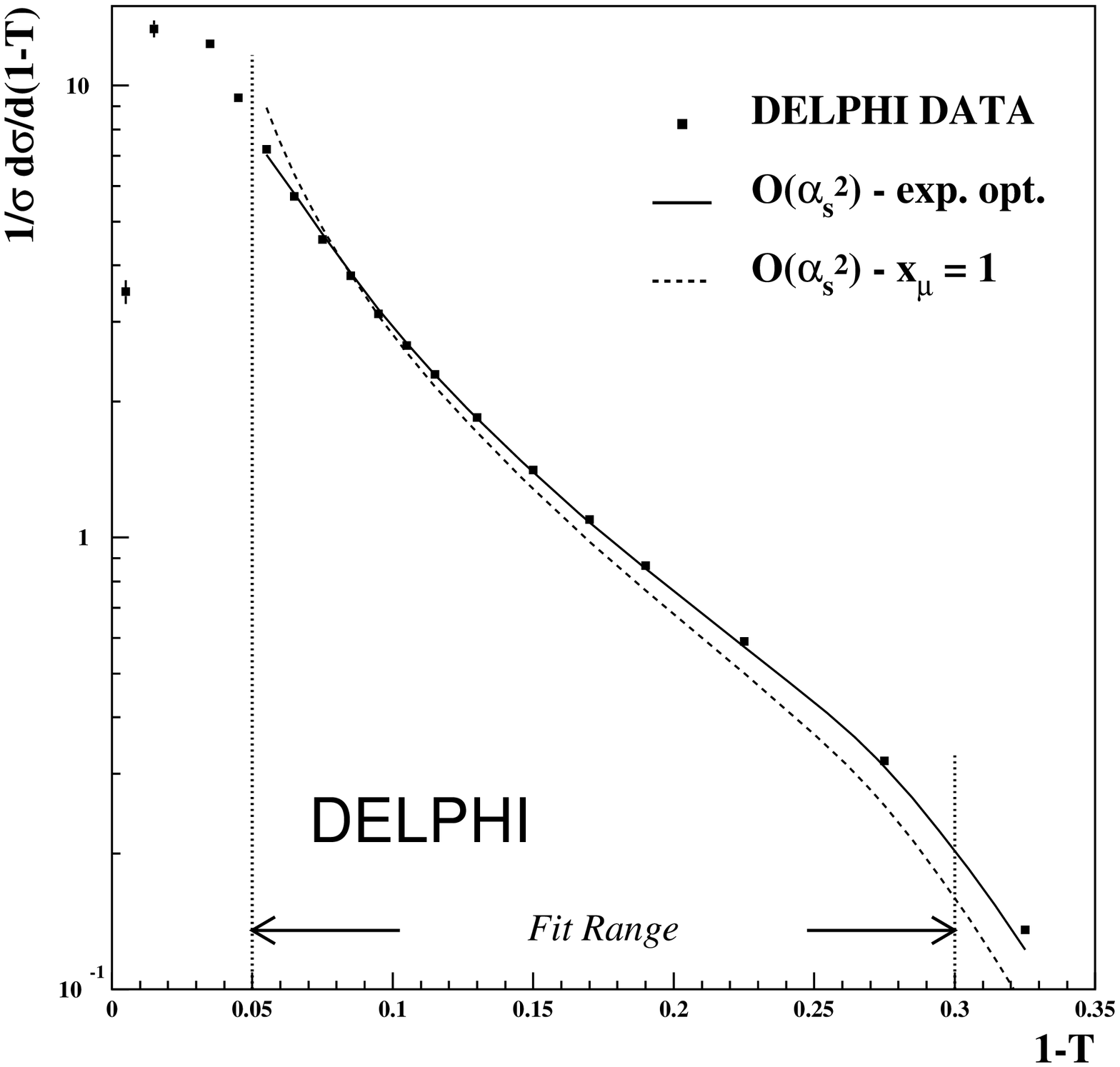,width=8.2 cm}}
\hspace{-0.2 cm}
\mbox{\epsfig{file=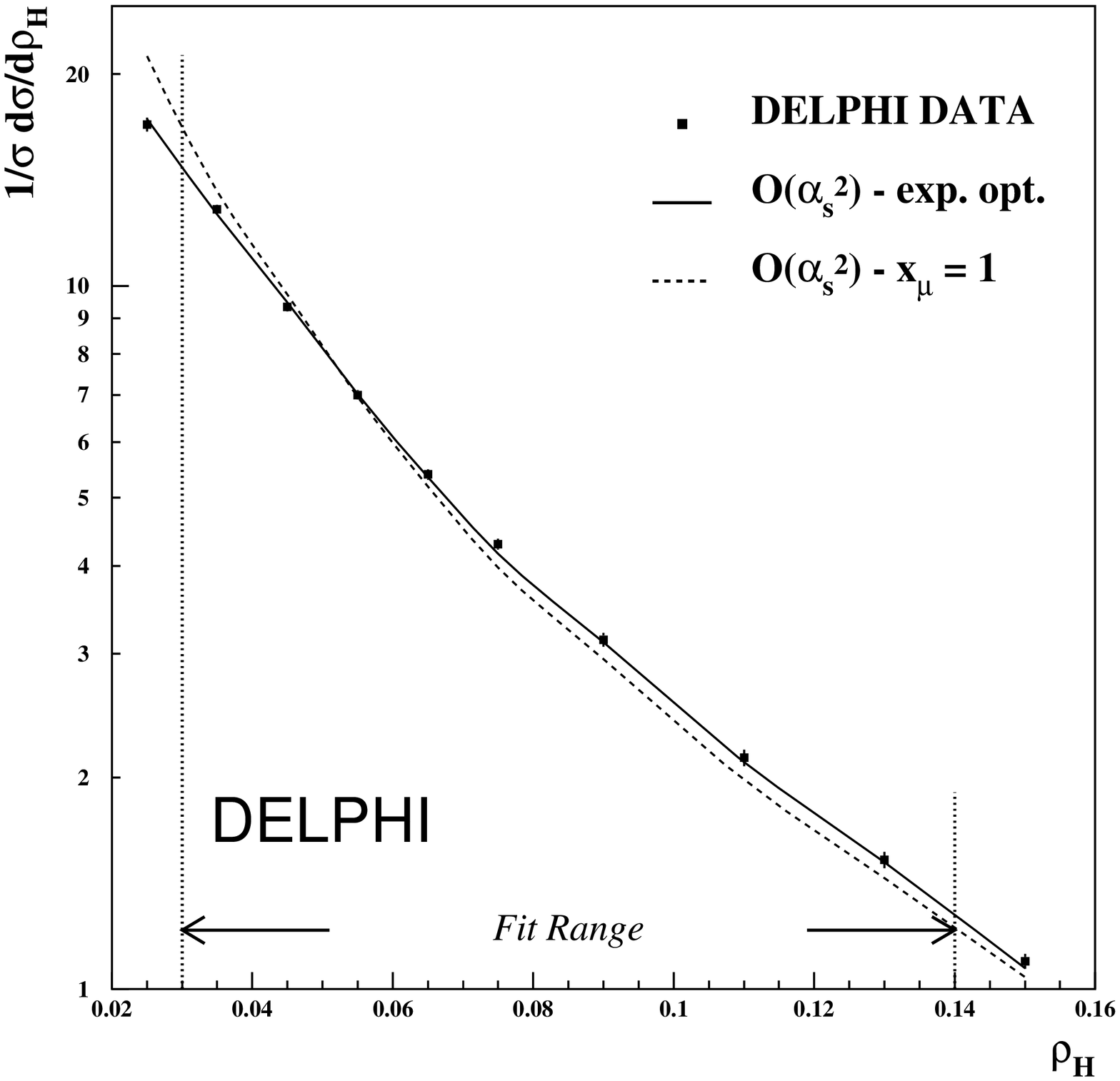,width=8.2 cm}}
\hspace{-0.2 cm}
\mbox{\epsfig{file=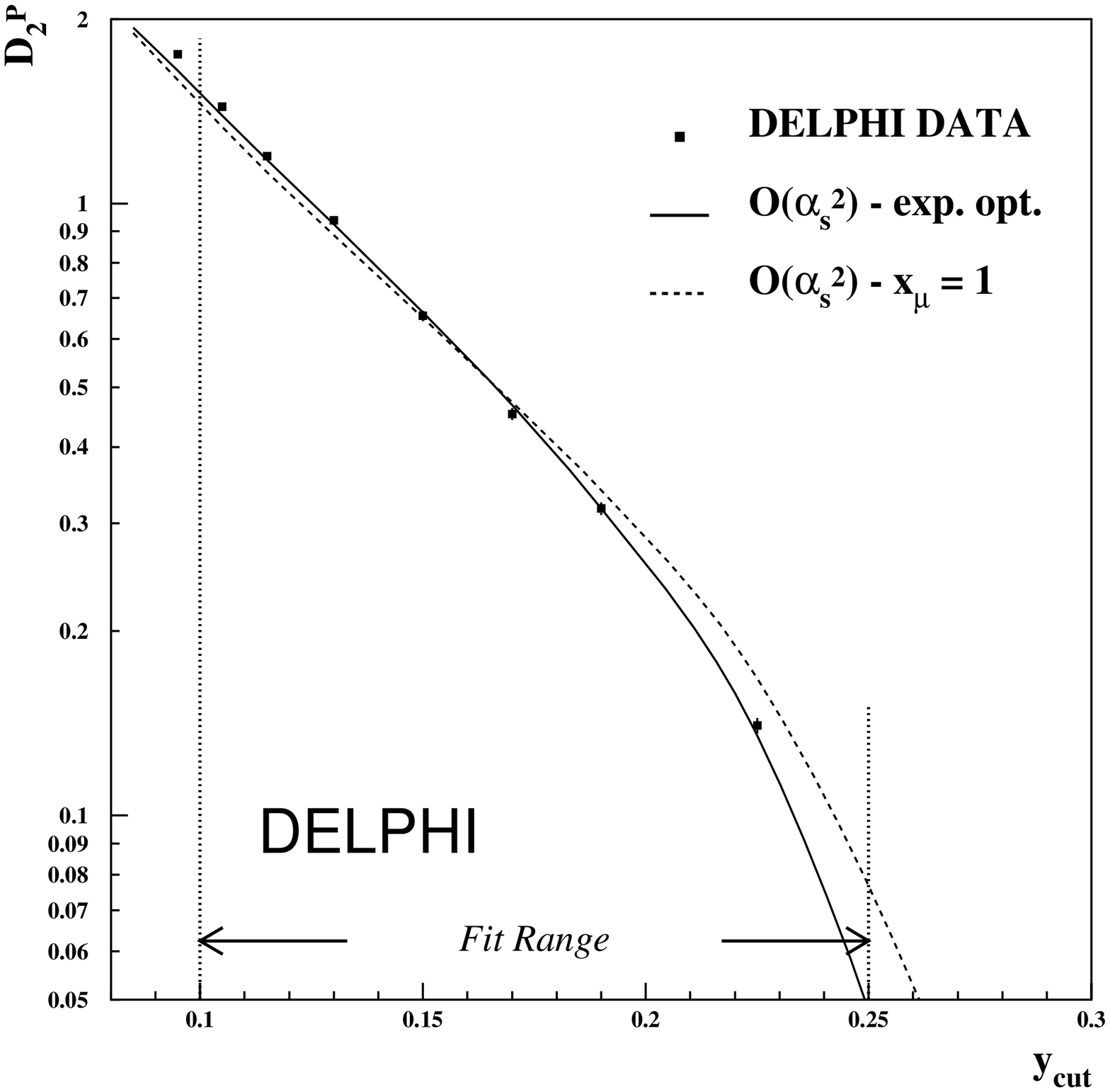,width=8.2 cm}}
\hspace{-1. cm}
\vspace{1. cm}
\caption[] {Comparison between \oass QCD fits with $ x_{\mu} = 1 $ and 
            experimentally optimized renormalization scales for the 
            observables $ 1-T $, $ \rho_H $ and $ D_2^{P} $. DELPHI
            data and theoretical predictions are shown averaged over
            $ \cos \vartheta_T $. The full (dotted) line correspond
            to fits with experimentally optimized renormalization
            scale values $ x_{\mu}^{exp} $ ($ x_{\mu} = 1 $), 
            respectively. The theoretical prediction is averaged 
            over the bin width and drawn through the bin center.
            }
\label{xmu_vgl}
\end{center}
\end{sidewaysfigure}

At this point a comparison with the results from applying a fixed 
renormalization scale value $ x_{\mu} = 1 $ seems appropriate.
Figure \ref{xmu_vgl} shows QCD fits to the data with experimentally optimized 
and with fixed renormalization scale values $ x_{\mu} = 1 $ 
for the observables $ 1-T $, $ \rho_H $ and $ D_2^{P} $.   
It can be seen that the slope of the experimental distributions cannot be 
described by the theoretical prediction applying $ x_{\mu} = 1 $. The same 
observation has also been made for other observables. Good description
of the data can in general only be achieved within the small kinematical 
region where the fit curve intersects with the data. Demanding
$ \chi^{2} / n_{df} \backsimeq 1 $ the thrust distribution for 
example can only be described within a maximum range of three bins in 1-T.
It should be noted, that in contrast to fits applying experimentally 
optimized scales, the stability of \asmz with respect to the choice of 
the fit range is in general quite poor. This observation is new and 
due to the fact that the data used in this analysis have both smaller
statistical and systematical errors than in previous publications.
(See also Section \ref{add-checks}) and \cite{MyThesis}.  \\

Combining the 18 individual results from the \asmz measurements applying
experimentally optimized renormalization scale values by using an 
unweighted average yields \\

\begin{center}
\asmz = 0.1170 $\pm$ 0.0025    \\
\end{center}

\nin whereas the corresponding average for the measurements using the fixed scales
$ x_{\mu} = 1 $ is \asmz $ = 0.1234 \pm 0.0154 $. For the experimentally 
optimized scales the scatter of the individual measurements is significantly reduced.\\

The consistency of the individual measurements using experimentally optimized scales
is clearly shown by the good $\chi^2 / n_{df} = 9.6 / 17 $  for the unweighted average.
This value is computed on the basis of the total uncertainty (experimental and 
hadronization uncertainty) without considering an additional renormalization scale error. 
For the average of the fixed scale $ x_{\mu} = 1 $ measurements 
$\chi^2 / n_{df} = 168 / 17 $, thus in this case the individual
measurements are clearly inconsistent with each other. This inconsistency may
be understood to arise from the imperfect description of the data for the fits with
$ x_{\mu} = 1 $, which cause \as, the parameter of the fit, not to be well defined.   \\

The idea behind the common analysis of such a large number of observables is 
to optimize the use of the information contained in the complex structure of
multi-hadron events. Errors due to the corrections for hadronization 
effects may be expected to cancel to some extent in the averaging procedure.
To test this expectation the analysis of each of the individual 18 observables
is repeated by performing hadronization corrections with all hadronization 
generators described in Section 3.2. This results in 7 times 18 individual
$\rm \alpha_s $ values. As a first test for each of the 18 observables the
unweighted average value of $ \rm \alpha_s $ from the seven models is 
evaluated. The average value of the 18 $ \rm \alpha_s $ values is 
\asmz = 0.1177 $ \pm $ 0.0029. 
In a second step for each of the 7 hadronization
models an unweighted average of the corresponding 18 $\rm \alpha_s $ values is 
calculated. Finally an unweighted average of the 7 average values
for the different hadronization models is computed resulting in 
\asmz = 0.1177 $\pm$ 0.0016. The result confirms that the scatter of the
average values due to different assumptions for hadronization corrections
is significantly smaller than the uncertainty of $ \pm 0.0025 $ of the  
mean value from 18 individual observables.  \\

The correlations between the \as values obtained from the
different observables must be taken into account in order 
to calculate their weighted average. Since the correlations 
are mostly unknown, the exact correlation pattern cannot be
worked out reliably. Therefore, we use a recently proposed method 
\cite{coraver}, which makes use of a robust estimation of the 
covariance matrix and has been used for example in \cite{waver-bethke}.
Here it is assumed that different measurements 
$\rm i $ and $\rm j $ are correlated with a fixed fraction $ \rhoe $ 
of the maximum possible correlation $\rm C_{ij}^{max} $:

\begin{equation}
   C_{ij} = \rhoe C_{ij}^{max} \quad i \neq j,
   \quad \mbox{with} \quad C_{ij}^{max} = \sigma_i \sigma_j  ~~~.
\end{equation}

For $ \rhoe = 0 $ the measurements are treated as uncorrelated,
for $ \rhoe = 1 $ as 100\% correlated entities. When $ \chi^2 < n_{df} $ 
the measurements are assumed to be correlated and 
the value $\rm \rho_{{\rm eff}} $ then is adjusted such that the 
$\chi^2 $ is equal to the number of degrees of freedom $ n_{df} $:

\begin{equation}
   \chi^2(\rhoe ) = \sum_{i,j} 
   (x_i - \overline{x}) (x_j - \overline{x}) (C^{-1})_{ij} = n - 1 =
   n_{df} ~~~.
\label{chisquare}
\end{equation}

When $ \chi^2 > n_{df} $ it is assumed that the errors of the
measurement are underestimated and will therefore be scaled until
$ \chi^2 = n_{df} $ is satisfied.  \\

Applying this method to the 18 observables studied,  
the weighted average (see also Fig. \ref {uaverexp}) yields: \\

\begin{center}
\asmz = 0.1168 $\pm$ 0.0026  \\
\end{center}

with $ \rm \rho_{eff} $ being 0.635. Both the central value
and its uncertainty are almost identical to the
unweighted average and the r.m.s. quoted above,
which in itself is a remarkable result. \\

It should be noted, that the method applied for the calculation
of the weighted average does not necessarily lead to 
the smallest possible error. In \cite{waver-bethke} it is shown
for example, that the error of the weighted average is increased 
if less significant measurements are included. Within this analysis
however, the \as measurements from all individual observables have been 
considered, regardless
of their significance. This is motivated by the fact
that the errors of the \as measurements quoted in
Table \ref{expres} and in the following subsections contain all 
uncertainties which can be evaluated from a careful experimental analysis.
However, the spread of the \as measurements may not be explainable 
by the  individual uncertainties alone, there may be 
additional uncertainties, which cannot be derived from a single observable.
Therefore the above averaging procedure has been applied and
robustness of the error estimate has been
preferred instead of minimizing the error.
Still this error estimate may not cover a possible general shift of the
measured average with respect to the true \asmz value. Apart from a better
theoretical understanding, today such a shift could only be inferred
by comparing to different types of calculation, like resummed 
or Pad\'{e} approximation, which are presented in Sections \ref{padesect}
and \ref{NLLAsect}. 


\subsection{Additional Cross Checks}
\label{add-checks}

An underlying assumption for the \oass QCD fits to the shape observables is
that the value of \as is approximately independent of a variation of the
renormalization scale within the fit range. 
To check this assumption, a cross check 
for the differential two jet rate observables has been performed following a 
suggestion in \cite{KRALA}. For these observables the QCD fits have been repeated, 
allowing the renormalization scale to vary proportionally to $ y_{cut} $, 
i.e. $ \mu^2 = x_{\mu} y_{cut} \sqrt{s} $. 
The differences in the \asmz determination have been found to be of the order 
of a few per mille for the individual jet rate observables and to be less than 
two per mille for the average of these observables. \\

A further investigation has been performed for all observables of Table \ref{results}.
The fit range listed in Table \ref{results} has been divided into two separate,
approximately symmetrical regions, allowing a maximum overlap of one bin.
\asmz has been determined applying experimentally optimized scales
for both regions independently. The fits were successful for all observables
except $ D_2^{Geneva} $, where the resulting fit ranges were too small to    
allow the fits to converge. Good agreement of the two \as values measured 
for each observable is found. 
In a further step, the two \asmz values have been combined for each observable
according to their statistical weight. These \as values have been
combined by calculating a weighted average as described before. The resulting
average value of $\alpha_s(M_{Z}^{2}) = 0.1168 \pm 0.0025 $ is identical    
to the value determined from the standard procedure.
No systematical trend of the two values of the renormalization scales found
for the two fit ranges (dominated by two respectively three jet events)
is observed. Further information on this study can be found in \cite{MyThesis}.
    
\clearpage

\clearpage

\section{Scale Setting Methods from Theory}
\label{theooptsect}

Several methods for the choice of an optimized value for the renormalization
scale have been suggested by theory. See e.g. \cite{theoambig} for 
an overview. In this section we compare the \asmz measurements, using
renormalization scales predicted by three different approaches with the
\asmz measurements using experimentally optimized scales: \\

({\it i}) The principle of minimal sensitivity (PMS) : Since all order 
predictions should be independent of the renormalization scale, Stevenson 
\cite{PMSScale} suggests choosing the scale to be least sensitive with 
respect to its variation, i.e. from the solution of

\begin{equation} 
\frac {\partial \sigma } {\partial x_{\mu} } = 0 \hspace{0.2 cm}.
\end{equation}      

({\it ii}) The method of effective charges (ECH) : The basic idea of this 
approach \cite{ECHScale} is to choose the renormalization scheme in such
a way that the relation between the physical quantity and the coupling is
the simplest possible one. In \oass, where the ECH approach is equivalent
to the method of fastest apparent convergence (FAC) \cite{ECHScale} the
scale is chosen in such a way that the second order term in Eq. 
(\ref{thseco}) vanishes:

\begin{equation} 
B (Y, \cos \vartheta_{T} ) 
+ \big( 2 \pi \beta_{0} \ln ( x_{\mu} ) - 2\big) 
A(Y,\cos\vartheta_{T}) = 0 \hspace{0.2 cm}.
\end{equation}      

({\it iii}) The method of Brodsky, Lepage and MacKenzie (BLM) \cite{BLMScale}: 
This method follows basic ideas in QED, where the renormalized 
electric charge is fully given by the vacuum polarization due to charged
fermion-antifermion pairs \cite{theoambig}. In QCD it is suggested to fix 
the scale with the requirement that all the effects of quark pairs be 
absorbed in the definition of the renormalized coupling itself. In \oass
this amounts to the requirement that $ x_{\mu} $ is chosen in such a way
that the flavour dependence $ n_{f} $ of the second order term in Eq. 
\ref{thseco} is removed:
 
\begin{equation} 
\left. \frac {\partial } {\partial n_{f} } \hspace {0.1 cm}
\Bigg\{ B (Y, n_{f}) + \big( 2 \pi \beta_{0} \ln ( x_{\mu} ) - 2 \big) 
A(Y)\Bigg\} \right|_{n_f=5} = 0 \hspace{0.2 cm}.
\end{equation}      

The results of the \asmz measurements for the individual observables
applying the different scale setting prescriptions are listed in table 
\ref{theores}. The weighted averages for the different methods yield: \\

({\it i}) PMS method :

\begin{center}
$ \rm \alpha_s(M_Z^2) = 0.1154 \pm 0.0045 
\hspace{1. cm} ( \chi^2 / n_{df} = 19 / 17 ) $
\end{center}

({\it ii}) ECH method :

\begin{center}
$ \rm \alpha_s(M_Z^2) = 0.1155 \pm 0.0044 
\hspace{1. cm} ( \chi^2 / n_{df} = 19 / 17 ) $
\end{center}

({\it iii}) BLM method :

\begin{center}
$ \rm \alpha_s(M_Z^2) = 0.1174 \pm 0.0068 
\hspace{1. cm} ( \chi^2 / n_{df} = 29 / 13 ) $
\end{center}

\nin to be compared with $ \rm \alpha_s(M_Z^2) = 0.1168 \pm 0.0026 $ 
( $ \chi^2 / n_{df} = 6.2 / 17 $) using the experimentally 
optimized scales.  

\pagebreak

\begin{sidewaysfigure} 
\begin{center}
\hspace{-1. cm}
\mbox{\epsfig{file=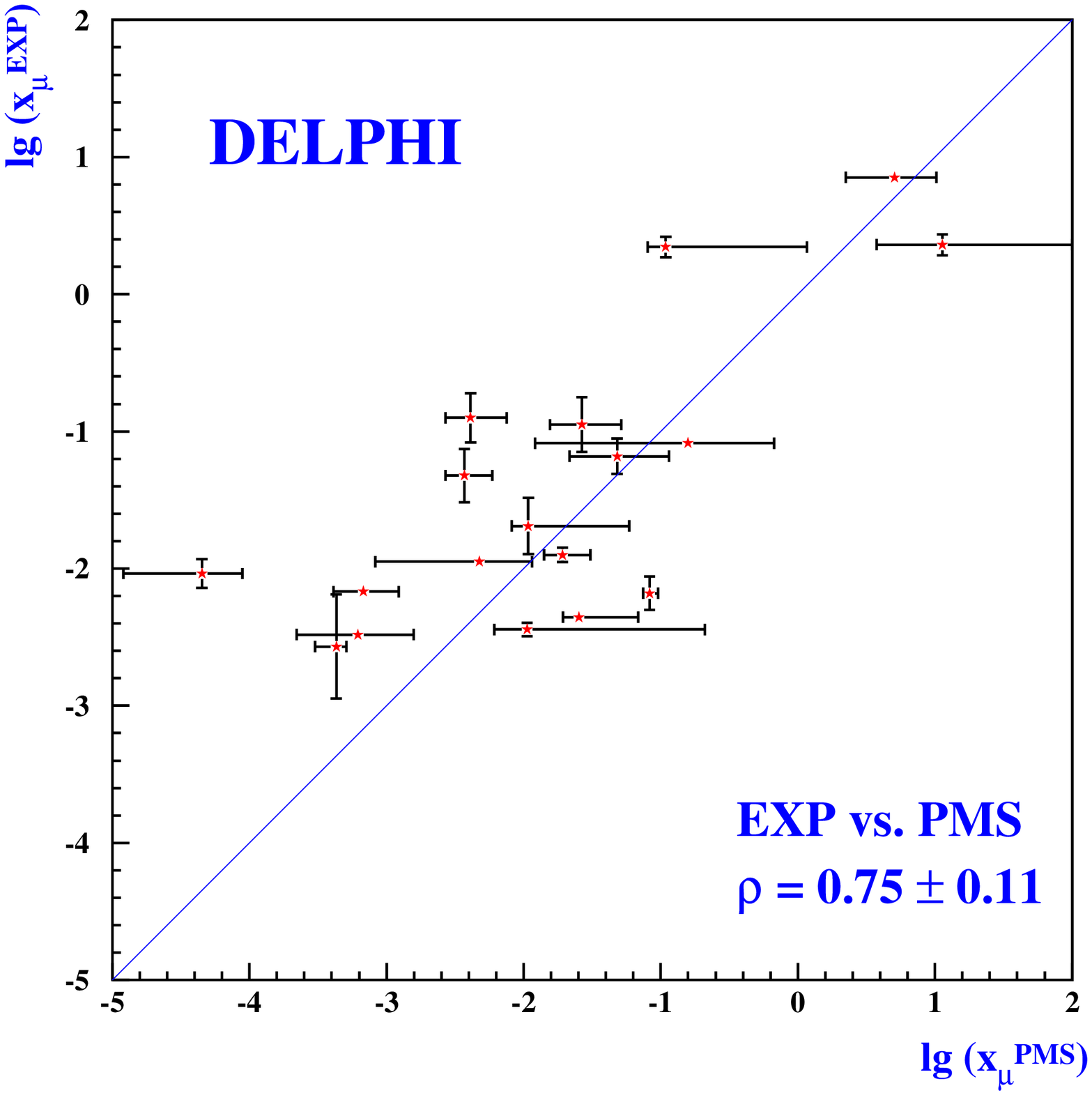,width=8.0cm}}
\hspace{-0.5 cm}
\mbox{\epsfig{file=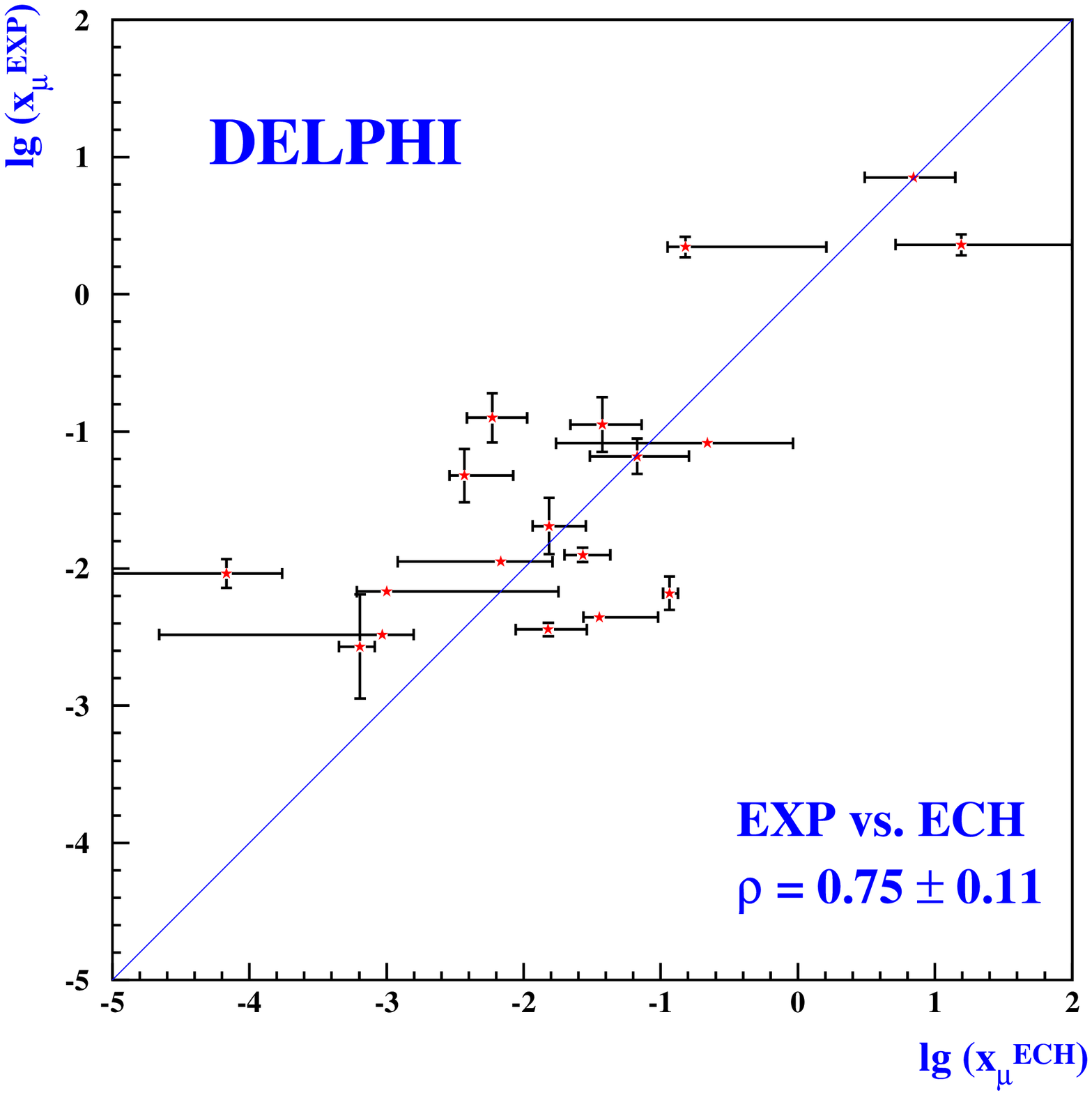,width=8.0cm}}
\hspace{-0.5 cm}
\mbox{\epsfig{file=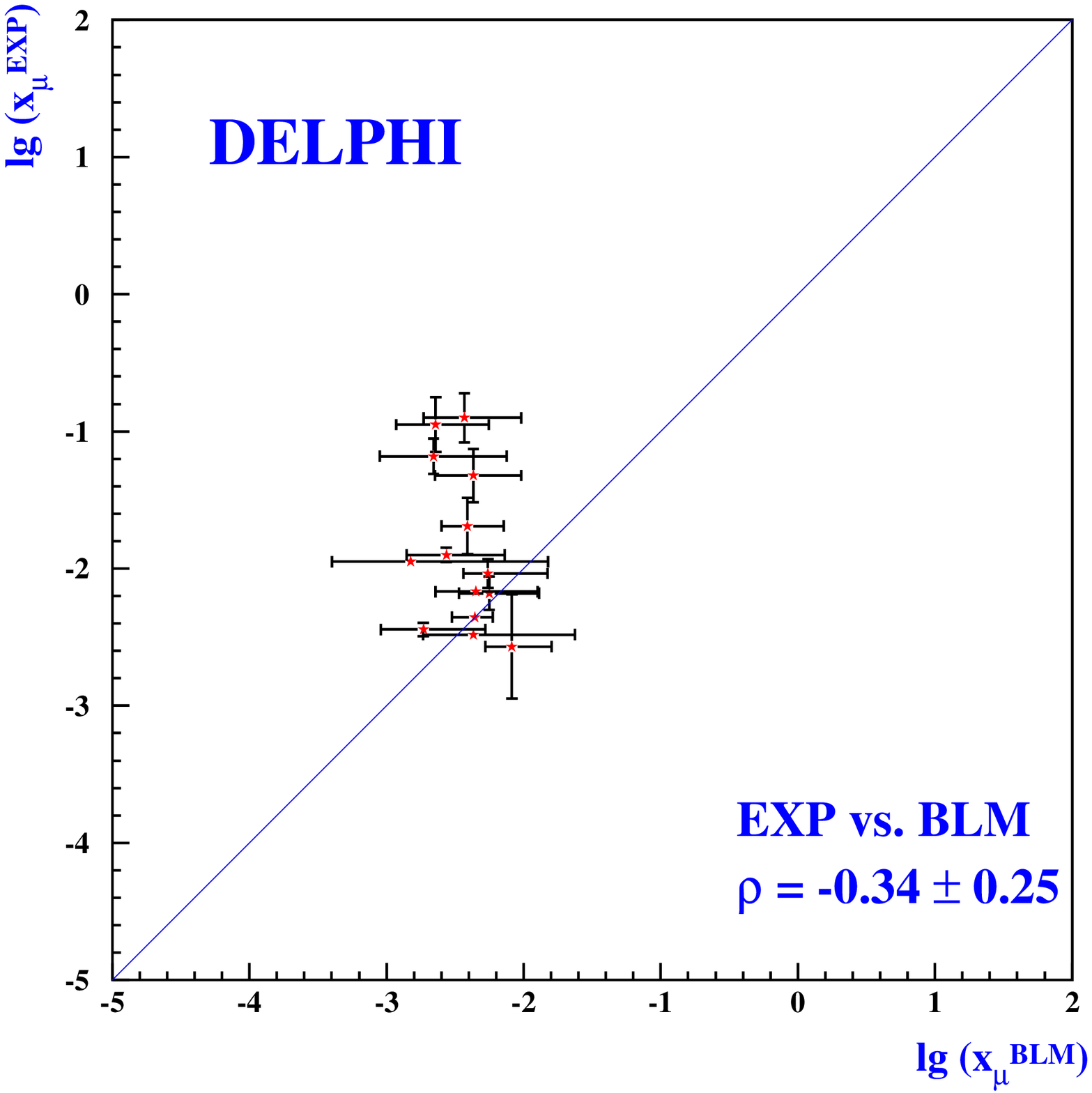,width=8.0cm}}
\hspace{-1. cm}
\vspace{1. cm}
\caption[] {Comparison between the logarithms of the experimentally optimized 
            renormalization scales and the logarithms of scales predicted by 
            PMS, ECH and the BLM method. The vertical error bars indicated 
            represent the uncertainties from the 2-parameter fits 
            in $ x_{\mu} $ and \as. The vertical bars represent the range
            of renormalization scale values for the theoretically motivated
            scale setting methods evaluated from the individual bins within
            the fit range of each distribution, whereas the central value has
            been derived by considering the full theoretical prediction within 
            the fit range.  
           }
\label{x_corel}
\end{center}
\end{sidewaysfigure}

The weighted averages for the different theoretical methods are in 
agreement with the average using experimentally optimized 
scales. The scatter of the individual measurements is lowest for
the experimentally optimized scales and highest for the BLM method.
Whereas for the experimentally optimized scales the fit values for
the individual \asmz measurements are perfectly consistent, the
consistency for the ECH and the PMS methods is only moderate. 
The results for the ECH and the PMS methods are very similar, the 
correlation $ \rho $ between ECH and PMS scales is almost 1.  
In the case of the BLM method the $ \chi^2 / n_{df} $ 
indicates that the individual \asmz measurements are inconsistent. 
Moreover, the fits using the scales predicted by the BLM method did not 
converge at all for the observables $ JCEF $, $ O $, $ \rho_D $ and 
$ D_2^{Geneva} $. Figure \ref{x_corel} shows the correlation between
the logarithms of the experimentally optimized scales and the logarithms 
of the scales predicted by the ECH, PMS and the BLM methods. For the ECH 
and the PMS method there are significant correlations of $ \rho = 0.75 \pm 0.11 $. 
In the case of BLM there is a slightly negative correlation of $ \rho = -0.34 \pm 0.25 $, 
compatible with zero within 1.5 $ \sigma $. 
Our results indicate that the ECH 
and the PMS methods are useful in the case where an experimental optimization 
can not be performed, whereas the BLM method does not seem to be suitable 
for the determination of \asmzx

\clearpage

\begin{table} [p]
\begin{center}
\begin{tabular} { l c c c c }
\hline
\hline
\hspace{0.1 cm}
Observable           & $\alpha_s^{EXP}(M_Z^2)$  & $\alpha_s^{PMS}(M_Z^2)$  & 
                       $\alpha_s^{ECH}(M_Z^2)$  & $\alpha_s^{BLM}(M_Z^2)$  \\
\hline
$\rm EEC             $ &  0.1142 &  0.1133 &  0.1135 &    0.1142 \\
$\rm AEEC            $ &  0.1150 &  0.1063 &  0.1064 &    0.1179 \\
$\rm JCEF            $ &  0.1169 &  0.1168 &  0.1169 &           \\
$\rm 1-T             $ &  0.1132 &  0.1101 &  0.1111 &    0.1133 \\
$\rm O               $ &  0.1171 &  0.1128 &  0.1124 &           \\
$\rm C               $ &  0.1153 &  0.1119 &  0.1124 &    0.1144 \\
$\rm B_{Max}         $ &  0.1215 &  0.1222 &  0.1217 &    0.1268 \\
$\rm B_{Sum}         $ &  0.1138 &  0.1023 &  0.1021 &    0.1118 \\
$\rm \rho_H          $ &  0.1215 &  0.1197 &  0.1198 &    0.1258 \\
$\rm \rho_S          $ &  0.1161 &  0.1154 &  0.1149 &    0.1169 \\
$\rm \rho_D          $ &  0.1172 &  0.1190 &  0.1203 &           \\
$\rm D_2^{E0}        $ &  0.1165 &  0.1145 &  0.1142 &   0.1143  \\
$\rm D_2^{P0}        $ &  0.1210 &  0.1204 &  0.1202 &   0.1232  \\
$\rm D_2^{P}         $ &  0.1187 &  0.1110 &  0.1108 &   0.1118  \\
$\rm D_2^{Jade}      $ &  0.1169 &  0.1137 &  0.1134 &   0.1137  \\
$\rm D_2^{Durham}    $ &  0.1169 &  0.1162 &  0.1159 &   0.1241  \\
$\rm D_2^{Geneva}    $ &  0.1178 &  0.1064 &  0.1171 &           \\
$\rm D_2^{Cambridge} $ &  0.1164 &  0.1164 &  0.1163 &   0.1124  \\
\hline
w. average             & $ 0.1168 \pm 0.0026 $ & $ 0.1154 \pm 0.0045 $ & 
                         $ 0.1155 \pm 0.0044 $ & $ 0.1174 \pm 0.0068 $ \\
$ \chi^2 / n_{df} $    & 6.2 / 17              & 19 / 17               & 
                         19 / 17               & 29 / 13               \\
\hline
\hline
\end{tabular}  
\end{center}
\caption[]{ Comparison of the \asmz values obtained using the different         
            methods for evaluating the renormalization scale suggested
            by theory. For each observable the \asmz values using 
            experimentally optimized scales and \asmz values for 
            the scales predicted by the
            PMS, ECH and BLM methods are shown. The errors for the \asmz 
            measurements are assumed to be identical for all methods
            (see table \ref{expres}).
            The weighted averages are calculated using $ \rhoe = 0.635 $ 
            and scaling the errors to yield $ \chi^2 / n_{df} = 1 $ in the 
            case of the PMS, ECH and the BLM methods (see text). The $ \chi^{2} $
            given for the averaging correspond to the values before adjusting
            $ \rhoe $ and rescaling the measurement uncertainties. 
            The fits using the scales predicted by BLM did not converge 
            for the observables JCEF, O, $ \rho_D $ and $ D_2^{Geneva} $. }    
\label{theores}
\end{table}


\clearpage

\section{Pad\'{e} Approximation}
\label{padesect}

An approach for estimating higher-order contributions to a perturbative
QCD series is based on Pad\'{e} Approximations. The Pad\'{e} Approximant $ [N/M] $ 
to the series

\begin{equation} 
S = S_0 + S_1 x + S_2 x^2 + \ldots  + S_n x^n
\end{equation} 

\nin is defined \cite{PAbas} by

\begin{equation} 
[N/M] \equiv \frac {a_0 + a_1 x + a_2 x^2 + \ldots + a_N x^N }
                   {1 + b_1 x + b_2 x^2 + \ldots + b_M x^M }
             \hspace{1. cm} ; \hspace{0.5 cm} N + M = n
\end{equation} 

\nin and

\begin{equation} 
\label{PaSet}
[N/M] = S + {\cal{O}} (x^{N+M+1}) \hspace{0.2 cm}.
\end{equation} 

\hspace{0.1 cm} \\

The set of equations (\ref{PaSet}) can be solved, and by consideration
of the terms of $ {\cal{O}} (x^{N+M+1}) $ one can obtain an estimate 
of the next order term $ S_{N+M+1} $ of the original series. 
This is called the PA method.
Furthermore, for an asymptotic series $ [N/M]$ can be taken as an 
estimate of the sum (PS) of the series to all orders. 
The PA method has been used successfully to estimate coefficients in statistical
physics \cite{PAbas}, and various quantum field theories including QCD
\cite{PAellis}. Justifications for some of these successes have been found
in mathematical theorems on the convergence and renormalization scale 
invariance of PAs \cite{PAellis}. In many cases the PAs yield 
predictions for the higher order coefficients in perturbative series with 
high accuracy, whereas this accuracy is not  expected for the lower order 
predictions such as \oasss. For the application of the \asmz determination from 
event shapes, the Pad\'{e} Approximation can serve as a reasonable estimate of the 
errors due to higher order corrections \cite{PAedisk}. \\

For each bin of our observables an estimate for the \oasss coefficient
$ C(y) $ can be derived from $ [0/1] $ with $ a_{0}=A $, $ b_{1}=-B/A $:

\begin{equation} 
C^{Pad\acute{e}}(y) = \frac { B^2(y) } { A(y) } .
\end{equation} 

\hspace{0.1 cm} \\

It should be noted that the PA predictions $ C^{Pad\acute{e}}(y) $ are positive
by construction which will result in large errors for kinematical regions 
where the \oasss contribution is negative. The fit range has therefore been 
determined in the following way: Starting from the same fit range as in
\oass, the fit has been accepted if  
$ \chi^2 / n_{df} \le 5 $. Otherwise the fit range was reduced bin by bin
until the fit yielded $ \chi^2 / n_{df} \le 5 $. \\

In addition to the \oasss fits in the Pad\'{e} Approximation, the PS method has been 
used as an estimate of the sum of the perturbative series and \asmz has been 
extracted by fitting the [0/1] approximation directly to the data. Here, 
the fit range has been chosen to be the same as for the fits in the \oasss Pad\'{e} 
approximation. The $ \chi^{2} $ dependence of the \asmz fits applying the Pad\'{e} 
Approximation as a function of the renormalization scale value $ x_{\mu} $ is 
quite small, especially for the PS method. For most of the observables 
\asmz and $ x_{\mu} $ could not be determined in a simultaneous fit. Therefore, 
the fits have been done choosing a fixed renormalization scale value $ x_{\mu} = 1 $. 
The uncertainty due to the scale dependence of \asmz has been estimated 
by varying $ x_{\mu} $ between 0.5 and 2.  \\

 
\begin{table} [b] 
\begin{center}
\begin{tabular} { l c c c c }
\hline
\hline

Observable        & Fit Range    & \asmz  & \das (Scale.) & \das(Tot.)  \\
\hline

$\rm EEC             $   &  $28.8^{\circ} - 151.2^{\circ}$ & 0.1189  &
                            $ \pm 0.0016 $ & $ \pm 0.0026 $          \\ 

$\rm AEEC            $   &  $25.2^{\circ} -  64.8^{\circ}$ & 0.1074  &
                            $ \pm 0.0030 $ & $ \pm 0.0056 $          \\ 

$\rm JCEF            $   &  $104.4^{\circ} - 169.2^{\circ}$ & 0.1169 &
                            $ \pm 0.0006 $ & $ \pm 0.0016 $           \\ 

$\rm 1-T             $   &  0.07 - 0.30                     & 0.1207 & 
                            $ \pm 0.0023 $ & $ \pm 0.0036 $           \\ 

$\rm O               $   &  0.24 - 0.32                     & 0.1098 &
                            $ \pm 0.0014 $ & $ \pm 0.0044 $           \\ 

$\rm C               $   &  0.32 - 0.72                     & 0.1208 &
                            $ \pm 0.0023 $ & $ \pm 0.0039 $           \\ 

$\rm B_{Max}         $   &  0.10 - 0.24                     & 0.1183 &  
                            $ \pm 0.0016  $ & $ \pm 0.0042 $          \\ 

$\rm B_{Sum}         $   &  0.14 - 0.18                    & 0.1127  &  
                            $ \pm 0.0016 $ & $ \pm 0.0068 $           \\ 

$\rm \rho_H          $   &  0.03 - 0.14                    & 0.1230  &
                            $ \pm 0.0015 $ & $ \pm 0.0036 $           \\ 

$\rm \rho_S          $   &  0.10 - 0.30                   & 0.1252   & 
                            $ \pm 0.0024 $ & $ \pm 0.0038 $           \\ 

$\rm \rho_D          $   &  0.07 - 0.30                   & 0.1045   &
                            $ \pm 0.0015 $ & $ \pm 0.0040 $           \\ 

$\rm D_2^{E0}        $   &  0.05 - 0.18                   & 0.1159   &  
                            $ \pm 0.0014 $ & $ \pm 0.0042 $           \\ 

$\rm D_2^{P0}        $   &  0.05 - 0.18                   & 0.1199   &
                            $ \pm 0.0011 $ & $ \pm 0.0034 $           \\ 

$\rm D_2^{P}         $   &  0.10 - 0.20                   & 0.1128   &
                            $ \pm 0.0008 $ & $ \pm 0.0030 $           \\ 

$\rm D_2^{Jade}      $   &  0.06 - 0.25                   & 0.1142   & 
                            $ \pm 0.0014 $ & $ \pm 0.0032 $           \\ 

$\rm D_2^{Durham}    $   &  0.015 - 0.16                  & 0.1170   & 
                            $ \pm 0.0009 $ & $ \pm 0.0023 $           \\ 

$\rm D_2^{Cambridge} $   &  0.011 - 0.18                  & 0.1164   &
                            $ \pm 0.0007 $ & $ \pm 0.0026 $           \\ 
\hline
average       &    & \spc $ 0.1168 \pm 0.0054 $ &
                  \multicolumn{2}{l} { \hspace{0.5 cm} \chindf = 30 / 16 } \\ 
                                      &                          \\ 
\hline
\hline
\end{tabular}
\end{center}
\caption[]{ Results on \asmz for QCD-Fits including the \oasss Term 
            in the Pad\'{e} Approximation (PA). For each of the observables 
          the fit range, \asmz , the uncertainty due to scale variation
          between $ 0.5 \le x_{\mu} \le 2 $ and the total uncertainty are
          shown. The experimental errors and the uncertainties due to the  
          hadronization corrections are assumed to be the same as
          for the \oass measurements.
          The weighted average is calculated using $ \rhoe = 0.635 $ 
          and scaling the errors to yield $ \chi^2 / n_{df} = 1 $ (see text).
          The $ \chi^{2} $ given for the averaging corresponds to the value before 
          adjusting $ \rhoe $ and rescaling the measurement uncertainties.
          The fit for $ D_2^{Geneva} $ did not converge. }
\label{paderes}
\end{table}

 
\begin{table} [b] 
\begin{center}
\begin{tabular} { l c c c }
\hline
\hline

Observable               & \asmz    & \das (Scale.)  & \das(Tot.)    \\
\hline

$\rm EEC             $   & 0.1147 &  $ \pm 0.0003 $ & $ \pm 0.0021 $ \\ 

$\rm AEEC            $   & 0.1070 &  $ \pm 0.0002 $ & $ \pm 0.0048 $ \\ 

$\rm JCEF            $   & 0.1169 &  $ \pm 0.0003 $ & $ \pm 0.0015 $ \\ 

$\rm 1-T             $   & 0.1165 &  $ \pm 0.0003 $ & $ \pm 0.0028 $ \\ 

$\rm O               $   & 0.1135 &  $ \pm 0.0003 $ & $ \pm 0.0042 $ \\ 

$\rm C               $   & 0.1150 &  $ \pm 0.0003 $ & $ \pm 0.0032 $ \\ 

$\rm B_{Max}         $   & 0.1196 &  $ \pm 0.0003 $ & $ \pm 0.0039 $ \\ 

$\rm \rho_H          $   & 0.1219 &  $ \pm 0.0004 $ & $ \pm 0.0033 $ \\ 

$\rm \rho_S          $   & 0.1161 &  $ \pm 0.0003 $ & $ \pm 0.0029 $ \\ 

$\rm \rho_D          $   & 0.1098 &  $ \pm 0.0003 $ & $ \pm 0.0037 $ \\ 

$\rm D_2^{E0}        $   & 0.1136 &  $ \pm 0.0003 $ & $ \pm 0.0040 $ \\ 

$\rm D_2^{P0}        $   & 0.1198 &  $ \pm 0.0003 $ & $ \pm 0.0032 $ \\ 

$\rm D_2^{P}         $   & 0.1124 &  $ \pm 0.0003 $ & $ \pm 0.0029 $ \\ 

$\rm D_2^{Jade}      $   & 0.1123 &  $ \pm 0.0003 $ & $ \pm 0.0029 $ \\ 

$\rm D_2^{Durham}    $   & 0.1164 &  $ \pm 0.0003 $ & $ \pm 0.0021 $ \\ 

$\rm D_2^{Cambridge} $   & 0.1162 &  $ \pm 0.0003 $ & $ \pm 0.0025 $ \\ 
\hline
average                  & \spc $ 0.1157 \pm 0.0037 $ &
                  \multicolumn{2}{l} { \hspace{0.5 cm} \chindf = 17 / 15 } \\ 
                                         &                          \\ 
\hline
\hline
\end{tabular}
\end{center}
\caption[]{ Results on \asmz for QCD-Fits applying the 
            Pad\'{e} Sum Approximation (PS). For each of the observables 
            \asmz , the uncertainty due to scale variation
            between $ 0.5 \le x_{\mu} \le 2 $ and the total uncertainty are
            shown. The experimental errors and the uncertainties due to the  
            hadronization corrections are assumed to be the same as
            for the \oass measurements.
            The weighted average is calculated using $ \rhoe = 0.635 $ 
            and scaling the errors to yield $ \chi^2 / n_{df} = 1 $ (see text).
            The $ \chi^{2} $ given for the averaging corresponds to the value 
            before adjusting $ \rhoe $ and rescaling the measurement uncertainties.
            The fits for $ B_{Sum} $ and $ D_2^{Geneva} $ did not converge. }
\label{pasures}
\end{table}

\begin{figure}      
\begin{center}
\hspace{-1. cm}
\mbox{\epsfig{file=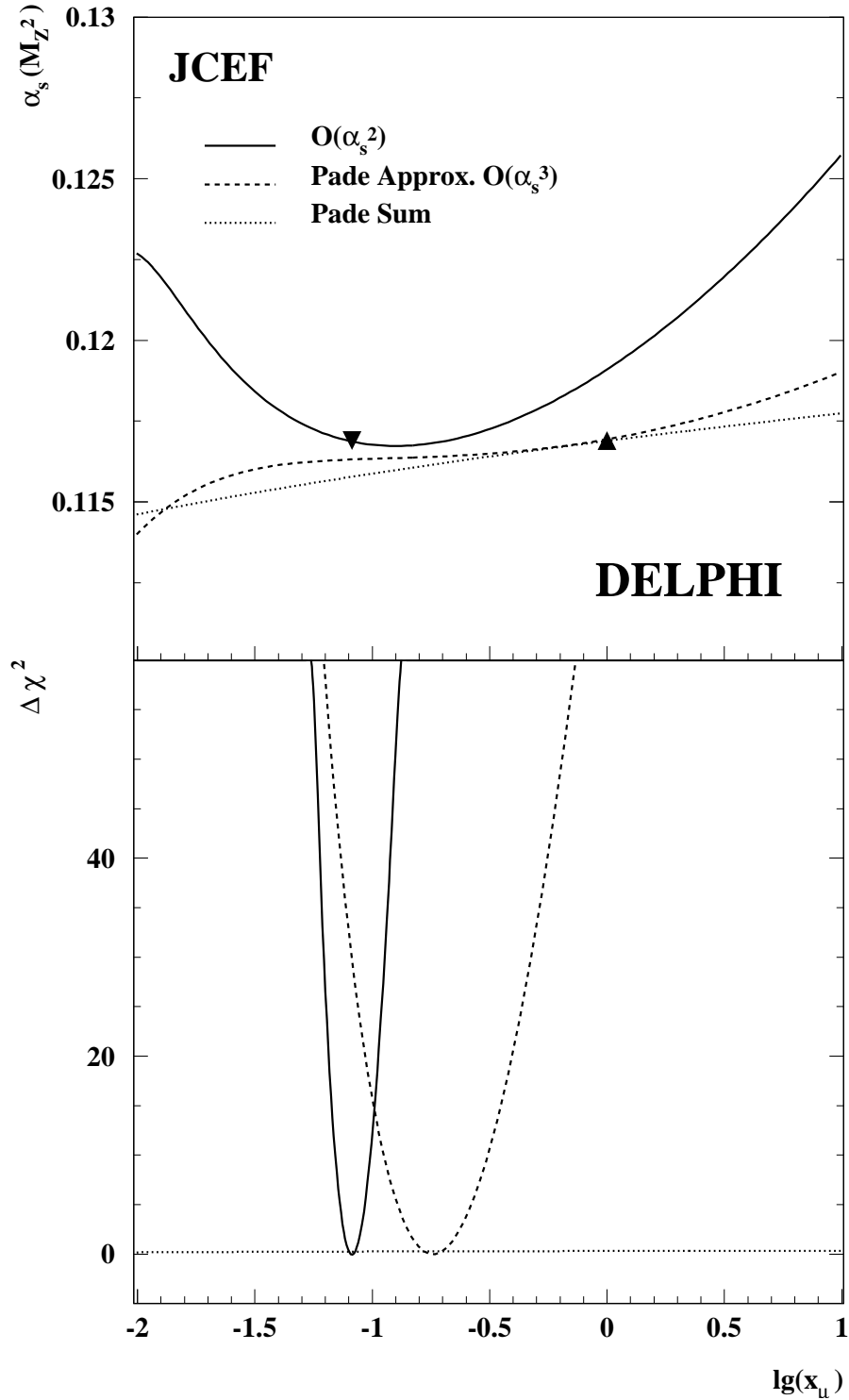, width=14.cm }}
\hspace{-1. cm}
\caption[]{ \asmz and $\rm \Delta \chi^2 = \chi^2 - \chi^2_{min} $
            for the distribution of the Jet Cone Energy Fraction 
            as a function of $ x_{\mu} $  
            from QCD fits applying \oass prediction, \oasss in 
            Pad\'{e} Approximation and the Pad\'{e} Sum Approximation. 
            Additionally, the $ \chi^2 $ minimum for the \oass fit
            and the renormalization scale value $ x_{\mu} = 1 $ 
            have been indicated in the \asmz curves. }
\label{Scale-pade}
\end{center}
\end{figure}

The fit results for the individual observables are listed in tables 
\ref{paderes} and \ref{pasures}. The fit applying PS to the $ B_{Sum} $ 
distribution did not converge for any fit range chosen. For the  
$ D_2^{Geneva} $ distribution, the fits did not converge for either method.    
Comparing the fit results of the \oasss fits in Pad\'{e} Approximation 
with the fit results in \oass applying $ x_{\mu} = 1 $, as given in table
\ref{fixres} , the scale dependence of 
\asmz is reduced for most of the observables, as one would expect from measurements 
using exact calculations in \oasssx For the PS method, the reduction of the
scale dependence is even larger. Here, \asmz is less scale dependent than
in the \oass fits for all observables considered. Figure \ref{Scale-pade} 
shows the scale dependence of \asmz applying the different QCD predictions 
to the distribution of the Jet Cone Energy Fraction as an example. 
There is almost no $ \chi^{2} $ dependence of the \asmz fits as a function of the 
renormalization scale for the PS prediction.
For the fits applying \oasss in the Pad\'{e} Approximation, the $ \chi^{2} $ dependence
is less than for the \oass prediction. However, the JCEF is one of the few 
observables, where a simultaneous fit of \asmz and $ x_{\mu} $ is possible.  \\
 
Assuming the same correlation as in the \oass fits, the weighted averages 
of \asmz over the observables used have been calculated as:

\begin{center}
 $ \rm \alpha_s(M_Z^2) = 0.1168 \pm 0.0054 $
\end{center}

\nin for the PA fits and

\begin{center}
 $ \rm \alpha_s(M_Z^2) = 0.1157 \pm 0.0037 $
\end{center}

\nin for the PS fits. The averages are in excellent agreement with the \oass 
value of $ \rm \alpha_s(M_Z^2) = 0.1168 \pm 0.0026 $ using optimized scales.
The scatter between the observables, however, 
is somewhat larger than in the \oass case.
The $ \chi^{2} / n_{df} $ for the average values is 30/16 and 17/15 
respectively.\\


\clearpage

\section{QCD in the Next-to-Leading Log Approximation}
\label{NLLAsect}

All orders resummed QCD calculations in the Next-to-Leading Log Approximation 
(NLLA) matched with \oass calculations have been used widely to measure \asmz 
from event shape observables \cite{DelPubs2,NLLAPubs}.  \\

For a generic event shape observable $ Y $ for which the theoretical
prediction can be exponentiated, the cumulative cross-section 
at the scale $\rm Q^2 \equiv s $ is defined by:
 
\begin{equation} 
   R(Y,\alpha_s)= \frac {1} {\sigma}
                  \int\limits_{0}^{Y}
                  \frac {d \sigma} {dY'} dY' 
\end{equation} 

\nin and can be expanded in the form

\begin{equation} 
   R(Y,\alpha_s)= C(\alpha_s)\exp {\Sigma(\alpha_s,L)} + F(\alpha_s,Y) \hspace{0.2 cm},
\end{equation} 

\nin where $ L \equiv \ln(1/Y) $ and

\begin{equation} 
   C(\alpha_s) = 1 + \sum\limits_{i=1}^{\infty} C_{i} \bar{\alpha}_s^i 
\end{equation} 

\begin{equation} 
\label{sigma}
   \Sigma (\alpha_s,L) = \sum\limits_{i=1}^{\infty} \bar{\alpha}_s^i
                         \sum\limits_{j=1}^{i+1} G_{ij} L^m 
\end{equation} 

\begin{equation} 
   F(\alpha_s,Y) = \sum\limits_{i=1}^{\infty} F_{i}(Y) \bar{\alpha}_s^i \hspace{0.2 cm}.
\end{equation} 

\nin The $ C_i $ are constant and the $ F_i(Y) $ vanish in the infrared 
limit $ Y \rightarrow 0 $. The factor $ \Sigma $ to be exponentiated
can be written

\begin{equation} 
   \Sigma(\alpha_s,L) = L f_{LL}(\alpha_s L ) + f_{NLL}(\alpha_s L) 
                      + \mbox{subleading terms ,} 
\end{equation} 

\nin where $ f_{LL} $ and $ f_{NLL} $ represent the leading and the next-to-leading
logarithms. They have been calculated for a number of observables, including
$ 1-T $ \cite{NLLAT}, $ C $ \cite{NLLAC}, $ B_{max} $\cite{NLLAB}, 
$ B_{sum} $\cite{NLLAB}, $ \rho_H $ \cite{NLLAR} and $ D_2^{Durham} $
\cite{NLLAD}, where the NLLA predictions for $ B_{max} $ and $ B_{sum} $  
entering into this analysis are the recently improved calculations by 
Yu. L. Dokshitzer et al. \cite{NLLABNEW}.  \\

Pure NLLA calculations can be used to measure \asmz in a limited kinematic 
region close to the infrared limit, where $ L $ becomes large. In order
to achieve a prediction where the kinematical range can be extended 
towards the 3 jet region, several procedures have been suggested
\cite{NLLAComb} to match the NLLA calculations with the calculations 
in \oassx  \\

The \oass QCD formula can be written in the integrated form:

\begin{equation} 
   R_{ \mbox{$\cal{O}$} (\alpha_s^2) }(Y,\alpha_s) 
   = 1 + \mbox{$\cal{A}$} (Y) \bar{\alpha}_s 
       + \mbox{$\cal{B}$} (Y) \bar{\alpha}_s^2 \hspace{0.2 cm},
\end{equation} 

\nin where $\cal A $ $ (Y) $ and $ \cal B $ $ (Y) $ are the cumulative forms  
of $ A(Y,\cos\vartheta_{T})$ and $ B(Y,\cos\vartheta_{T}) $ in 
Eq. (\ref{thseco}), integrated over $ \vartheta_{T} $. Together with 
the first and second order part of Eq. (\ref{sigma})

\begin{equation} 
   g_1(L) = G_{12} L^2 + G_{11} L
\end{equation} 
   
\begin{equation} 
   g_2(L) = G_{23} L^3 + G_{22} L^2 + G_{21} L \hspace{0.2 cm},
\end{equation} 

\nin the $ \ln R $ matching scheme can be defined as: 

\begin{equation} 
   \ln R(Y,\alpha_s) = \Sigma(\alpha_s,L) 
                     + H_1(Y) \bar{\alpha}_s + H_2(Y) \bar{\alpha}_s^2 \hspace{0.2 cm},   
\end{equation} 

\nin where

\begin{equation} 
   H_1(Y) = \mbox{$\cal{A}$} (Y) - g_1(L) 
\end{equation} 

\begin{equation} 
   H_2(Y) = \mbox{$\cal{B}$} (Y) 
          - \frac {1} {2} \mbox{$\cal{A}$}^2 (Y) - g_2(L)  . 
\end{equation} 

When combining \oass predictions with NLLA calculations one has to take
into account that the resummed terms do not vanish at the upper kinematic
limit $ Y_{max} $ of the event shape distributions. In order to correct 
for this, the resummed logarithms are redefined 
\cite{NLLAJets} by:

\begin{equation}
\label{NLLA-redef} 
   L = \ln ( 1/Y - 1/Y_{max} + 1 ) .
\end{equation} 

Several other matching schemes can be defined which differ in the treatment
of the subleading terms, thus introducing a principal ambiguity in the
matching procedure. The $ \ln R $ matching scheme has become
the preferred one, because it includes the $ C_2 $ and the $ G_{21} $
coefficients implicitly and uses only those NLLA terms which are known
analytically. It yields the best description of the
data in terms of $ \chi^2 / n_{df} $ in most cases \cite{DelPubs2}.
Therefore, we quote the results for the matched predictions
using the $ \ln R $ matching scheme and use the $ R $ and $ R-G_{21} $
matching schemes as defined e.g. in \cite{DelPubs2} for the estimation
of the uncertainty due to the matching ambiguity.


\subsection{Measurement of \bfasmz using pure NLLA predictions}


To measure \asmz from pure NLLA calculations the fit range has to be
restricted to the extreme 2-jet region, where L becomes large
and the resummed logarithms dominate. We define $ \omega $ as the ratio 
of the resummed logarithms to the non-exponentiating second order 
contributions as follows :

\begin{equation} 
   \omega = \frac { \Sigma(\alpha_s,L) } 
                 { H_1 (Y) \bar{\alpha}_s + H_2 (Y) \bar{\alpha}_s^2 } 
\end{equation} 

In addition to the fit range criteria listed in Section \ref{exoptsect}
we require the minimum of the ratio $ \omega $ over the fit range not
to fall below 5 for the fits in pure NLLA. This leads to 
the fit ranges listed in table \ref{nllafitran}. For the observable
$ D_2^{Durham} $ the ratio $ \omega $ remains small even for small 
values of $ y_{cut} $. No fit range can be found where the resummed
logarithms dominate the prediction. Therefore $ D_2^{Durham} $ has not
been used for the fits in pure NLLA.  \\


\begin{table} [t]
\begin{center}
\begin{tabular}{ l c c  }
\hline
\hline

Observable           & Fit Range (NLLA) & Fit Range (matched)  \\
\hline
$ \rm 1-T $           & 0.04 - 0.09     & 0.04 - 0.30       \\
$ \rm C $             & 0.08 - 0.16     & 0.08 - 0.72       \\
$ \rm B_{max} $       & 0.02 - 0.04     & 0.02 - 0.24       \\
$ \rm B_{sum} $       & 0.06 - 0.08     & 0.06 - 0.24       \\
$ \rm \rho_H $        & 0.03 - 0.06     & 0.03 - 0.30       \\
$ \rm D_2^{Durham} $  &                 & 0.015 - 0.16      \\
\hline
\hline
\end{tabular}  
\end{center}
\caption[]{Fit range for the observables in pure NLLA and
           matched NLLA fits. The observable $ D_2^{Durham} $
           has not been used for pure NLLA fits, since no fit range can 
           be found, where the resummed logarithms dominate the 
           predictions (see text). }
\label{nllafitran}
\end{table}


Contrary to the \oass predictions an optimization of the renormalization scale
(or more precisely an optimization of the renormalization scheme)
cannot be easily performed  for the resummed NLLA predictions \cite{scaledisk}. 
Therefore the scale was fixed to $ x_{\mu} = 1 $. The uncertainty due 
to the scale dependence of \asmz
was estimated by varying the scale $x_{\mu}$ between 0.5 and 2.\\

The fit results for the individual observables are listed in table 
\ref{nllares}. The weighted average of \asmz for the 5 observables is \\

\begin{center}
\asmz = $ 0.116 \pm 0.006 $ \\
\end{center}

\nin which is in excellent agreement with the average value for the \oass fits
of \asmz = $ 0.1168 \pm 0.0026 $.

\vspace{4 cm}

\begin{table} [b]
\begin{center}
\begin{tabular} { l c c c c c c }
\hline
\hline

Observable      & \asmz & \das (exp.) & \das (had.) &
                  \das (scal.) & \das (tot.) &  \chindf        \\
\hline
$ \rm 1-T $     & 0.120 & 
                  $ \pm 0.001 $  &    
                  $ \pm 0.004 $  &   
                  $ \pm 0.004 $  &        
                  $ \pm 0.006 $  & 
                  0.59             \\
 $ \rm C $      & 0.116 & 
                  $ \pm 0.002 $  &   
                  $ \pm 0.003 $  &     
                  $ \pm 0.004 $  &        
                  $ \pm 0.006 $  & 
                  0.53            \\
$ \rm B_{max} $ & 0.111 & 
                  $ \pm 0.004 $  &  
                  $ \pm 0.003 $  &     
                  $ \pm 0.002 $  & 
                  $ \pm 0.006 $  &  
                  2.37            \\

$ \rm B_{sum} $ & 0.116 & 
                  $ \pm 0.003 $  &  
                  $ \pm 0.004 $  &     
                  $ \pm 0.002 $  & 
                  $ \pm 0.006 $  &  
                  1.24          \\

$ \rm \rho_H $  & 0.117 &
                  $ \pm 0.004 $  &  
                  $ \pm 0.006 $  &  
                  $ \pm 0.004 $  & 
                  $ \pm 0.009 $  &  
                  0.43            \\
\hline
average         & $ 0.116 \pm 0.006 $ &
                  \multicolumn{5}{l}{
                  \hspace{1.0 cm} \chindf = 1.2 / 4  
                  \hspace{1.0 cm} $ \rhoe $ = 0.71   }         \\
\hline
\hline
\end{tabular}  
\end{center}
\caption[]{ Results for the \asmz fits in pure NLLA for the individual
            observables together with the individual sources of 
            uncertainties and the $ \chi^2 / n_{df} $ for the NLLA fits. 
            The total error on \asmz listed, is the quadratic sum of the 
            experimental error (statistical and systematic uncertainty), 
            the uncertainty due to the hadronization correction and the
            uncertainty due to the scale dependence of \asmz. 
            Also listed is the weighted average of \asmz for the 
            5 observables together with the $ \chi^2 / n_{df} $ for the
            averaging procedure and the correlation parameter 
            $ \rhoe $. The $ \chi^{2} $ corresponds to the value before 
            readjusting according to Eq. \ref{chisquare}.    }
\label{nllares}
\end{table}

\pagebreak

\subsection{Measurement of \bfasmz using NLLA predictions matched with \oass }

For the QCD fits using NLLA theory matched with \oass predictions the 
fit range has been chosen as the combined fit range for the pure NLLA and
the \oass fits. The results for the individual observables in the
$\ln R $ matching scheme are listed in table \ref{lnrres}. The additional
uncertainty due to the matching ambiguity has been estimated as the
maximum difference of \asmz in the $ \ln R $ matching scheme and the
two alternative matching schemes, $ R $ and $ R-G_{21} $. The 
dependence on the choice of $ Y_{max} $, 
i.e. the value of the upper kinematic limit for the shape observables 
used for the redefinition of the resummed logarithms according 
to Eq. \ref{NLLA-redef}, has been studied by repeating the fits 
with the value of $ Y_{max} $ being reduced by 10 \%. The resulting variation 
in \asmz has been found to be small, the maximum change is about 1 \%. \\

The average value of \asmz in the $ \ln R $ matching scheme is  

\begin{center}
\asmz = $ 0.119 \pm 0.005 $ \\
\end{center}

\nin which is in good agreement with the average value for the \oass fits
of \asmz = $ 0.1168 \pm 0.0026 $.  \\


\begin{table} [b]
\begin{center}
\begin{tabular}{ l c c c c c c c }
\hline
\hline
Observable \spc      & \spc \asmz &
                       \spc \das \spv (exp.)  &
                       \spa \das \spv (had.)  &
                       \spa \das \spv (scal.) &
                       \spa \das \spv (mat.)  & 
                       \spa \das \spv (tot.)  &
                       \spa \chindf \\
\hline
$ \rm 1-T $          & \spc  0.124 & 
                       \spc $ \pm  0.002  $    & 
                       \spa $ \pm  0.003  $   & 
                       \spa $ \pm  0.004  $   & 
                       \spa $ \pm  0.003  $   & 
                       \spa $ \pm  0.007  $     & 
                       \spa   9.5 \\
$ \rm C $            & \spc 0.120 & 
                       \spc $ \pm 0.002  $ &     
                       \spa  $ \pm 0.002  $     & 
                       \spa  $ \pm 0.004  $      & 
                       \spa  $ \pm 0.004  $       & 
                       \spa  $ \pm 0.007  $     & 
                       \spa    15.2 \\
$ \rm B_{max} $      & \spc 0.113 & 
                       \spc $ \pm 0.002  $   & 
                       \spa  $ \pm 0.002  $     & 
                       \spa  $ \pm 0.003  $      & 
                       \spa  $ \pm 0.003  $       & 
                       \spa  $ \pm 0.005  $     & 
                       \spa    8.4 \\
$ \rm B_{sum} $      & \spc 0.122 &  
                       \spc $ \pm 0.002  $   & 
                       \spa  $ \pm 0.003  $     & 
                       \spa  $ \pm 0.004  $      & 
                       \spa  $ \pm 0.005  $       & 
                       \spa  $ \pm 0.008  $     & 
                       \spa    11.9 \\
$ \rm \rho_H $       & \spc 0.119 & 
                       \spc $ \pm 0.002  $   & 
                       \spa  $ \pm 0.002  $     & 
                       \spa  $ \pm 0.003  $      & 
                       \spa  $ \pm 0.005  $       & 
                       \spa  $ \pm 0.007  $     & 
                          1.33 \\
$ \rm D_2^{Durham} $ & \spc 0.121 & 
                       \spc $ \pm 0.001 $    & 
                       \spa  $ \pm 0.002 $  &    
                       \spa  $ \pm 0.002 $       & 
                       \spa  $ \pm 0.005 $        & 
                       \spa  $ \pm 0.006 $      & 
                       \spa    1.70 \\
\hline
average         & \spc $ 0.119 \pm 0.005 $ &
                  \multicolumn{5}{l}{
                  \hspace{1.0 cm} \chindf = 2.3 / 5  
                  \hspace{1.0 cm} $ \rhoe $ = 0.57   }         \\
\hline
\hline
\end{tabular}  
\end{center}
\caption[]{ Results of the QCD fits in the $\ln R $  matching scheme for
            the individual observables together with the individual sources 
            of uncertainties and the $ \chi^2 / n_{df} $ for the \asmz fits. 
            The total error is the quadratic sum of the 
            experimental error (statistical and systematic uncertainty), 
            the uncertainty due to the hadronization correction, the
            uncertainty due to the scale dependence of \asmz and the 
            uncertainty due to the matching ambiguity. 
            Also listed is the weighted average of \asmz for the 
            6 observables together with the $ \chi^2 / n_{df} $ for the
            averaging procedure and the correlation parameter 
            $ \rhoe $. The $ \chi^{2} $ corresponds to the value before 
            readjusting according to Eq. \ref{chisquare}.    }
\label{lnrres}
\end{table}


Looking at the individual fit results, one finds from the $\chi^2 / n_{df} $ 
that most of the shape distributions cannot be successfully described in a 
fit range expected to apply for the combined theory. 
The \asmz values are 
higher than for the fits in pure NLLA for all observables considered. In the 
case of $ 1-T $, $ C $ and $ B_{sum} $ the measured \asmz values are even 
above the values for both the pure NLLA fits and the \oass fits using 
experimentally optimized scales, where one naively might expect the matched 
predictions to be a kind of `average' of the individual theories.  \\ 

\begin{sidewaysfigure} 
\begin{center}
\hspace{-1. cm}
\mbox{\epsfig{file=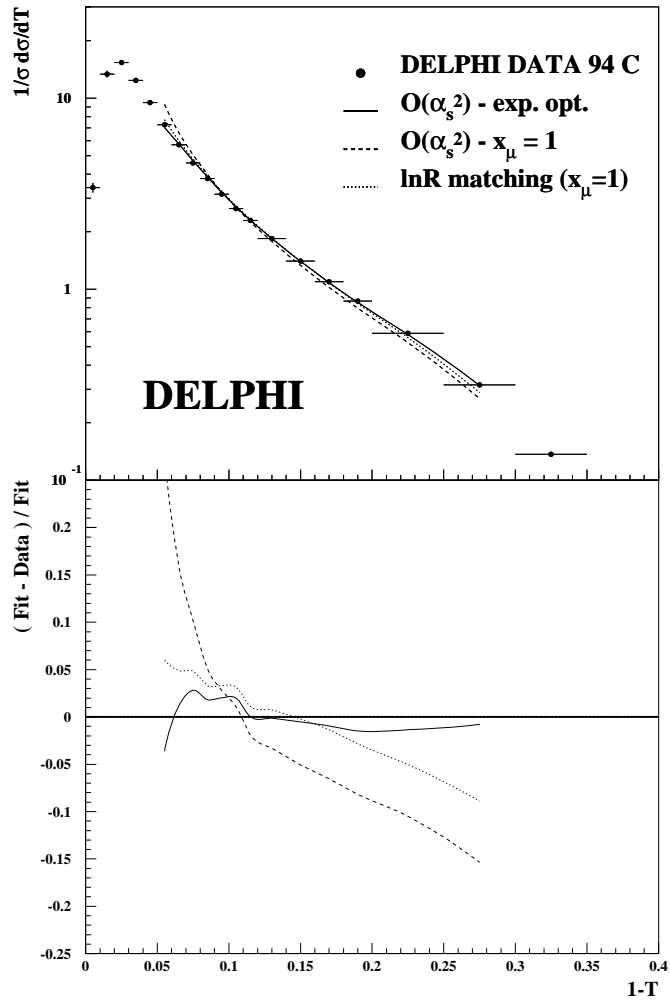,width=10.0cm,height=14.cm}}
\mbox{\epsfig{file=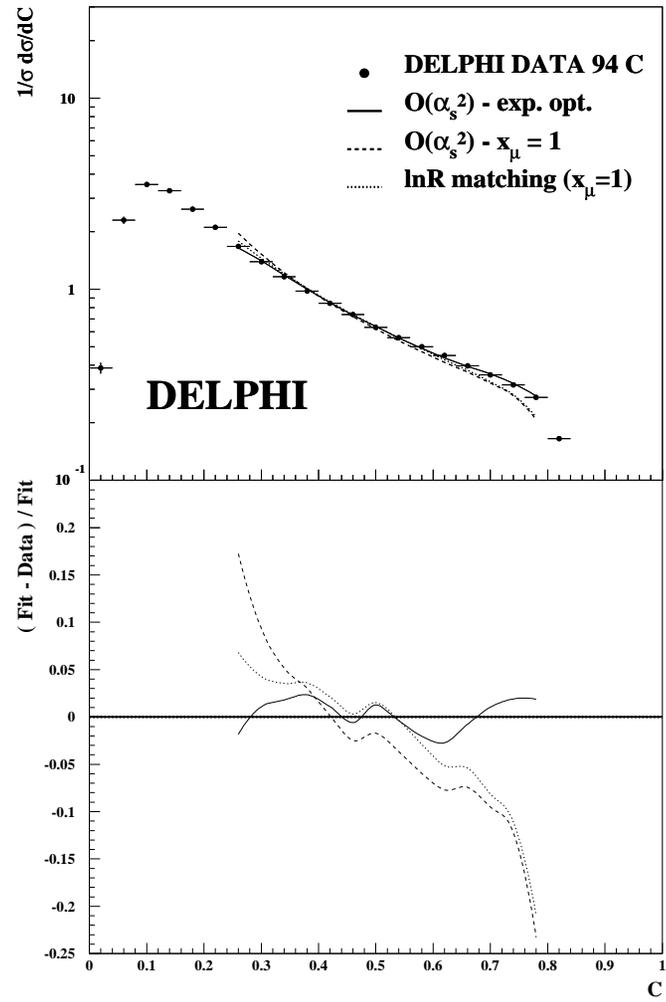,width=10.0cm,height=14.cm}}
\hspace{-1. cm}
\caption[]{ {\it left part:}
            Comparison of DELPHI data with three different QCD Fits: 
            {\it i }) \oass using an experimentally optimized 
            renormalization scale, {\it ii}) \oass using a fixed 
            renormalization scale $ x_{\mu} = 1 $ and
            {\it iii}) $ \ln R $ matched NLLA ($ x_{\mu}=1 $) 
            for the thrust Distribution.
            The lower part shows the relative difference (Fit-Data)/Fit.
            Whereas the \oass curve describes the data over the whole
            fit range, the slope of the curves for the fixed scale 
            and $ \ln R $ matching show a similar 
            systematic distortion with respect to the data.
            {\it right part:}
            The same for the C-Parameter. Here the distortion 
            is even stronger.   }
\label{lnrvgl}
\end{center}
\end{sidewaysfigure}

In order to investigate this result further it is instructive
to compare the theoretical predictions of the shape distributions for 
the different methods with the data distributions. Figure \ref{lnrvgl} 
shows experimental distributions for $ 1-T $ and $ C $ in comparison with the 
fitted curves for three different types of QCD fits, namely \oass using
experimentally optimized scales, \oass using a fixed renormalization scale
$ x_{\mu} $ = 1 and the fits in the $\ln R $ matching scheme. 
For the fits in \oass using experimentally optimized scales, the data are
described well over the whole fit range. For the fits in \oass using a fixed
renormalization scale and the fits in the $\ln R $ matching scheme, 
we find only a poor description and the slope of both 
curves show a similar systematic distortion with respect to the data. 
In the case of \oass applying $ x_{\mu} = 1 $ the distortion
arises from the wrong choice of the renormalization scale. Since the
scale value for the matched predictions is also chosen to be 
$ x_{\mu} = 1 $, the similarity of the curves indicates that the 
subleading and non-logarithmic terms originating from the \oass part
of the matched theory and introduced using the scale value 
$ x_{\mu} = 1 $ dominate the $ \ln R $ predictions. 
It should be noted that the matched theory requires a renormalization
scale value of $\cal O $ (1). Unlike the \oass case, 2 parameter fits in
\asmz and $ x_{\mu} $ do not converge for most of the observables; for such 
low scale values as in \oass the data can not be described at all in the 
matched theory.
It seems that the combination of all orders resummed predictions and 
terms only known in \oass results in a systematic shift in \asmz due
to the impossibility of choosing an appropriate renormalization scale 
value. Although the average values for the \oass fits, the fits in pure 
NLLA and the fits in the $\ln R $ matching scheme are in good agreement, 
the matched results should be considered to be less reliable than those of the
\oass and pure NLLA analyses due to the systematic deviation of the prediction 
to most of the data distributions (see e.g. Figure \ref{lnrvgl} and 
Table \ref{lnrres}).
    

\clearpage 

\section{Heavy Quark Mass Effects}
\label{qmassect}

Studies of the influence of quark mass effects on jet 
cross-sections \cite{impr-bmass} have shown that these effects can be important
in the study of event shape observables and hence in the measurement
of the strong coupling. For a natural mixture of quark flavours, the 
influence of the quark masses on the measurement of \asmz from event shapes 
is expected to be $ \sim 1 \% $ \cite{vancouver}. 
In order to derive a high precision value of \asmz, the \oass measurement
with experimentally optimized scales has been refined by considering  
the influence of $ b $ quark mass effects in leading order: 


\begin{table} [b]
\begin{center}
\begin{tabular} { l c c c }
\hline
\hline

Observable            &  \asmz        &  \das (Mass)  &  \das (Tot.)     \\       
\hline

$\rm EEC          $    &  0.1145       &  $\pm$ 0.0001 &  $\pm$ 0.0028    \\

$\rm AEEC         $    &  0.1149       &  $\pm$ 0.0001 &  $\pm$ 0.0111    \\

$\rm JCEF         $    &  0.1180       &  $\pm$ 0.0007 &  $\pm$ 0.0018    \\

$\rm 1-T          $    &  0.1136       &  $\pm$ 0.0001 &  $\pm$ 0.0036    \\  

$\rm O            $    &  0.1184       &  $\pm$ 0.0006 &  $\pm$ 0.0057    \\  

$\rm C            $    &  0.1156       &  $\pm$ 0.0002 &  $\pm$ 0.0036    \\  

$\rm B_{Max}      $    &  0.1223       &  $\pm$ 0.0002 &  $\pm$ 0.0041    \\  

$\rm B_{Sum}      $    &  0.1144       &  $\pm$ 0.0003 &  $\pm$ 0.0053    \\  

$\rm \rho_H       $    &  0.1216       &  $\pm$ 0.0001 &  $\pm$ 0.0060    \\  

$\rm \rho_S       $    &  0.1160       &  $\pm$ 0.0001 &  $\pm$ 0.0029    \\  

$\rm \rho_D       $    &  0.1196       &  $\pm$ 0.0008 &  $\pm$ 0.0039    \\  

$\rm D_2^{E0}     $    &  0.1165       &  $\pm$ 0.0002 &  $\pm$ 0.0044    \\  

$\rm D_2^{P0}     $    &  0.1207       &  $\pm$ 0.0002 &  $\pm$ 0.0033    \\  

$\rm D_2^{P}      $    &  0.1186       &  $\pm$ 0.0001 &  $\pm$ 0.0046    \\  

$\rm D_2^{Jade}   $    &  0.1182       &  $\pm$ 0.0005 &  $\pm$ 0.0041    \\  

$\rm D_2^{Durham} $    &  0.1172       &  $\pm$ 0.0003 &  $\pm$ 0.0026    \\  

$\rm D_2^{Geneva} $    &  0.1216       &  $\pm$ 0.0013 &  $\pm$ 0.0310    \\  

$\rm D_2^{Cambridge} $ &  0.1176       &  $\pm$ 0.0006 &  $\pm$ 0.0026    \\  

\hline
average                & \spb $ 0.1174 \pm 0.0026 $ &
                  \multicolumn{2}{l} { \hspace{0.3 cm} \chindf = 6.60 / 17 
                                       \hspace{0.5 cm} $\rhoe $ = 0.615 }  \\ 
\hline
\hline
\end{tabular}
\end{center}
\caption[]{ Results of the refined \oass measurement of \asmz including 
            $ b $ quark mass effects in leading order. The central value
            of \asmz quoted is the average of \as derived from applying 
            the $ b $ pole mass $ M_b $ and the $ b $ running mass 
            $ m_b(M_Z) $ definition. $ \Delta \alpha_s(Mass) = 
            \left| \alpha_s(M_b) - \alpha_s(m_b(M_Z)) \right| / 2 $ has been 
            taken as an estimate of the uncertainty due to quark mass effects
            of higher orders.  
            The total error displayed is the quadratic sum of the 
            experimental uncertainty, the hadronization  uncertainty, 
            the scale uncertainty and the uncertainty due to quark mass  
            effects. The $ \chi^{2} $ given, corresponds to the value before 
            readjusting according to Eq. \ref{chisquare}. }
\label{alphas-b}
\end{table}


\begin{equation} 
     \frac { 1 } { \sigma_{tot} } 
     \frac { d^{2} \sigma (Y, \cos \vartheta_T ) } 
           { dY d \cos \vartheta_T } =     
     \bar{\alpha}_s ( \mu^2 ) \cdot  
     A (Y, \cos \vartheta_T ) \big[ 1 + \Delta m \big] +
     \bar{ \alpha_s}^2 ( \mu^2 ) \cdot 
     B (Y, \cos \vartheta_{T}, x_{\mu} )  
\label{thmass}
\end{equation} 

\pagebreak
\nin where

\begin{equation} 
\Delta m = 
\frac{\displaystyle \sum_{q,m_b\ne 0} d^2 \sigma_q (Y,\cos \theta_T)/
      \sum_{q,m_b\ne 0} \sigma_q}
     {\displaystyle \sum_{q,m_b=0} d^2 \sigma_q (Y,\cos \theta_T)/
      \sum_{q,m_b=0} \sigma_q} - 1
\label{qmassratio}
\end{equation}     

\hspace{1. cm} \\

The coefficients $ \Delta m $ for the 18 observables studied have been 
computed numerically \cite{german} for two different definitions of the 
b-quark mass: The $ b $ pole mass of 
$ M_b \thickapprox 4.6 \hspace{0.1 cm} {\rm GeV/c^2} $ 
and the $ b $ running mass at the $ M_Z $, 
$ m_b(M_Z) \thickapprox 2.8 \hspace{0.1 cm} {\rm GeV/c^2} $ \cite{impr-bmass}. 
All definitions of the quark mass are equivalent to leading order; 
differences are entirely due to higher orders in $ \alpha_s $.  
\asmz has been determined applying both mass definitions. 
The average of \asmz derived from the pole mass definition and the running mass 
definition has been taken as the central value, and half the difference 
has been taken as an estimate of the uncertainty due to higher order mass effects. 
The results for the individual observables are listed in table \ref{alphas-b}. 


\section{Summary}    

From 1.4 Million hadronic $\rm Z_0 $ decays recorded with the DELPHI detector
and reprocessed with improved analysis software, the distributions of 18
infrared and collinear safe observables have been precisely measured
at various values of the polar angle $\rm \vartheta_T $ of the thrust axis
with respect to the beam direction. The $\rm \vartheta_T $ dependence of all
detector properties has been taken fully into account to achieve the best
possible experimental precision. In order to compare with QCD 
calculations in \oass, hadronization corrections are evaluated
from precisely tuned fragmentation models.  \\

The precise data are used to measure \asmz applying a number of different
methods described in the literature. The most detailed studies have been
performed in second order pertubative QCD. Fits taking explicit account 
of the \as dependent event orientation as predicted by QCD with 
experimental acceptance corrections less than $ \sim 25 \% $ and 
hadronization corrections less than $ \sim 40 \% $ yield the result 
that the data can be surprisingly well described in \oass by using 
a common value of \asmz with a small uncertainty.  \\  

Taking account of the correlation among the observables an average value 
of $\alpha_s(M_Z^2) = 0.1168 \pm 0.0026 $ is obtained from the data. 
The consistency of the individual \asmz determinations
from the 18 shape distributions is achieved by using the different values 
of the renormalization scales as obtained from the individual fits,
i.e. applying the so called experimental optimization method.
The significance of the fits is improved due to the large number
of data points per distribution. Thus, definite results concerning 
the choice of the optimal renormalization scale value become possible.
It should be pointed out that for most of the investigated observables
the scale dependence of \as is very small in the vicinity of the
experimentally optimized scale. The quoted error
of \asmz includes the uncertainty due to a variation of the 
experimentally optimized scale in the range between 
$ 0.5 \cdot x_{\mu}^{exp} $ and $ 2. \cdot x_{\mu}^{exp} $. \\

An analysis with a fixed renormalization scale value of $ x_{\mu} = 1 $
yields an unacceptable description of the data for many observables
and leads to a wide spread of the \asmz values. In contrast to fits 
applying experimentally optimized scales, the stability of \asmz 
with respect to the choice of the fit range is in general quite poor.
Due to the improved accuracy of the data, systematic differences 
between the \oass analysis using experimentally optimized and 
fixed renormalization scale values $ x_{\mu} = 1 $ become visible. 
These differences propagate also into the matched NLLA-analysis.   \\

To check the reliability of the \as results obtained from the 
experimentally optimized scales three further approaches for
choosing an optimized value of the renormalization scale have been 
investigated: The principle of minimal sensitivity (PMS), the method
of effective charges (ECH), and the method of Brodsky, Lepage and
MacKenzie (BLM). The weighted averages of \as from the three methods
are in excellent agreement with the weighted average of \as obtained
from the experimental optimization, but their scatter is larger.
The scatter is largest for BLM. A significant correlation between the 
renormalization scale values evaluated with ECH and PMS with the 
experimentally optimized scale values is observed. No such correlation 
exists for the BLM scales. \\

Further approaches to estimate the influence of higher order contributions
to the perturbative QCD series are based on Pad\'{e} approximants. 
The $[0/1]$ Pad\'{e} approximant 
has been used as an estimate of the sum of the perturbative series as well as
for the extrapolation of the unknown \oasss coefficients for the 18 distributions.
In both studies the renormalization scale has been set
to $ x_{\mu} = 1 $. Again the average values of \asmz are consistent with the
average value from the experimental scale optimization in \oass.  \\

 While all above mentioned determinations of \asmz use fixed order 
perturbation theory the last part of the paper describes measurements of 
\asmz using all orders resummed calculations in the next-to-leading 
logarithmic approximation (NLLA). In a first step, pure NLLA predictions
have been confronted with the data in a limited fit range where the ratio 
of the resummed next-to-leading logarithms to the non-exponentiating 
\oass contributions is large. The very good agreement between the 
average value of \asmz obtained from the pure NLLA fits with a 
renormalization scale value $ x_{\mu} =1 $ and the \oass fits using 
experimentally optimized scales is remarkable. In a further step
NLLA matched to \oass calculations have been applied. The corresponding 
average value of \asmz is again consistent with the \oass result though
the \as values from all investigated observables are systematically 
higher. More importantly, the application of matched NLLA to the high
precision data reveals that the trend of the data deviates in a systematic 
fashion from the predictions of the matched theory. This problem has not
previously been observed. The matched results should be considered 
to be less reliable than those of the \oass and pure NLLA analyses.\\ 

The \as values derived from the different approaches considered
are in very good agreement. The \oass analysis applying a simultaneous  
fit of $\alpha_s$ and $ x_{\mu} $ to the experimental data yields superior 
results in all respects. \\

In a final step the influence of heavy quark mass 
effects on the measurement of \asmz has been studied. The weighted average
of \asmz from the \oass measurements using experimentally optimized
renormalization scale values and corrected for the $b-\rm{mass}$ to
leading order yields  \\ 

\begin{center}
 $ \rm \alpha_s(M_Z^2) = 0.1174 \pm 0.0026 $.
\end{center}

The result is consistent with the result of a previous DELPHI publication
\cite{DelPubs1} if compared with partonshower hadronization corrections. \\

Among the observables studied, the Jet Cone Energy Fraction (JCEF) \cite{JCEF}
naturally reveals some superior properties. First, the size of the hadronization 
correction is extremely small, the average correction within the applied fit range 
being only about $ 3.5 \% $. 
Furthermore, the second order contribution to the cross-section is quite small. 
Within the applied fit range, 
the average ratio of the second to first order contributions is 
$ \left< r_{NLO} \right> \sim 25 \%  $ if $ x_{\mu} $ is fixed at 1 and
only $ \left< r_{NLO}  \right> \sim 6 \% $ if the experimentally
optimized scale value is applied. This indicates a good convergence behavior of the 
corresponding perturbative series. Also the scale 
dependence of \asmz derived from JCEF is very small. Both experimental 
and theoretical uncertainties are smallest for the \asmz measurement from the
JCEF distribution. \\


\begin{table} [tbp]
\begin{center}
\begin{tabular}{l c c c}
\hline
\hline  
prediction                     & \asmz                 & $ x_{\mu} $ & \chindf \\ 
\hline
\oass exp. opt. scale          & $ 0.1169 \pm 0.0017 $ & 0.0820      & 1.05 \\
ECH/FAC                        & $ 0.1169 \pm 0.0017 $ & 0.2189      & 2.75 \\
PMS                            & $ 0.1168 \pm 0.0017 $ & 0.1576      & 1.91 \\
Pad\'{e} \oasss (fixed scale)      & $ 0.1169 \pm 0.0016 $ & 1.0         & 3.12 \\
Pad\'{e} Sum (fixed scale)         & $ 0.1169 \pm 0.0015 $ & 1.0         & 3.05 \\
Pad\'{e} \oasss (exp. opt. scale)  & $ 0.1164 \pm 0.0015 $ & 0.1814      & 2.45 \\
\oass (fixed scale)            & $ 0.1191 \pm 0.0024 $ & 1.0         & 7.7  \\
\hline
\oass exp. opt. scale          & $ 0.1180 \pm 0.0018 $ &             &      \\
+ LO quark mass effects        &                       &             &      \\ 
\hline
\hline
\end{tabular}
\end{center}
\caption{Summary of \asmz measurements from the distribution of the 
         Jet Cone Energy Fraction (JCEF). }
\label{jcefsummar}
\end{table}


Table \ref{jcefsummar} shows a summary of \as measurements from 
the JCEF distribution for the different methods. The \oass fit
of \asmz applying a fixed renormalization scale value $ x_{\mu} = 1 $ clearly 
fails to describe the data with \chindf = 965/125.
However, even for this method, the deviation of \asmz 
from the value obtained using the experimentally optimized scale 
value is only about 2 \%. All other approaches yield nearly identical 
\as values within a few per mille. The deviation of the \as values
derived from the different methods, which can serve as an estimate of the theoretical 
uncertainty due to missing higher order terms \cite{asstatus}, is clearly smaller
than the uncertainty of $ \pm 0.0008 $ derived from the variation of the 
renormalization scale value. Due to the outstanding qualities of this observable,
the JCEF is considered as best suited for a precise determination of \asmzx 
After correcting the measured value for heavy quark mass effects, the final result is \\

\begin{center}
$ \rm \alpha_s(M_Z^2) = 0.1180 \pm 0.0006 (exp.) \pm 0.0013 (hadr.) 
                               \pm 0.0008 (scale) \pm 0.0007 (mass)$.
\end{center}

Comparing this result with other recent precision measurements of \asmz, there is
very good agreement with the determination 
of \asmz = $ 0.1174 \pm 0.0024 $  from 
Lattice Gauge Theory \cite{LGTres} and the recent result from an NNLO analysis of 
$ ep $ deep inelastic scattering data 
of \asmz $ = 0.1172 \pm 0.0024 $ \cite{epdisres}.
The result is also in good agreement with the result from the LEP electroweak
working group of \asmz $ = 0.119 \pm 0.004 $ \cite{ewres} from the standard 
model fit to the full set of electroweak precision data.  
Compared with the most recent result from spectral functions in hadronic tau decays, \as 
is smaller than the central value of \asmz $ = 0.1219 \pm 0.0020 $
quoted in \cite{taures}, but in very good agreement with the value of 
\asmz $ = 0.1169 \pm 0.0017 $ derived in \cite{taures} 
by using an alternative analysis method considering renormalon chains.

\pagebreak

\section*{Acknowledgements}

We thank M. Seymour for providing us with the EVENT2 generator and for 
useful discussions. We further thank P. Aurenche, S. Catani, J. Ellis 
and P. Zerwas for critical comments and stimulating discussions.\\
 We are greatly indebted to our technical 
collaborators, to the members of the CERN-SL Division for the excellent 
performance of the LEP collider, and to the funding agencies for their
support in building and operating the DELPHI detector.\\
We acknowledge in particular the support of \\
Austrian Federal Ministry of Science and Traffics, GZ 616.364/2-III/2a/98, \\
FNRS--FWO, Belgium,  \\
FINEP, CNPq, CAPES, FUJB and FAPERJ, Brazil, \\
Czech Ministry of Industry and Trade, GA CR 202/96/0450 and GA AVCR A1010521,\\
Danish Natural Research Council, \\
Commission of the European Communities (DG XII), \\
Direction des Sciences de la Mati$\grave{\mbox{\rm e}}$re, CEA, France, \\
Bundesministerium f$\ddot{\mbox{\rm u}}$r Bildung, Wissenschaft, Forschung 
und Technologie, Germany,\\
General Secretariat for Research and Technology, Greece, \\
National Science Foundation (NWO) and Foundation for Research on Matter (FOM),
The Netherlands, \\
Norwegian Research Council,  \\
State Committee for Scientific Research, Poland, 2P03B06015, 2P03B03311 and
SPUB/P03/178/98, \\
JNICT--Junta Nacional de Investiga\c{c}\~{a}o Cient\'{\i}fica 
e Tecnol$\acute{\mbox{\rm o}}$gica, Portugal, \\
Vedecka grantova agentura MS SR, Slovakia, Nr. 95/5195/134, \\
Ministry of Science and Technology of the Republic of Slovenia, \\
CICYT, Spain, AEN96--1661 and AEN96-1681,  \\
The Swedish Natural Science Research Council,      \\
Particle Physics and Astronomy Research Council, UK, \\
Department of Energy, USA, DE--FG02--94ER40817. \\

\newpage
